%% file: main.tex
\documentclass[11pt]{article}

\pdfoutput=1

\usepackage{a4wide}
\usepackage[margin=1.0in]{geometry} 
\usepackage[utf8]{inputenc}
\usepackage{stmaryrd} 
\usepackage{amsmath}
\usepackage{amssymb}
\usepackage{xspace}
\usepackage{graphicx}
\usepackage{amsthm}
\usepackage{complexity}
\usepackage{bm}
\usepackage{enumitem}
\usepackage{tikz}
\usetikzlibrary{positioning, shapes, calc}
\usepackage{subcaption} 
\usepackage{multirow} 
\usepackage{thm-restate}

 \usepackage{mathptmx}      

\theoremstyle{definition}
\newtheorem{theorem}{Theorem}[section]

\newtheorem{corollary}[theorem]{Corollary}

\newtheorem{definition}[theorem]{Definition}
\newtheorem{example}[theorem]{Example}

\newtheorem{lemma}[theorem]{Lemma}
\newtheorem{proposition}[theorem]{Proposition}
\newtheorem{remark}[theorem]{Remark}

\newcommand{\defeq}{\stackrel{\scriptscriptstyle\text{def}}{=}}

\newcommand{\ie}{\text{i.e.}\xspace}

\newcommand{\etal}{\text{et al.}\xspace} 

\newcommand{\N}{\mathbb{N}}                    
\newcommand{\Z}{\mathbb{Z}}                    
\newcommand{\set}[1]{\left\{#1\right\}}        

\newcommand{\zero}{\mathbf{0}}

\newcommand{\multiset}[1]{\Lbag#1\Rbag}        

\newcommand{\pop}[1]{\mathrm{Pop}(#1)}         
\renewcommand{\PP}{\mathcal{P}}                  
\newcommand{\trans}[1]{\xrightarrow{#1}}       
\newcommand{\NN}{\mathcal{N}}                  
\newcommand{\Conf}{\textit{Conf}}               
\newcommand{\Step}{\mathit{Step}}        
\newcommand{\send}[1]{\xrightarrow{#1 +}}  
\newcommand{\MS}{S}                     
\newcommand{\Cons}[1]{Con_{#1}}   
\newcommand{\Stab}[1]{St_{#1}}   

\newcommand{\prezm}{\mathit{pre}_z} 
\newcommand{\postzm}{\mathit{post}_z} 
\newcommand{\prestar}{\mathit{pre}^*}
\newcommand{\poststar}{\mathit{post}^*}
\newcommand{\prezmstar}{\prezm^*}
\newcommand{\postzmstar}{\postzm^*}

\newcommand{\compl}[1]{\overline{#1}}

\newcommand{\vect}[1]{\vec{#1}}

\newcommand{\unorm}[1]{\|{#1}\|_u}
\newcommand{\lnorm}[1]{\|{#1}\|_l}
\newcommand{\sem}[1]{\llbracket{#1}\rrbracket}

\newcommand{\pass}[2]{\textit{off}[#1,#2]}
\newcommand{\act}[2]{\textit{on}[#1,#2]}
\newcommand{\stable}[2]{\textit{at}[#1,#2]}
\newcommand{\switch}[4]{\textit{move}[#1,#2,#3,#4]}

\newcommand{\B}{\mathcal{B}}
\newcommand{\increment}{\texttt{increment}}
\newcommand{\one}{\texttt{one}}
\newcommand{\free}{\texttt{free}}
\newcommand{\token}{\texttt{token}}
\newcommand{\final}{\texttt{final}}

\usepackage[final]{commenting} 

\declareauthor{je}{Javier}{blue}
\declareauthor{mr}{Mikhail}{red!70}
\declareauthor{cwk}{Chana}{green!60!black}
\declareauthor{sj}{Stefan}{orange!90!black}

\authorcommand{je}{comment}
\authorcommand{mr}{comment}
\authorcommand{cwk}{comment}
\authorcommand{sj}{comment}

\begin{document}

\title{The Complexity of Verifying Population Protocols \thanks{Previous versions of some of the results of this paper were published in \cite{EsparzaGMW18} and \cite{EsparzaRW19}. } \thanks{Funded by the European Research Council (ERC) under the European Union’s Horizon 2020 research and innovation programme under grant agreement No 787367 (PaVeS)}}

\author{Javier Esparza\footnote{\texttt{esparza@in.tum.de} (corresponding author), Technical University of Munich (Germany), ORCID: 0000-0001-9862-4919}
\and Stefan Jaax\footnote{\texttt{jaax@in.tum.de}, Technical University of Munich (Germany), ORCID: 0000-0001-5789-8091}
\and Mikhail Raskin\footnote{\texttt{raskin@in.tum.de}, Technical University of Munich (Germany), ORCID: 0000-0002-6660-5673}
\and Chana Weil-Kennedy\footnote{\texttt{chana.weilkennedy@in.tum.de}, Technical University of Munich (Germany), ORCID: 0000-0002-1351-8824}}

\date{\today}

\maketitle

\begin{abstract}
Population protocols [Angluin \etal, PODC, 2004] are a model of distributed computation in which indistinguishable, finite-state agents interact in pairs to decide if their initial configuration, i.e., the initial number of agents in each state, satisfies a given property.
In a seminal paper Angluin \etal classified population protocols according to their communication mechanism, and conducted an exhaustive study of the expressive power of each class, that is, of the properties they can decide [Angluin \etal, Distributed Computing, 2007].
In this paper we study the correctness problem for population protocols, \ie, whether a given protocol decides a given property.
A previous paper [Esparza \etal, Acta Informatica, 2017] has shown that the problem is decidable for the main population protocol model, but at least as hard as the reachability problem for Petri nets, which has recently been proved to have non-elementary complexity.
Motivated by this result, we study the computational complexity of the correctness problem for all other classes introduced by Angluin \etal, some of which are less powerful than the main model.
Our main results show that for the class of observation models the complexity of the problem is much lower, ranging from $\Pi_2^p$ to \PSPACE.
%
\end{abstract}

\section{Introduction}\label{sec:intro}
\input{intro}

\section{Protocol Models}\label{sec:models}
\input{prelim}
\subsection{A Unified Model}
\label{subsec:unified}
\input{unified-model}
\subsection{Immediate Delivery Models}\label{subsec:immdeliv}
\input{immediate-delivery}
\subsection{Delayed Delivery Models}\label{subsec:deldeliv}
\input{delayed-delivery}
\subsection{Expressive Power and Correctness Problem}\label{sec:expressivity}
\input{expressivity}
\subsection{Correctness as a Reachability Problem}
\label{sec:formalizing-correctness}
\input{formalizing-correctness}

\section{Lower Bounds for Observation Models}\label{sec:observation-hardness}
\input{intro-observation-hardness}
\subsection{Correctness of IO Protocols is \PSPACE-hard}
\label{sec:io-hardness}
\input{io-hardness}
\subsection{Correctness of DO Protocols is $\Pi_2^p$-hard}
\label{sec:mfdo-do-hardness}
\input{do-hardness}

\section{Reachability in Observation Models: The Pruning and Shortening Theorems}\label{sec:reach-obs}
\input{intro-reach-obs}

\subsection{An auxiliary model: Message-Free Delayed-Observation Protocols}
\label{sec:mfdo}
\input{mfdo-intro-JE}
\subsection{Pruning Theorems for IO and MFDO Protocols}
\label{sec:pruning}
\input{new-pruning-setting-JE}
\input{pruning-theorems}
\subsection{Shortening Theorem for MFDO Protocols}
\label{sec:shortening}
\input{shortening-q4-MR}

\subsection{Pruning and Shortening Theorems for DO Protocols}
\label{sec:pruning-and-shortening-DO}
\input{pruning-and-shortening-DO}
\section{Set Reachability in Observation Models: The Closure Theorems}\label{sec:sym-reach-obs}
\label{sec:small-instance}
\input{small-instance-section}
\section{Upper Bounds for Observation Models} \label{sec:coNP}
%
\input{correctness-intro-CWK}
\subsection{Correctness of IO Protocols is in \PSPACE}
\label{sec:PSPACE}
\input{io-correctness-sans-reachability-CWK}
\subsection{Correctness of DO Protocols is in $\Pi_2^p$}
\input{mfdo-correctness-CWK}


\section{Hardness and Decidability of Correctness for Transmission-Based Models}
\label{sec:transmission-decidability}
\subsection{Correctness of Transmission-Based Models is TOWER-Hard}\label{sec:transmission}
\input{sec-tower-hardness-transition}

\subsection{Decidability of Correctness for PP and DT Protocols}
\label{sec:decidability}
\input{section-decidability-CWK}
\subsection{Correctness in Probabilistic Models}
\label{sec:decidability-probabilistic}
\input{decidability-probabilistic}

\section{Related models and approaches}
\label{sec:related}
\input{related}

\section{Conclusion}
\label{sec:conclusion}
\input{conclusion}

\section*{Acknowledgements}
Pierre Ganty and Rupak Majumdar co-authored one of the publications (reference \cite{EsparzaGMW18}) on which part of this paper is based.
We thank them for very useful discussions.
We also thank the anonymous reviewers for helpful feedback.

\bibliographystyle{plain}
\bibliography{references,ref}

\appendix
\input{appendix-obs-3-1}

\input{appendix-obs-3-2}

\input{appendix-obs-4-2}

\input{appendix-obs-4-3}

\input{app-lower-bounds}
\input{app-decidability}
\end{document}

%% file: intro.tex
Population protocols are a theoretical model for the study of ad hoc
networks of tiny computing devices without any infrastructure~\cite{AADFP04,AADFP06}. The
model postulates a ``soup'' of indistinguishable, finite-state agents that behave
identically. Agents repeatedly interact in pairs, changing their states according to a joint transition function.
A global fairness condition ensures that every global configuration that is reachable infinitely often is also reached infinitely often.
The purpose of a population protocol is to allow
agents to collectively compute some information about their initial configuration,
defined as the function that assigns to each local state the number of agents
that initially occupy it. For example, assume that initially each agent picks a
boolean value by choosing, say, $q_0$ or $q_1$ as its initial state.
The (many) \emph{majority protocols} described in the
literature allow the agents to eventually reach a stable consensus on
the value chosen by a majority of the agents. More formally, let $x_0$
and $x_1$ denote the initial numbers of agents in states $q_0$ and $q_1$;
majority protocols compute the predicate $\varphi(x_0,
x_1) \colon \N \times \N \to \{0, 1\}$ given by $\varphi(x_0, x_1)
= (x_1 \geq x_0)$. Throughout the paper, we use the term ``predicate'' as an abbreviation
for ``function from $\N^k$ to $\{0,1\}$ for some $k$''.

The expressive power of population protocols (that is, which predicates they can compute), and their efficiency (how fast they can compute them) have been both extensively studied (see e.g.~            \cite{AlistarhAEGR17,AlistarhAG18,AlistarhG18,ElsasserR18}). In a seminal paper~\cite{AAER07}, Angluin \etal  showed that population protocols can compute exactly the predicates definable in Presburger arithmetic. In the same publication, they observed that while the two-way communication discipline of the standard population protocol model is adequate for natural computing applications, where agents represent molecules or cells that communicate by means of physical encounters, it is less so when agents represent electronic devices, where communication usually takes place by asynchronous message-passing, and information flows only from the sender to the receiver. For this reason, they also conducted a thorough investigation of the expressive power of the population protocol model when two-way communication is replaced by           one-way communication.            They classified one-way communication models into \emph{transmission} models, where the sender is allowed to change its state as a result of sending a message, and \emph{observation} models, where it is not.  Intuitively, in observation models the receiver observes the state of the sender, who may not even be aware that it is being observed. Further, they distinguished between \emph{immediate delivery} models,  where a   send event and its corresponding receive event                occur simultaneously,  \emph{delayed delivery} models, where delivery may take time, but receivers are always willing to receive any message,  and \emph{queued delivery} models, where delivery may take time, and                        receivers may choose to postpone incoming messages until they have sent a message themselves.  This results in five one-way models: immediate and delayed observation, immediate and delayed transmission, and queued transmission.
Angluin \etal showed that no one-way model is more expressive than the two-way model, and some of them are strictly less expressive. In fact, they characterized the expressive power of each model in terms of natural classes of Presburger predicates.

In this paper we investigate the correctness problem for population protocols, that is, the problem of deciding
if a given protocol computes a given Presburger predicate. For each possible input, deciding if the protocol reaches a consensus only requires to inspect one of these finite transition systems, and can be done automatically using a model checker.
This approach has been followed in \cite{DBLP:conf/tase/PangLD08,DBLP:conf/cav/SunLDP09,DBLP:conf/sss/ChatzigiannakisMS10,guidelines}, but it only proves the correctness of a protocol for a finite number of (typically small) inputs.
 The question whether the protocol reaches the right consensus for \emph{all} inputs remained open until 2015, when Esparza \etal showed that the problem is decidable \cite{EsparzaGLM17}. However, in the same paper they proved that the correctness problem is at least as hard as the reachability problem for Petri nets. This problem, which was known to be EXPSPACE-hard since the 1970s \cite{Lipton}, has recently been shown to be TOWER-hard \cite{CLLLM19}, where TOWER is the union of the classes of problems solvable in $k$-EXPTIME for every $k \geq 0$. Motivated by this high complexity of the two-way model, we examine the
complexity of the problem for the one-way models studied in \cite{AAER07}.  We show that, very satisfactorily, for observation models the complexity decreases dramatically. In our two main positive results, we prove that correctness is $\Pi_2^p$-complete for the delayed observation model, and \PSPACE-complete for the immediate observation model,
when predicates are specified as quantifier-free formulas of Presburger arithmetic\footnote{Since Presburger arithmetic admits quantifier elimination, the quantifier-free fragment is as expressive as the full language, if one adds divisibility predicates with constant divisor.}.
Surprisingly, we show that this is also the complexity of checking that the protocol is correct for one single given input. So, loosely speaking, in observation models checking correctness for one input and for all infinitely many possible inputs has the same complexity.


In the second part of the paper we present negative results on the transmission models: In all of them, correctness is at least as hard as the reachability problem for Petri nets, and thus TOWER-hard. Further, for the delayed delivery and queued delivery models the single input case is already TOWER-hard, while for the immediate transmission model the single-input problem is \PSPACE-complete. On the positive side, we show that the decidability proof of \cite{EsparzaGLM17} can be easily extended to the immediate and delayed transmission models, but not to the queued transmission model. In fact, for the queued transmission model we leave the decidability of the correctness problem as an open question. However, we also show that this question is less relevant for queued models than for the others. Indeed, in this model the fairness condition of \cite{AAER07} bears no immediate relation to the probabilistic interpretation of population protocols used in the literature in order to study their efficiency.  Table \ref{tab:complexity} summarizes the results and shows their places in the paper.

The paper is organized as follows. Section \ref{sec:models} recalls the protocol models introduced in \cite{AAER07}. Section  \ref{sec:observation-hardness} presents our lower bounds for observation models.
Sections \ref{sec:reach-obs}, \ref{sec:sym-reach-obs} and \ref{sec:coNP}, the most involved part of the paper, prove the results leading to the upper bounds for observation models. 
Section \ref{sec:transmission-decidability} contains the decidability and TOWER-hardness results for transmission-based models.
Section \ref{sec:related} gives a brief overview of the most closely related models and approaches that we are aware of.

Previous versions of some of the results of this paper were published in \cite{EsparzaGMW18} and \cite{EsparzaRW19}.

\mr{Maybe we need a more specific relative-contribution w.r.t. conference proceedings here}

\begin{table*}[ht]
\caption{Decidability and complexity results}
\begin{center}
\begin{tabular}{|l|l|l|l|l| }
\hline
 \multicolumn{3}{|c|}{\textbf{ Communication Model}} & \multicolumn{1}{c|}{\textbf{ Single-input corr.}} & \multicolumn{1}{c|}{\textbf{ All-inputs corr.}} \\
\hline
\multirow{7}{5em}{ One-way}& \multirow{2}{6em}{ Observation} & \ Immediate & \ \PSPACE-complete\ &  \ \PSPACE-complete\ \\ \cline{3-5}
& & \ Delayed\ \ & \ $\Pi_2^p$-complete & \ $\Pi_2^p$-complete \\ \cline{2-5} \cline{2-5}
& \multirow{5}{6.5em}{ Transmission} & \multirow{2}{5.5em}{ Immediate} & \multirow{2}{8.5em}{ \PSPACE-complete} & \multirow{2}{8em}{ TOWER-hard,\\ \ decidable}  \\
& & & &  \\ \cline{3-5}
& & \multirow{2}{4em}{ Delayed} & \multirow{2}{8em}{ TOWER-hard,\\ \ decidable} &  \multirow{2}{8em}{ TOWER-hard,\\ \  decidable}  \\
& & & & \\ \cline{3-5}
& & \ Queued\  & \ TOWER-hard\  & \ TOWER-hard \ \\  \cline{1-5}
\multirow{2}{5em}{ Two-way } & \multirow{2}{6.5em}{ Transmission}  & \multirow{2}{5.5em}{  Immediate } &  \multirow{2}{10.5em}{ \PSPACE-complete \cite{EsparzaGLM17}} & \multirow{2}{8em}{ TOWER-hard,\\ \ decidable \cite{EsparzaGLM17}}\\
& & & &\\ \cline{1-5}
\end{tabular}
\label{tab:complexity}
\end{center}
\end{table*}

%% file: prelim.tex
After some preliminaries (Section \ref{subsec:prelim}), we recall the definitions of the models introduced by Angluin \etal in \cite{AAER07} (Sections \ref{subsec:unified} to \ref{subsec:deldeliv}),
formalize the correctness problem (Section \ref{sec:expressivity}), and rephrase it in two different ways as a reachability problem (Section \ref{sec:formalizing-correctness}).

\subsection{Multisets and populations}
\label{subsec:prelim}

A \emph{multiset} on a finite set \(E\) is a mapping \(C \colon E \rightarrow \N\), i.e.                   \(C(e)\) denotes the number of occurrences of an element \(e \in E\) in \(C\).
Operations on \(\N\) are extended to multisets by defining them componentwise on each
element of \(E\). We define in this way the sum $C_1 + C_2$, comparison $C_1 \leq C_2$,
or maximum $\max \{C_1, C_2\}$ of two multisets $C_1, C_2$.
Subtraction, denoted $C_1 - C_2$, is allowed only if $C_1 \geq C_2$.
We let $|C|\defeq\sum_{e\in E} C(e)$ denote the total number of occurrences of elements in $C$,
also called the \emph{size} of $C$.
We sometimes write multisets using set-like notation. For example,  both $\multiset{a, 2 \cdot
  b}$ and $\multiset{a, b, b}$ denote the multiset $C$ such that $C(a) = 1$, $C(b) = 2$ and $C(e) = 0$ for every $e \in E \setminus
\{a, b\}$. Sometimes we use yet another representation, by assuming a total order $e_1 \prec e_2 \prec \cdots \prec e_n$ on $E$, and representing a multiset $C$ by the vector $(C(e_1), \ldots, C(e_n))\in \N^n$.

A \emph{population} $P$ is a multiset on a finite set $E$ with at least two elements, \ie $P(E)\geq 2$.
The set of all populations on $E$ is denoted $\pop{E}$.

%% file: unified-model.tex
We recall the unified framework for protocols introduced by Angluin \etal in \cite{AAER07},
which allows us to give a generic definition of the predicate computed by a protocol.

\begin{definition}
\label{def:unified}
A \emph{generalized protocol} is a quintuple $\PP = (\Conf, \Sigma, \Step, I, O)$ where
\begin{itemize}
\item $\Conf$ is a countable set of \emph{configurations}.
\item $\Sigma$ is a finite \emph{alphabet} of \emph{input symbols}. The elements of
$\pop{\Sigma}$ are called \emph{inputs}.
\item $\Step \subseteq \Conf \times \Conf$ is a reflexive \emph{step relation}, capturing when a first configuration can reach another one in one step.
\item $I \colon \pop{\Sigma} \rightarrow \Conf$ is an \emph{input function} that assigns to every
input an initial configuration.
\item $O \colon \Conf \rightarrow \{0,1\}$ is a partial \emph{output function} that assigns an output  to each configuration on which it is defined.
\end{itemize}
\end{definition}
We write $C \trans{} C'$ and $C \trans{*} C'$ to denote $(C, C') \in \Step$ and $(C, C') \in \Step^*$, respectively. 
We say $C'$ is \emph{reachable} from $C$ if $C \trans{*} C'$.
An \emph{execution} of $\PP$ is a (finite or infinite) sequence of configurations $C_0, C_1, \ldots$ such that $C_j \trans{} C_{j+1}$ for every $j \geq 0$.  Observe that, since we assume that the step relation is reflexive, all maximal executions (i.e., all executions that cannot be extended) are infinite.

An execution $C_0, C_1, \ldots$ is \emph{fair} if for every $C \in \Conf$  the following property holds: If there exist infinitely many indices $i \geq 0$ such that $C_i \trans{*} C$, then there exist infinitely many indices $j \geq 0$ such that $C_j = C$. In words, in fair sequences every configuration which can be reached infinitely often is reached infinitely often.

%
%
%

A fair execution $C_0, C_1, \ldots $ \emph{converges to $b \in \{0,1\}$} if there exists an
index $m \geq 0$ such that for all $j \geq m$ the output function is defined on $C_j$
and $O(C_j) =b$.
A protocol outputs $b \in \{0,1\}$ for input $a \in \pop{\Sigma}$ if every fair execution starting at $I(a)$ converges to $b$. A protocol \emph{computes} a predicate $\varphi \colon \pop{\Sigma} \rightarrow \{0,1\}$ if it outputs $\varphi(a)$ for every input $a \in \pop{\Sigma}$.

The \emph{correctness problem} for a class of protocols consists of deciding for a given protocol
$\PP$ in the class, and a given predicate $\varphi \colon \pop{\Sigma} \rightarrow \{0,1\}$, where $\Sigma$ is the alphabet of $\PP$, whether $\PP$ computes $\varphi$. The goal of this paper is to determine
the decidability and complexity of the correctness problem for the classes of protocols introduced by Angluin \etal in \cite{AAER07}.

In the rest of the section we formally define the six protocol classes studied by Angluin \etal, and summarize the results of \cite{AAER07} that characterize the predicates they can compute.
Angluin \etal distinguish between models in which agents interact directly with each other, with zero-delay, and models in which agents interact through messages with possibly non-zero transit time.
We describe them in Sections \ref{subsec:immdeliv} and \ref{subsec:deldeliv}, respectively.

%% file: immediate-delivery.tex
In immediate interaction models, a configuration only needs to specify the current state of each agent. In delayed models, the configuration must also specify which messages are in transit. 
Angluin \etal study three immediate delivery models.

\paragraph{Standard Population Protocols (PP).}
Population protocols describe the evolution of a population of finite-state agents. Agents are indistinguishable, and interaction is two-way. When two agents meet, they exchange full information about their current states, and update their states in reaction to this information.

\begin{definition}
\label{def:standard-pp}
A \emph{standard population protocol} is a quintuple $\PP=(Q,\delta,\Sigma, \iota, o)$ where 
$Q$ is a finite set of states, $\delta \colon Q^2\rightarrow Q^2$ is the transition function,
$\Sigma$ is a finite set of input symbols, $\iota:\Sigma\rightarrow Q$ is the initial state mapping, and
$o \colon Q\rightarrow \{0,1\}$ is the state output mapping.
\end{definition}

Observe that $\delta$ is a total function, and so we assume that every pair of agents can interact, although the result of the interaction can be that the agents do not change their states.
Every standard population protocol determines a protocol in the sense of Definition \ref{def:unified} as follows, where $C, C' \in \pop{Q}$, $D \in \pop{\Sigma}$, and $b \in \{0,1\}$:
\begin{itemize}
\item the configurations are the populations over $Q$, that is,  $\Conf = \pop{Q}$; 
\item $(C,C') \in \Step$ if there exist states $q_1,q_2,q_3,q_4 \in Q$ such that $\delta(q_1,q_2)= (q_3,q_4)$, $C \geq \multiset{q_1, q_2}$, and $C'=C- \multiset{q_1, q_2} +\multiset{q_3, q_4}$.
        The inequality cannot be omitted because some of the states can coincide.
        \je{I've changed to multiset notation, I think it is clearer.}
\item $I(D)=\sum_{\sigma \in \Sigma} D (\sigma)\iota(\sigma)$; in other words, if the input $D$ contains $k$ copies of $\sigma \in \Sigma$, then the configuration $I(D)$ places $k$ agents in the state $\iota(\sigma)$;
\item $O(C)=b$ if $o(q)=b$ for all $q \in Q$ such that $C(q)>0$; in other words, $O(C)=b$ if in the configuration $C$ all agents are in states with output $b$. We often call a configuration $C$ satisfying this property a \emph{$b$-consensus}.
\end{itemize}

%

The two other models with immediate delivery are one-way. They are defined as subclasses of the standard population protocol model.

\paragraph{Immediate Transmission Protocols (IT).}
In these protocols, at each step an agent (the sender) sends its state to another agent (the receiver). Communication is immediate, that is, sending and receiving happen in one atomic step. The new state of the receiver depends on both its old state and the old state of the sender, but the new state of the sender depends only on its own old state, and not on the old state of the receiver. Formally:

\begin{definition}
        A standard population protocol $\PP=(Q,\delta,\allowbreak \Sigma,\allowbreak \iota, o)$ is an \emph{immediate transmission protocol} if there exist two functions $\delta_1: Q\rightarrow{}Q,\ \delta_2 \colon Q^2 \rightarrow Q$ satisfying $\delta(q_1,q_2)=(\delta_1(q_1), \delta_2(q_1,q_2))$ for every $q_1,q_2\in Q$.
\end{definition}

\paragraph{Immediate Observation Protocol (IO).}
In these protocols, the state of a first agent can be observed by a second agent, which updates its state using this information. Unlike in the immediate transmission model, the first agent does not
update its state (intuitively, it may not even know that it has been observed). Formally:

\begin{definition}
A standard population protocol $\PP=(Q,\delta,\allowbreak \Sigma,\allowbreak \iota, o)$ is an \emph{immediate observation protocol} if there exists a function $\delta_2 \colon Q^2 \rightarrow Q$ satisfying     
$\delta(q_1,q_2)=(q_1, \delta_2(q_1,q_2))$ for every $q_1,q_2\in Q$.
\end{definition}

\paragraph{Notation.} 
We sometimes write $q_1,q_2 \rightarrow q_3,q_4$ for $\delta(q_1,q_2)=(q_3, q_4)$. 
In the case of  IO protocols we sometimes write $q_2 \trans{q_1} q_4$ for $\delta(q_1,q_2)=(q_1, q_4)$, and say that the agent moves from $q_2$ to $q_4$ by observing $q_1$.

%% file: delayed-delivery.tex
In delayed delivery models agents communicate by sending and receiving messages. The set of messages that can be sent (and received) is finite. Messages are sent to and received from one
single pool of messages; in particular, the sender does not choose the recipient of the message.
The pool can contain an unbounded number of copies of a message.
Agents update their state after sending or receiving a message. Angluin \etal define the following three delayed delivery models.

\paragraph{Queued Transmission Protocols (QT).}  The set of messages an agent is willing to receive depends on its current state. In particular, in some states the agent may not be willing to receive any message.

\begin{definition}
\label{def:queued-transmission}
A \emph{queued transmission protocol} is a septuple $\PP=(Q,M,\delta_s,\delta_r,\Sigma, \iota, o)$ where
$Q$ is a finite set of states,
$M$ is a finite set of messages,
        $\delta_s \colon Q\rightarrow M\times Q$ is the partial send function,
        \cwk{why "partial" send?}
$\delta_r \colon Q\times M \rightharpoonup Q$ is the partial receive function,
$\Sigma$ is a finite set of input symbols,
$\iota \colon \Sigma\rightarrow Q$ is the initial state mapping, and
$o \colon Q\rightarrow \{0,1\}$ is the state output mapping.
\end{definition}

Every queued transmission protocol determines a protocol in the sense of Definition \ref{def:unified} as follows, where $C, C' \in \pop{Q}$, $D \in \pop{\Sigma}$, and $b \in \{0,1\}$:
\begin{itemize}
\item  the configurations are the populations over $Q\cup M$, that is,  $\Conf = \pop{Q\cup M}$;
\item $(C, C') \in \Step$ if there exist states $q_1,q_2$ and a message $m$ such that
\begin{itemize}
\item  $\delta_s(q_1)= (m,q_2)$, $C \geq \multiset{q_1}$, and
$C'=C- \multiset{q_1}+\multiset{m, q_2}$; or
\item $\delta_r(q_1,m)= q_2$, $C \geq \multiset{q_1,m}$, and
$C=C'- \multiset{q_1, m} + \multiset{q_2}$.
\end{itemize}
\item $I(D)=\sum_{\sigma \in \Sigma} C(\sigma)\iota(\sigma)$; notice that since $\iota$ does not map symbols of $\Sigma$ to $M$, the configuration $I(D)$ has no messages;
\item $O(C)=b$ if $o(q)=b$ for all $q \in Q$ such that $C(q)>0$.
\end{itemize}




\paragraph{Delayed Transmission Protocols (DT).}
DT protocols are the subclass of QT protocols in which, loosely speaking, agents can never refuse to receive a message. This is modeled by requiring the receive transition function to be total.

\begin{definition}
\label{def:queued-transmission}
A queued transmission protocol $\PP$ is a \emph{delayed transmission protocol} if
its receive function $\delta_r$ is a total function.
\end{definition}

\paragraph{Delayed Observation Protocols (DO).}

Intuitively, DO protocols are the subclass of DT-protocols in which ``sender'' and ``receiver'' actually means ``observee'' and ``observer''. This is modeled by forbidding the sender to change its state when it sends a message (since the observee many not even know it is being observed).

\begin{definition}
\label{def:queued-transmission}
Let $\PP=(Q,M,\delta_s,\delta_r,\Sigma, \iota, o)$ be a queued transmision protocol. $\PP$ is a \emph{delayed observation protocol} if $\delta_r$ is a total function and
for every $q \in Q$ the send funtion $\delta_s$ satisfies $\delta_s(q)=(m,q)$ for some $m\in M$.
\end{definition}

\paragraph{Notation.}
We write $q_1 \send{m} q_2$ when $\delta_s(q_1)=(m, q_2)$, and $q_1 \trans{m-} q_2$ when $\delta_r(q_1,m)= q_2$, denoting that the message $m$ is added to or removed from the pool of messages.  In the case of DO protocols, we sometimes write simply $q_1 \send{m}$.

The following fact follows immediately from the definitions, but is very important.

\paragraph{Fact.}  In immediate delivery protocols (PP, IT, IO),  if $C \trans{*} C'$ then $|C|=|C'|$. Indeed, in these models configurations are elements of $\pop{Q}$, and so the size of a configuration is the total number of agents, which does not change when transitions occur. In particular, for every configuration $C$ the number of configurations reachable from $C$ is finite.

In delayed delivery protocols (QT, DT, DO), configurations are elements of $\pop{Q \cup M}$,
and so the size of a configurations is the number of agents \emph{plus} the number of messages sent but not yet received. Since transitions can increase or decrease the number of messages, the number of configurations reachable from a given configuration can be infinite.

\medbreak
Table \ref{tab:transition} summarizes the different transition functions and restrictions of the models.

\input{transition-table}

%% file: transition-table.tex
\begin{table*}[ht]
\caption{Transition functions and restrictions of the five models.}
\begin{center}
\begin{tabular}{|l|l|l|}
\hline
 \ Standard Population Protocol (PP) & $\PP=(Q,\delta,\allowbreak \Sigma,\allowbreak \iota, o)$  & $\delta \colon Q^2\rightarrow Q^2$ \\
\hline
\multirow{3}{12em}{Immediate Transmission (IT)} & 
\multirow{2}{8em}{$\PP=(Q,\delta,\allowbreak \Sigma,\allowbreak \iota, o)$} &  
\multirow{2}{16em}{$\exists \delta_1, \delta_2$ such that $\forall q_1,q_2\in Q,$ \\ $\delta(q_1,q_2)=(\delta_1(q_1), \delta_2(q_1,q_2))$} \\
& & \\
         \cline{2-3}
        & \multicolumn{2}{|c|}{New sender state does not depend on receiver state}
\\
\hline
\multirow{3}{12em}{Immediate Observation (IO)} & 
\multirow{2}{8em}{$\PP=(Q,\delta,\allowbreak \Sigma,\allowbreak \iota, o)$} &  
\multirow{2}{16em}{$\exists \delta_2$ such that $\forall q_1,q_2\in Q,$ \\ $\delta(q_1,q_2)=(q_1, \delta_2(q_1,q_2))$} \\
& & \\
         \cline{2-3}
        & \multicolumn{2}{|c|}{Sender state does not change}
\\
\hline
        \multirow{3}{12em}{Queued Transmission (QT)} & 
$\PP=(Q,M,\delta_s,\delta_r,\Sigma, \iota, o)$ &  
$\delta_s \colon Q\rightarrow M\times Q$, $\delta_r \colon Q\times M \rightharpoonup Q$ \\
         \cline{2-3}
        & \multicolumn{2}{|c|}{\multirow{2}{24em}{An agent can send a message, or receive one. \\ The set of messages it can receive depends on its state.}} \\
        & &
        \\
\hline
        \multirow{2}{12em}{Delayed Transmission (DT)} & 
$\PP=(Q,M,\delta_s,\delta_r,\Sigma, \iota, o)$ &  
$\delta_s \colon Q\rightarrow M\times Q$, $\delta_r \colon Q\times M \rightarrow Q$  \\
         \cline{2-3}
        & \multicolumn{2}{|c|}{Each agent can receive every message}
        \\
\hline
\multirow{4}{12em}{Delayed Observation (DO)} & 
\multirow{2}{12em}{$\PP=(Q,M,\delta_s,\delta_r,\Sigma, \iota, o)$} &  
\multirow{2}{16em}{$\delta_s \colon Q\rightarrow M\times Q$, $\delta_r \colon Q\times M \rightarrow Q$ \\ 
$\forall q\in Q, \delta_s(q)=(m,q)$ for some $m\in M$} \\
& & \\
         \cline{2-3}
        & \multicolumn{2}{|c|}{\multirow{2}{20em}{Each agent can receive every message \\ and the sender state does not change.}}
        \\
        & & 
        \\
\hline
\end{tabular}
\label{tab:transition}
\end{center}
\end{table*}

%% file: expressivity.tex
Let $\Sigma = \{\sigma_1, \ldots, \sigma_n\}$ be a finite alphabet. We introduce the class of predicates $\varphi \colon \pop{\Sigma} \rightarrow \{0,1\}$ definable in Presburger arithmetic, the first-order theory of addition. 

A population $P \in \pop{\Sigma}$ is completely characterized by the number $k_i$ of occurrences of each input symbol $\sigma_i$ in $P$, and so we can identify $P$ with the vector $(k_1, \ldots, k_n)$. A predicate $\varphi \colon \pop{\Sigma} \rightarrow \{0,1\}$ is a \emph{threshold} predicate if there are coefficients $a_1, \ldots, a_n, b \in \Z$ such that $\varphi(k_1, \ldots, k_n) = 1$ if{}f $\sum_{i=1}^n a_i \cdot k_i < b$. The class of Presburger predicates is the closure of the threshold predicates under boolean operations and existential quantification. By the  well-known result that Presburger arithmetic has a quantifier elimination procedure, a predicate is Presburger if{}f it is a boolean combination of threshold and \emph{modulo} predicates,  defined as the predicates of the form $\sum_{i=1}^n a_i \cdot k_i \equiv_c b$ (see e.g. \cite{Cooper72}).  Abusing language, we call a boolean combination of threshold and modulo terms a \emph{quantifier-free Presburger predicate}.

In \cite{AAER07}, Angluin \etal characterize the predicates computable by the six models of protocols we have introduced. Remarkably, all the classes compute only Presburger predicates. More precisely:
\begin{itemize}
\item DO computes the boolean combinations of predicates of the form $x \geq 1$, where $x$ is an input symbol. This is the class of predicates that depend only on the presence or absence of each input symbol. 
\item IO computes the boolean combinations of predicates of the form $x \geq c$, where $x$ is an input symbol and $c \in \N$. 
\item IT and DT compute the Presburger predicates that are similar to a boolean combination of modulo predicates for sufficiently large inputs; for the exact definition of similarity we refer the reader to \cite{AAER07}.
\item PP and QT compute exactly the Presburger predicates.
\end{itemize}

The results of \cite{AAER07} are important in order to define the correctness problem. The inputs to the problem are a protocol \emph{and} a predicate. The protocol is represented by giving its sets of places, transitions, etc. However, we still need a finite representation for Presburger predicates. 
There are three possible candidates: full Presburger arithmetic, quantifier-free Presburger arithmetic, and semilinear sets. Semilinear sets are difficult to parse by humans, and no paper on population protocols uses them to describe predicates. Full Presburger arithmetic is very succinct, but its satisfiability problem lies between \textsf{2-}\NEXP\ and \textsf{2-}\EXPSPACE\  \cite{FischerR74,Ber80,Haase18}. Since the satisfiability problem can be easily reduced to the correctness problem,  choosing full Presburger arithmetic would ``mask'' the complexity
of the correctness problem in the size of the protocol for several protocol classes. This leaves quantifier-free Presburger arithmetic, which also has several advantages of its own. First,  standard predicates studied in the literature (like majority, threshold, or remainder predicates) are naturally expressed without quantifiers. Second, there is a synthesis algorithm for population protocols that takes a quantifier-free Presburger predicate as input and outputs a population protocol (not necessarily efficient or succinct) that computes it \cite{AADFP04,AADFP06}; a recent, more involved algorithm even outputs a protocol with polynomially many states in the size of the predicate 
\cite{BlondinEGHJ20}. Third, the satisfiability problem for quantifier-free Presburger predicates is ``only'' NP-complete, and, as we shall see, the complexity in the size of the protocol will always be higher for all protocol classes.

Taking these considerations into account, we formally define the correctness problem as follows:

\begin{quote}
\underline{Correctness problem} 

\medskip
\textbf{Given}:  A protocol $\PP$ over an alphabet $\Sigma$, belonging to one of the six classes PP, DO, IO, DT, IT, QT;  a quantifier-free Presburger predicate $\varphi$ over $\Sigma$. \\
\textbf{Decide:} Does $\PP$ compute the predicate represented by $\varphi$?
\end{quote}

We also study the correctness problem over a single input. We refer to it as the single-instance correctness problem and define it in the following way:

\begin{quote}
\underline{Single-instance correctness problem} 

\medskip
\textbf{Given}: A protocol $\PP$ over an alphabet $\Sigma$ and with initial state mapping $\iota$, belonging to one of the six classes PP, DO, IO, DT, IT, QT;  an input $D \in \pop{\Sigma}$,
and a boolean $b$. \\
\textbf{Decide:} Do all fair executions of $\PP$ starting at $I(D)$ converge to $b$ ?
\end{quote}

%% file: formalizing-correctness.tex
In the coming sections we will obtain matching upper and lower bounds on the complexity of the correctness problem for different protocol classes. The upper bounds are obtained by reducing the correctness problem to two different reachability problems. The reductions require the protocols to be \emph{well behaved}. We first define well-behaved protocols, and then present the two reductions.

\paragraph{Well-behaved protocols.} Let $\PP$ be a generalized protocol. 
A configuration $C$ of $\PP$ is a \emph{bottom configuration} if $C \trans{*} C'$ implies $C' \trans{*} C$ for every configuration $C'$. In other words, $C$ is a bottom configuration if it belongs to a bottom strongly connected component (SCC) of the configuration graph of the protocol.
\je{I think this is the first time we mention SCC, so I added (SCC)}

\begin{definition}
A generalized protocol       is \emph{well-behaved} if every fair execution contains a bottom configuration.
\end{definition}

We show that all our protocols are well behaved, with the exception of queued-transmission
protocols. Essentially, this is the reason why the decidability of the
correctness problem for QT is still open.

\begin{lemma}
\label{lm:wellbehaved}
Standard population protocols (PP) and delayed-transmission protocols (DT) are well behaved. 
\end{lemma}
\begin{proof}
In standard population protocols, if $C \trans{*} C'$ then $|C|=|C'|$.  It follows that for every configuration $C\in \Conf$ the set of configurations reachable from $C$ is finite. So every fair execution eventually visits a bottom configuration.
	
In delayed-transmision protocols, the size of a configuration is equal to the number of agents plus the number of messages in transit.
So there is no bound on the size of the configurations reachable from a given configuration $C$, and in particular the set of configurations reachable from $C$ can be infinite. 
However, since agents can always receive any message, for every configuration $C$ there is at least one reachable configuration $Z$ without any message in transit.
Since the number of such configurations with a given number of agents is finite, for every fair execution $\pi = C_0, C_1, \ldots$ there is a configuration $Z$ without messages in transit such that $C_i=Z$ for infinitely many $i$. By fairness, every configuration $C'$ reachable from $Z$ also appears infinitely often in $\pi$, and so every configuration $C'$ reachable from $Z$ verifies $C' \trans{*} Z$.
So $Z$ is a bottom configuration. 
 \end{proof}

Since IT and IO are subclasses of PP and DO is a subclass of DT, the proof is valid for IT, IO, and DT as well. The following example shows that queued-transmission protocols are not necessarily well-behaved.

\begin{example}
	\label{ex:fair}
Consider a queued-transmission protocol in which an agent in state $q$ can send a message $m$, staying in $q$.
Assume further that no agent can ever receive a message $m$ (because, for example, there are no receiving transitions for it).
Then any execution in which the agent in state $q$ sends the message $m$ infinitely often and never receives any messages is fair:
Indeed, after $k$ steps the system can only reach configurations with at least $k$ messages, and so no configuration is reachable from infinitely many configurations in the execution.
Since this fair execution does not visit any bottom configuration, the protocol is not well-behaved.
Moreover, if $q$ is the only state of the protocol, there are no bottom configurations at all.
\end{example}

\paragraph{Characterizing correctness of well-behaved protocols.}

We start with a useful lemma valid for arbitrary protocols.

\begin{lemma}[\cite{AAER07}]
\label{lem:little-new}
Every finite execution of a generalized protocol can be extended to a fair execution.
\end{lemma}
\begin{proof}
Let $\Conf$ be the set of configurations of the protocol, and let $\pi$ be a finite execution. Fix an infinite sequence $\rho = C_0, C_1, \ldots$ of configurations such that every configuration of $\Conf$  appears infinitely often in $\rho$. Define the infinite execution $\pi_0 \, \pi_1 \,  \pi_2 \ldots$ and the infinite subsequence $C_{i_0}, C_{i_1}, C_{i_2} \ldots$ of $\rho$ inductively as follows.  
For $i=0$, let $\pi_0 := \pi$ and $C_{i_0} := C_0$. For every $j \geq 0$, let $\pi_0 \, \ldots \,  \pi_j \,\pi_{j+1}$ be any execution leading to the first configuration of $\rho$ after $C_{i_j}$ that is reachable from the last configuration of $\pi_0 \, \ldots \,  \pi_j$. It is easy to see that $\pi_0 \, \pi_1 \,  \pi_2 \ldots$ is fair.
 \end{proof}

\noindent Now we introduce some notations. Let $\PP=(\Conf, \Sigma, \Step,\allowbreak I, O)$ be a generalized protocol,
and let $\varphi$ be a predicate.
\begin{itemize}
\item The sets of predecessors and successors of a set $\mathcal{M}$ of configurations of $\PP$ are defined as follows:
$$\begin{array}{rcl}
\prestar(\mathcal{M})  & \defeq & \{ C' \in \Conf \mid \exists C \in \mathcal{M} \, . \, C' \xrightarrow{*} C \} \\
\poststar(\mathcal{M}) & \defeq & \{ C \in \Conf \mid \exists C' \in \mathcal{M} \, . \, C' \xrightarrow{*} C \}
\end{array}$$
\item For every $b \in \set{0,1}$, we define $\Cons{b} \defeq O^{-1}(b)$, the set of configurations with output $b$.
We call $\Cons{b}$ the set of $b$-consensus configurations.

\item For every $b \in \{0, 1\}$, we let $\Stab{b}$ denote the set of configurations $C$ such that every configuration reachable from $C$ (including $C$ itself) has output $b$. 
$\Stab{}$ stands for \emph{stable output}.  It follows easily from the definitions of $\prestar$ and $\poststar$ that
$$\Stab{b} = \compl{\prestar\left(\compl{\Cons{b}}\right)} \ ,$$
\noindent where $\compl{\mathcal{M}} \defeq \Conf \setminus \mathcal{M}$ for every set of configurations $\mathcal{M} \subseteq \Conf$. Indeed, the equation states that a configuration belongs to $\Stab{b}$ if{}f it cannot reach any configuration with output $1-b$, or with no output.  
\item For every $b \in \{0, 1\}$, we define $I_b \defeq \{I(D) \mid D \in \pop{\Sigma} \wedge \varphi(D) = b \}$.
In other words, $I_b$ is the set of initial configurations for which $\PP$ should output $b$ in order to compute $\varphi$.
\end{itemize}

\begin{proposition}
\label{prop:corr-2}
Let $\PP=(\Conf, \Sigma, \Step, I, O)$ be a well-behaved generalized protocol and
let $\varphi$ be a predicate. $\PP$ computes $\varphi$ if{}f 
$$\poststar(I_b) \subseteq \prestar(\Stab{b}) $$
\noindent holds for every $b \in \{0,1\}$.
\end{proposition}
\begin{proof}
Assume that $\poststar(I_b) \subseteq \prestar(\Stab{b})$ holds for $b \in \{0,1\}$.
Let $\pi=C_0, C_1, \ldots$ be a fair execution with $C_0 \in I_b$ for some $b \in \{0,1\}$. 
We show that $\pi$ converges to $b$.
Protocol $\PP$ is well-behaved, so $\pi$ contains a bottom configuration $C$ of a bottom SCC $B \subseteq \mathcal{B}$.
By assumption, we know that $\Stab{b}$ is reachable from $C$, so there exists $C' \in \Stab{b}$ such that $C \trans{*} C'$.
This entails $C' \in B$.
Since for all $D \in \Stab{b}$, if $D \trans{*} D'$ then $D' \in \Stab{b}$, we obtain that $B \subseteq \Stab{b}$.
Every configuration of $\Stab{b}$ is a $b$-consensus so $\pi$ converges to $b$.

Assume that $\PP$ computes $\varphi$, \ie that every fair execution starting in $I_b$ converges to $b$ for $b \in \set{0,1}$. 
Let us show that $\poststar(I_b) \subseteq \prestar(\Stab{b})$ holds.
Consider $C \in \poststar(I_b)$. 
There exists $C_0 \in I_b$ such that $C_0 \trans{*} C$ and, by Lemma \ref{lem:little-new}, this finite execution can be extended to a fair infinite execution $\pi$. 
Since $\PP$ is well-behaved, the execution contains a bottom configuration $C'$ of a bottom SCC $B \subseteq \mathcal{B}$.
If $B \subseteq \Stab{b}$ then $C \in \prestar(\Stab{b})$ and our proof is done.
Suppose this is not the case, \ie $B \cap \compl{\Stab{b}} \neq \emptyset$.
This means that there is a configuration $\hat{C} \notin \Cons{b}$ that is in $B$. 
It is thus reachable from any configuration of $\pi$ and so by fairness it is reached infinitely often. 
Thus $\pi$ does not converge to $b$, contradicting the correctness assumption.
 \end{proof}

\paragraph{A second characterization.}

Proposition \ref{prop:corr-2} is useful when it is possible
to compute adequate finite representations of the sets $\poststar(I_b)$
and $\prestar(\Stab{b})$. We will later see that this is the case for IO and DO protocols.
Unfortunately, such finite representations have not yet been found for PP or for transmission
protocols. For this reason, our results for these classes will be based on a second characterization.

Let $\PP=(\Conf, \Sigma, \Step, I, O)$ be a well-behaved generalized protocol, and let
$\B$ denote the set of bottom configurations of $\PP$. Further, for every $b \in \{0,1\}$, 
let $\B_b$ denote the set of configurations $C \in \B$ such that every configuration $C'$ reachable from $C$ satisfies $O(C')=b$. Equivalently, $\B_b \defeq \B \cap \Stab{b}$.  

Observe that every fair execution of a well-behaved protocol eventually gets trapped in a bottom strongly-connected component of the configuration graph and, by fairness, visits all its configurations infinitely often. Further, if any configuration of the SCC belongs to $\B_b$, then all of them belong to $\B_b$. This occurs independently of whether the SCC contains finitely or infinitely many configurations.

\begin{restatable}{proposition}{PropCorrectness}
\label{prop:corr}
Let $\PP$ be a well-behaved generalized protocol and let $\varphi$ be a predicate. 
$\PP$ computes $\varphi$ if{}f for every $b \in \{0,1\}$ the set $\B \setminus \B_b$ is not reachable from $I_b$.
\end{restatable}
\begin{proof}
Assume that $\B \setminus \B_b$ is reachable from $\varphi^{-1}(b)$ for some $b \in \{0,1\}$. Then there exists an input $a \in \pop{\Sigma}$ and an execution $C_0, C_1, \ldots, C_i$ such that $\varphi(a)=b$, $I(a) = C_0$, and $C_i \in \B \setminus \B_b$. 
By Lemma \ref{lem:little-new} the execution can be extended to a fair execution $C_0, C_1, \ldots$. Since $C_{i+k} \trans{*} C_i$ for every $k \geq 0$, the execution visits $C_i$ and all its successors infinitely often.
Since $C_i \notin \B_b$, the execution does not converge to $b$. So $\PP$ does not compute $\varphi$. 

Assume that $\PP$ does not compute $\varphi$. Then there exists an input $a \in \pop{\Sigma}$, a boolean $b \in \{0,1\}$, and a fair execution $\pi=C_0, C_1, \ldots$ such that $\varphi(a)=b$ and $I(a) = C_0$, but $\pi$ does not converge to $b$. Since $\PP$ is well-behaved, $\pi$ contains a configuration $C_i \in \B$. Since $\pi$ does not converge to $b$, there is $j > i$ such that $O(C_j)$ is undefined, or defined but different from $b$. Since $C_j$ belongs to the same SCC as $C_i$, we have $C_i \notin \B_b$.
 \end{proof}

%% file: intro-observation-hardness.tex
We prove that the correctness problem is \PSPACE-hard for IO protocols and $\Pi_2^p$-hard for DO protocols, and that these results also hold for the single-instance problem.

%% file: io-hardness.tex
We prove that the single-instance correctness and correctness problems for IO protocols are \PSPACE-hard by reduction from the acceptance problem for bounded-tape Turing machines.
We show that the standard simulation of bounded-tape Turing machines
by  1-safe Petri nets, as described for example in \cite{ChengEP95,Esparza96},
can be modified to produce an IO protocol.
This can be done for IO protocols but not for DO protocols: the simulation of the Turing machine relies on the fact that a transition will only occur in an IO protocol if an agent observes another agent in a certain state \emph{at the present moment}.

We fix a deterministic Turing machine $M$ with set of control states $Q$, alphabet $\Sigma$ containing the empty symbol $\text{\textvisiblespace}$, and partial transition function $\delta \colon  Q\times\Sigma \to Q\times\Sigma\times D$ ($D=\{-1,+1\}$). Let $K$ denote an upper bound on the number of tape cells visited by the computation of $M$ on empty tape. We assume that $K$ is provided with $M$ in unary encoding.

The \emph{implementation} of $M$ is the IO protocol $\PP_M$ described below.
Strictly speaking, $\PP_M$ is not a complete protocol, only two sets of states and transitions.
The rest of the protocol, which is slightly different for the single-instance correctness and the correctness problems, is described in the proofs.

\smallskip
 \noindent \textbf{States of $\PP_M$.} The protocol  $\PP_M$ contains two sets of \emph{cell states}
        and \textit{head states} modeling the state of the tape cells and the head, respectively. The cell states are:
        \begin{itemize}
        \item $\pass{\sigma}{n}$ for each $\sigma\in \Sigma$ and $1\leq n\leq K$.
        An agent in $\pass{\sigma}{n}$ denotes that cell $n$ contains symbol $\sigma$, and the cell is ``off'', i.e., the head is not on it.
        \item $\act{\sigma}{n}$ for each $\sigma\in \Sigma$ and $1\leq n\leq K$, with analogous intended meaning.
        \end{itemize}
        The head states are:
        \begin{itemize}
        \item $\stable{q}{n}$ for each $q\in Q$ and $1\leq n\leq K$.  An agent in $\stable{q}{n}$ denotes that the head is in control state $q$ and at cell $n$.
        \item $\switch{q}{\sigma}{n}{d}$ for each $q\in Q$, $\sigma\in\Sigma$,
        $1\leq n\leq K$
        and every $d\in D$ such that $1\leq n+d\leq K$.
        An agent in $\switch{q}{\sigma}{n}{d}$ denotes that head is in control state $q$, has left cell $n$ after writing symbol $\sigma$ on it, and is currently moving in the direction given by $d$.
        \end{itemize}
        Finally, the protocol also contains two special states \textit{observer} and $success$. Intuitively,
          $\PP_M$ uses them to detect that $M$ has accepted.

       \noindent \textbf{Transitions of $\PP_M$.} Intuitively, the implementation of $M$ contains a  set of \emph{cell transitions} in which a cell observes the head and changes its state, a set of \emph{head transitions} in which the head observes a cell. Each of these sets contains transitions of two types.
 The set of cell transitions contains:
        \begin{itemize}
        \item Type  \textbf{1a}: A transition
        $\pass{\sigma}{n} \trans{\stable{q}{n}} \act{\sigma}{n}$ for every state $q \in Q$, symbol $\sigma \in \Sigma$, and cell $1 \leq n \leq K$. \\
        The $n$-th cell, currently \textit{off}, observes that the head is on it, and switches itself \textit{on}.
        \item Type \textbf{1b}: A transition
        $\act{\sigma}{n} \trans{\switch{q}{\sigma'}{n}{d}} \pass{\sigma'}{n}$ for every $q \in Q$, $\sigma \in \Sigma$, and $1 \leq n \leq K$ such that $1 \leq n+d \leq K$. \\
        The $n$-th cell, currently \textit{on}, observes that the head has left after writing $\sigma'$, and switches itself \textit{off} (accepting the character the head intended to write).
        \end{itemize}
        The set of head transitions contains:
        \begin{itemize}
        \item Type \textbf{2a}: A transition $$\stable{q}{n} \trans{\act{\sigma}{n}} \switch{\delta_Q(q,\sigma)}{\delta_\Sigma(q,\sigma)}{n}{\delta_D(q,\sigma)}$$ for every $q \in Q$, $\sigma \in \Sigma$, and $1 \leq n \leq K$ such that $1\leq n+\delta_D(q,\sigma)\leq K$. \\
        The head, currently on cell $n$, observes that the cell is \textit{on}, writes the new symbol on it, and leaves.
        \item Type \textbf{2b}: A transition $\switch{q}{\sigma}{n}{d} \trans{\pass{\sigma}{n}}\stable{q}{n+d}$ for every $q \in Q$, $\sigma \in \Sigma$, and $1 \leq n \leq K$ such that $1 \leq n+d \leq K$.\\
        The head, currently moving, observes that the old cell has turned \textit{off}, and places itself on the new cell.
        \end{itemize}

Figure \ref{figure-turing-io} graphically represents some of the states and transitions
of $\PP_M$; the double arcs indicates the states being observed. We define the configuration of $\PP_M$ that corresponds to a given configuration of the Turing machine.

\begin{restatable}{figure}{FigureTuringIO}
\centering
\includegraphics[width=0.7\linewidth]{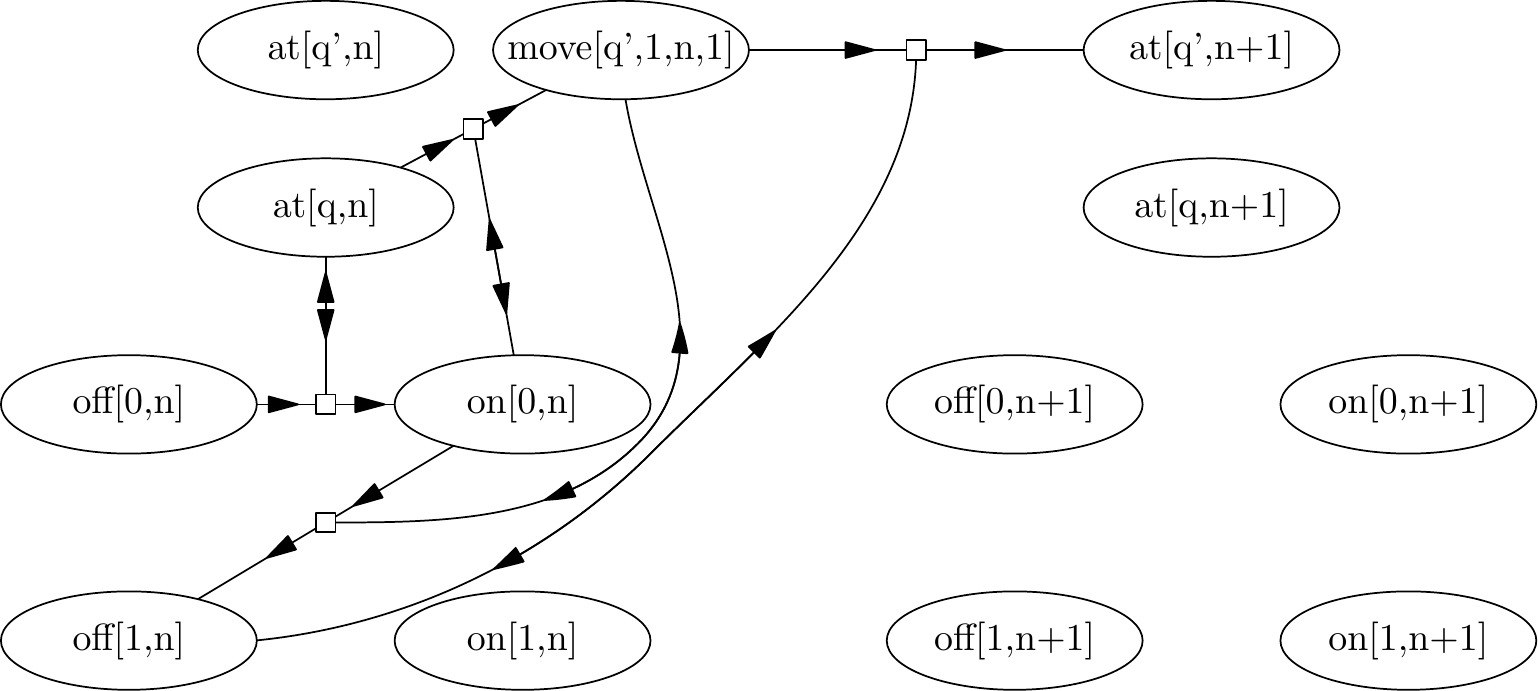}
\caption{Some of the states and transitions involved in modelling a Turing machine}
\label{figure-turing-io}
\end{restatable}

\begin{definition}
\label{def:turing-configuration}
        Given a configuration $c$ of $M$ with control state $q$, tape content $\sigma_1\sigma_2\cdots \sigma_K$, and head on cell $n \leq K$, let $C_c$ be the configuration that puts one agent in $\pass{\sigma_i}{i}$ for each $1\leq i\leq K$,  one agent in $\stable{q}{n}$, and no agents elsewhere.
\end{definition}

Theorem \ref{thm:simul} below formalizes the relation between the Turing machine $M$ and its implementation $\PP_M$.

\begin{restatable}{theorem}{theoremSimulationStep}\label{thm:simul}
For every two configurations $c, c'$ of $M$ that write at most $K$ cells:  $c \trans{} c'$ if{}f $C_c \trans{t_1t_2t_3t_4} C_{c'}$ in $\PP_M$ for some transitions $t_1, t_2, t_3, t_4$ of types \textbf{1a}, \textbf{2a}, \textbf{1b}, \textbf{2b}, respectively.
\end{restatable}
\begin{proof}
        By Lemma~\ref{lemma:modellingmarkingevolution},
        for all $c$ there is either zero or one possibility for the
sequence $t_1,t_2,t_3,t_4$ starting in $C_c$.
        It is easy to see from the definition of steps configuration $\switch{\cdot}{\cdot}{\cdot}{\cdot}$
        states that if such a sequence exists, it results in $c'$ such that $c\trans{}c'$.
        If such a sequence doesn't exist, the failure must occur when trying to populate
        a  $\switch{\cdot}{\cdot}{\cdot}{\cdot}$ state.
        In that case the configuration~$c$ must be blocked, either by
        the transition being undefined or by going out of bounds.
 \end{proof}

Now we can finally prove the \PSPACE\ lower bound.

\begin{restatable}{theorem}{ThmIOCorrectnessHard}\label{thm:io-correctness-hard}
The single-instance correctness and correctness problems for IO protocols are \PSPACE-hard.
\end{restatable}
\begin{proof}
By reduction from the following problem: Given a polynomially space-bounded deterministic Turing machine $M$ with two distinguished states $q_{acc}, q_{rej}$, such that the computation of $M$ on empty tape ends when the head enters for the first time $q_{acc}$ or $q_{rej}$ (and one of the two occurs), decide whether $M$ accepts, i.e., whether the computation on empty tape reaches $q_{acc}$. The problem is known to be \PSPACE-hard.

\medskip\noindent\textbf{Single-instance correctness.} We construct a protocol $\PP$ and an input $D_0$ such that $M$ accepts on empty tape if{}f all fair executions of $\PP$ starting at the configuration $I(D_0)$ converge to 1.

\smallskip\noindent \textit{Definition of $\PP$.} Let $\PP_M$ be the IO protocol implementation of $M$. We add two states to $\PP_M$, called \textit{observer} and \textit{success}. We also add transitions allowing an agent in state \textit{observer} to move to \textit{success} by observing any agent in a state of the form $\stable{q_{acc}}{i}$, as well as transitions allowing an agent in \textit{success} to ``attract'' agents in all other states to \textit{success}:
\begin{itemize}
	\item[(i)] $\textit{observer} \trans{\stable{q_{acc}}{i}} \textit{success}$ for every $1 \leq i \leq K$, and
	\item[(ii)] $q \trans{\textit{success}} \textit{success}$ for every $q \neq \textit{success}$.
	\end{itemize}
\noindent Further, we set the output function to 1 for the state \textit{success}, and to 0 for all other states.  Finally, we choose the alphabet of input symbols of $\PP$ as $\{1,2, \ldots, K+2\}$, and define the input function as follows: $\iota(i) = \pass{\text{\textvisiblespace}}{i}$ for every $1 \leq i \leq K$; $\iota(K+1) =\stable{q_0}{0}$; and $\iota(K+2) =\textit{observer}$.

\smallskip\noindent \textit{Definition of $D_0$.} We choose $D_0$ as the input satisfying $D_0(i)=1$ for every input symbol of $\PP$.  It follows that $I(D_0)$ is the configuration of
$\PP$ corresponding to the initial configuration of $M$ on empty tape.
By Theorem \ref{thm:simul}, the fair executions of $\PP$ from $I(D_0)$ simulate the execution of $M$ on empty tape.

\smallskip\noindent \textit{Correctness of the reduction.} If $M$ accepts, then, since $\PP$ simulates the execution of $M$ on empty tape, every fair execution of $\PP$ starting at $I(D_0)$ eventually puts an agent in a state of the form $\stable{q_{acc}}{i}$. This agent stays there until the agent in state \textit{observer} eventually moves to \textit{success} (transitions of (i)), after which all agents are eventually attracted to \textit{success} (transitions of (ii)). So all fair computations of $\PP$ starting at $I(D_0)$ converge to 1.
If $M$ rejects, then no computation of $\PP$ starting at $I(D_0)$ (fair or not) ever puts an agent in \textit{success}. Since all other states have output 0,  all fair computations of $\PP$
starting at $I(D_0)$ converge to 0.

\medskip\noindent\textbf{Correctness.}
Notice that the hardness proof for single-instance correctness establishes PSPACE-hardness already for restricted instances $(\PP, D)$ satisfying $D(q) \in \{0, 1\}$ for every state $q$. Call this restricted variant the $0/1$-single-instance correctness problem for IO.
We claim that the $0/1$-single-instance correctness problem for IO is polynomial-time reducible to the correctness problem for IO. By PSPACE-hardness of the $0/1$-single-instance correctness problem for IO, the claim entails PSPACE-hardness for the latter.

Let us now show the claim. Given an IO protocol $\PP$ and some configuration $D$ for the $0/1$-instance-correctness problem, we provide a polynomial-time construction of an IO protocol $\PP'$ such that $\PP'$ computes the constant predicate $\varphi(\vec{x}) = 0$ if and only if every fair run of $\PP$ starting in $D$ stabilizes to $0$. It is well known that, given two protocols $\PP_1$ and $\PP_2$ with $n_1$ and $n_2$ states and computing two predicates $\varphi_1$ and $\varphi_2$, it is possible to construct a third protocol computing $\varphi_1 \wedge \varphi_2$, often called the \emph{synchronous product}, whose states are pair of states of $\PP_1$ and $\PP_2$, and has therefore $O(n_1 \cdot n_2)$ states (see e.g. \cite{AADFP06}).  We define $\PP'$ as the synchronous product of $\PP$ with a protocol $\PP_D$ that computes whether the input is equal to $D$. The output function of $\PP'$ maps the product state $(q_1, q_2)$ to $1$ if and only if both $q_1$ and $q_2$ map to output $1$ in their respective protocols.
Thus, a fair run of $\PP'$ stabilizes to $1$ if and only if the input configuration equals $D$ and $\PP$ stabilizes to $1$ for input $D$, which is precisely the case if $(\PP, D)$ is a positive instance for the $0/1$-single-instance problem.

It remains to show that $\PP_D$ is polynomial-time constructible. Such a protocol is well-known, but we repeat the definition. Let $D = (d_1, \ldots, d_m)$ with $d_i \in \{0, 1\}$, and let $i_1 \leq i_2 \leq \ldots \leq i_k$ be the maximal sequence of indices satisfying $d_{i_j} = 1$ for every $j$. Since every population has at least two agents, we have $k \geq 2$. We first construct an IO protocol $\PP_\psi$ that computes the predicate $\psi = d_{i_1} \geq 1 \land d_{i_2} \geq 1 \land \ldots \land d_{i_k} \geq 1$, using $m + k - 1$ states: The states of $\PP_\psi$ are $Q_\PP \uplus \{2, \ldots, k\}$ where $Q_\PP$ is the set of states of $\PP$. The input mapping of $\PP_\psi$ is identical to the input mapping of $\PP$. Let $q_{i_j}$ denote the state that corresponds to the entry $d_{i_j}$ in $D$. The transitions of $\PP_\psi$ are given by
\begin{align*}
    & q_{i_2} \trans{q_{i_1}} 2 && \\
    & q_{i_j} \trans{j-1} j && \text{ for every } 1 < j \leq k, \\
    & q \trans{k} k && \text{ for every state } q.
\end{align*}

All states except $k$ shall map to output $0$. It is readily seen that $\PP_\psi$ computes $\psi$. Further notice that the predicate $\vec{x} = D$ is equivalent to $\psi \land |\vec{x}| \leq k$. Moreover, it is well-known that the right conjunct $|\vec{x}| \leq k$ is computable with $k$ states in an immediate observation protocol (see e.g. \cite{AADFP06}), and thus we can define $\PP_D$ as the synchronous product of the protocol $\PP_\psi$ with the protocol that computes $|\vec{x}| \leq k$, using $\poly(k)$ states. This completes the proof.

\end{proof}

%% file: do-hardness.tex
We show that the single-instance correctness and the correctness problems are $\Pi_2^p$-hard for DO protocols, where $\Pi_2^p= \text{co\NP}^{\text{co\NP}}$ is one of the two classes at the second level of the polynomial hierarchy \cite{Stockmeyer1976}. Consider the natural complete problem for $\Sigma_2^p$: Given a boolean circuit $\Gamma$ with inputs $\mathbf{x}=(x_1, \ldots, x_n)$ and $\mathbf{y}= (y_1, \ldots, y_m)$, is there a valuation of $\mathbf{x}$ such that for every valuation of $\mathbf{y}$ the circuit outputs 1? We call the inputs of $\mathbf{x}$ and $\mathbf{y}$ \emph{existential} and \emph{universal}, respectively.
Given $\Gamma$ with inputs $\mathbf{x}$ and $\mathbf{y}$, we construct  in polynomial time a DO protocol $\PP_\Gamma$ with input symbols $\{x_1, \ldots, x_n\}$ that computes the false predicate, i.e., the predicate answering  $0$ for all inputs, if{}f $\Gamma$ does \emph{not} satisfy the property above. This shows that the correctness problem for DO protocols is
$\Pi_2^p$-hard. A little modification of the proof shows that single-instance correctness is also $\Pi_2^p$-hard.

The section is divided in several parts. We first introduce basic notations about boolean circuits.
Then we sketch a construction that, given a boolean circuit $\Gamma$, returns  a \emph{circuit evaluation protocol} $\widehat{\PP}_\Gamma$ that nondeterministically chooses values for the input nodes, and simulates an execution of $\Gamma$ on these inputs. In a third step we add some states and transitions to $\widehat{\PP}_\Gamma$ to produce the final DO protocol $\PP_\Gamma$. The fourth and final step proves the correctness of the reduction.

\paragraph{Boolean circuits.} A boolean circuit $\Gamma$ is a directed acyclic graph. The nodes of $\Gamma$ are either \emph{input nodes}, which have no incoming edges, or \emph{gates}, which have at least one incoming edge. A gate with $k$ incoming edges is labeled by a boolean operation of arity $k$. We assume that $k$ is bounded by some constant. This assumption is innocuous since it is well known that every boolean function can be implemented using a combination of gates of constant arity. The nodes with outgoing edges leading to a a gate $g$ are called the \emph{arguments} of $g$. There is a distinguished \emph{output gate} $g_o$ without outgoing edges.  We assume that every node is connected to the output gate by at least one path.

A circuit configuration assigns to each input node a boolean value, $0$ or $1$, and to each gate a
value, $0$, $1$, or $\square$, where $\square$ denotes that the value has not yet been computed
and so it is still unknown. A configuration is initial if it assigns $\square$ to all gates. The step relation between circuit configurations is defined as usual: a gate can change its value to the result of applying the  boolean operation to the arguments; if at least one of the arguments has value $\square$, then  by definition the result of the boolean operation is also $\square$.

\paragraph{The protocol $\widehat{\PP}_\Gamma$.} Given a circuit $\Gamma$ with output node $g_o$, we define the circuit evaluation protocol $\widehat{\PP}_\Gamma=(Q, M, \delta_s, \delta_r,\allowbreak \Sigma,\allowbreak  \iota,\allowbreak  o)$. As mentioned above, $\widehat{\PP}_\Gamma$ nondeterministically chooses input values for $\Gamma$, and simulates an execution on them.

\smallskip \noindent \textbf{States.}
The set $Q$ of states contains all tuples  $(n, v_n, \arg, v_o)$, where:
\begin{itemize}
\item $n$ is a node of $\Gamma$ (either an input node or a gate);
\item $v_n \in \{0,1,\square\}$ represents the current opinion of the agent about the value of $n$;
\item $\arg \in \{0,1,\square\}^k$, where $k$ is the number of arguments of $n$, represents the current  opinion of the agent about the values of the arguments of $n$ (if $n$ is an input node then $\arg$ is the empty tuple);
\item $v_o \in \{0, 1, \square\}$ represents the current opinion of the agent about the value of the output gate $g_o$.
\end{itemize}

\smallskip \noindent \textbf{Alphabet, input and output functions.} The set $\Sigma$ of input symbols is the set of nodes of $\Gamma$.  The initial state mapping $\iota$ maps each node $n$ to the state $\iota(n) := (n, \square, (\square, \ldots, \square), \square)$, \ie, to
the state with node $n$, and with all values still unknown. The output function is defined by
$$o(n, v_n, \arg, v_o):= \text{if $v_o \neq \square$ then $v_o$ else $0$} \ .$$
\noindent Intuitively, agents have opinion $1$ if they think the circuit outputs $1$, and $0$ if they think the circuit outputs $0$ or has not yet produced an output.

\smallskip \noindent \textbf{Messages.}   The set $M$ of messages contains all pairs $(n, v)$, where $n$ is a node, and $v \in \{0,1,\square\}$ is a value.

\smallskip \noindent \textbf{Transitions.}  An agent in state $(n, v_n, \arg, v_o)$ can
\begin{itemize}
\item Send the message $(n, v_n)$, \ie, an agent can send its node and its current opinion on the
value of the node.
\item Receive a message $(m, v_m)$, after which the agent updates its state as follows:
\begin{itemize}
\item[(1)] If $n$ is an input node and $v_n =  \square$, then if $m=n$ the agent moves to
 state $(n, 0, \arg, v_o)$, \ie, updates its value to $0$, and if $m = g_o$ it moves
to state $(n, 1, \arg, v_o)$, \ie, updates its value to $1$. 
Intuitively, this is an artificial but simple way of ensuring that each input node nondeterministically chooses a value, $0$, or $1$, depending on whether it first receives a message from itself, or from the output node.
\footnote{Alternatively, one could include an explicit rule for this non-deterministic behavior. We choose to model it this way to preserve the deterministic definition of the DO model introduced by Angluin \etal in \cite{AAER07}.}
\item[(2)] If $n$ is a gate and $m$ is an argument of $n$, then the agent moves to $(n, v_n', \arg', v_o)$, where $\arg'$ is the result of
updating the value of $m$ in $\arg$ to $v_m$, and $v_n'$ is the result of applying the boolean operation of the gate to $\arg$.
\item[(3)] If $n$ is any node, $m = g_o$, and $v_m \neq \square$, then the agent moves to $(n, 0, \arg, v_m)$, \ie, it updates its opinion of the output of the circuit to $v_m$.
\end{itemize}
\end{itemize}
Notice that if an agent is initially in state $\iota(n)$, then it remains forever in states having $n$ as node. So it makes sense to speak of \emph{the} node of an agent.

\smallskip Let us examine the behaviour of $\widehat{\PP}_\Gamma$ from the initial configuration $C_0$ that puts exactly one agent in state $\iota(n)$ for every node $n$. The executions of $\widehat{\PP}_\Gamma$ from $C_0$ exactly simulate the executions of the circuit. Indeed, the transitions of (1) ensure that each input agent (\ie, every agent whose node is an input node) eventually chooses a value, $0$ or $1$. The transitions of (2) simulate the computations of the gates. Finally, the transitions of (3) ensure that every node eventually updates its opinion of the value of $g_o$ to the value computed by $\Gamma$ for the chosen input. The following lemma, proved in the appendix, formalizes this.

\begin{restatable}{lemma}{LemmaCircuitEvaluation}
        \label{lm:circuit-evaluation-correct}
Let $\Gamma$ be a circuit and let $\widehat{\PP}_\Gamma$ be its evaluation protocol. Let $C_0$ be the initial configuration that puts exactly one agent in state $\iota(n)$ for every node $n$. A fair execution starting at $C_0$ eventually reaches a configuration $C$ where each input agent is in a state with value $0$ or $1$, and these values do not change afterwards. The tail of the execution starting at $C$ converges to a stable consensus equal to the output of $\Gamma$ on these assigned inputs.
\end{restatable}

Observe, however,  that  $\widehat{\PP}_\Gamma$ also has initial configurations whose executions may not simulate any execution of $\Gamma$. For example, this is the case of an initial configuration that puts two agents in state $\iota(n)$ for some node $n$, and the executions in which one of these agents chooses input $0$ for $n$, and the other input $1$. It is also the case of an initial configuration that puts zero agents in state $\iota(n)$ for some node $n$.  Observe further that $\widehat{\PP}_\Gamma$ can only select values for the inputs, and simulate an execution of $\Gamma$. We need a protocol that selects values for the existential inputs, and can then repeatedly simulate the circuit for different values of the universal inputs. These two problems are solved by appropriately extending $\widehat{\PP}_\Gamma$ with new states and transitions.

\paragraph{The protocol $\PP_\Gamma$.} We add a new state and some transitions to $\widehat{\PP}_\Gamma$ in order to obtain the final protocol $\widehat{\PP}_\Gamma$.
\begin{itemize}
\item Add a new \emph{failure state} $\bot$ with $o(\bot)=0$  to the set of states $Q$, and
a new message $m_\bot$ to the set of messages $M$.
\item Add the following send and receive transitions:
\begin{itemize}
\item[(4)] An agent in state $\bot$ can send the message $m_\bot$.
\item[(5)] An agent in state $\bot$ that receives any message (including $m_\bot$)
stays in state $\bot$; an agent (in any state, including $\bot$) that receives $m_\bot$ moves to state $\bot$. \\
(In particular, if some agent ever reaches state $\bot$, then all agents eventually
 reach state $\bot$ and stay there, and so the protocol converges to $0$.)
\item[(6)] If an agent in state $(n, v_n, \arg, v_o)$, where $n$ is an existential input node and $v_n \neq \square$, receives a message $(n, v_n')$ such that $v_n \neq v_n' \neq \square$, then the agent moves to state $\bot$. \\
(Intuitively, if an agent discovers that another agent has chosen a different value for the same   existential input, then the agent moves to $\bot$, and so, by the observation above, the protocol converges to $0$.)
\item[(7)] If an agent in state $(n, v_n, \arg, v_o)$, where $n$ is a universal input node and $v_n \neq \square$, receives a message $(g_o, 1)$, then the agent moves to state $(n,1-v_n, \arg, v_o)$. \\
(Intuitively, this allows the protocol to flip the values of any universal inputs whenever the output gate takes value 1.)
\end{itemize}
\end{itemize}

\paragraph{Proof of the reduction.} We claim that $\PP_\Gamma$ does not compute the false predicate (\ie, the predicate that answers $0$ for every input) if{}f $\exists \vec{x} \forall \vec{y} \Gamma(\vec{x}, \vec{y}) = 1$, that is, if there is a valuation of the existential inputs of $\Gamma$ such that, for every valuation of the universal inputs, $\Gamma$ returns $1$. Let us sketch the proof of the claim.  We consider two cases:

\smallskip
\noindent \textit{$\exists \vec{x} \forall \vec{y} \Gamma(\vec{x}, \vec{y}) = 1$ is true.}
Let $C_0$ be the initial configuration that puts exactly one agent in state $\iota(n)$ for every node $n$. We show that not every fair execution from $C_0$ converges to $0$, and so that $\PP_\Gamma$ does not compute the $0$ predicate.

Let $\vec{x}_0$ be a valuation of $\vec{x}$ such that $\forall \vec{y} \Gamma(\vec{x}_0, \vec{y}) = 1$.  The execution proceeds as follows: first, the agents for the inputs of $\vec{x}$ receive messages, sent either by themselves or by the output node, that make them choose the values of $\vec{x}_0$. An inspection of the transitions of $\PP_\Gamma$ shows that these values cannot change anymore. Let $C$ be the configuration reached after the agents have received the messages. Since $\Gamma(\vec{x}_0, \vec{y}) = 1$ holds for every $\vec{y}$, by Lemma \ref{lm:circuit-evaluation-correct} every configuration $C'$ reachable from $C$ can reach a consensus
of $1$. Indeed, it suffices to first let the agents receive all messages of $C'$ (which does not change the values of the existential inputs), then let the agents for $\vec{y}$ that still have value $\square$  pick a boolean value (nondeterministically), and then let all agents simulate the circuit.  Since $\Gamma(\vec{x}_0, \vec{y}) = 1$ holds for every $\vec{y}$, after the simulation the node for $g_o$ has value 1. Finally, we let all agents move to states satisfying $v_o=1$.

\smallskip
\noindent \textit{$\exists \vec{x} \forall \vec{y} \Gamma(\vec{x}, \vec{y}) = 1$ is false.}
This case requires a finer analysis. We have to show that $\PP_\Gamma$ computes the false predicate, i.e., that every fair execution from every initial configuration converges to $0$. By fairness, it suffices to show that for every initial configuration $C_0$ and for every configuration $C$ reachable from $C_0$, it is possible to reach from $C$ a stable consensus of $0$.

Thanks to the $\bot$ state, which is introduced for this purpose, configurations $C$ in which two agents for the same existential input node choose inconsistent values eventually reach the configuration with all agents in state $\bot$, which is a stable consensus of $0$. Thanks to the assumption that every node is connected to the output gate by at least one path, configurations $C$ in which there are no agents for some node cannot reach any configuration in which some agent populates a state with $v_o=1$, and so $C$ itself is a stable consensus of $0$. So, loosely speaking, configurations in which the agents pick more than one value, or can pick no value at all, for some existential input eventually reach a stable consensus of $0$.

Consider the case in which, for every node $n$, the configuration $C$ has at least one agent in a state with node $n$. By fairness, $C$ eventually reaches a configuration $C'$ at which each agent for an existential input has chosen a boolean value, and we can assume that all agents for the same input choose the same value. This fixes a valuation $\vec{x}_0$ of the existential inputs. Recall that this valuation cannot change any more, since the protocol has no transitions for that. By assumption, there is $\vec{y}_0$ such that $\Gamma(\vec{x}_0, \vec{y}_0) = 0$.
We sketch how to reach a stable consensus of $0$ from $C'$. First, let the agents consume all
messages of $C'$, and let $C''$ be the resulting configuration. If $C''$ cannot reach any configuration with circuit output $1$, then the configuration reached after informing each agent about the value of $g_o$ is a stable consensus of $0$, and we are done.
Otherwise, starting from such a configuration with output $1$, let the agents send and receive the appropriate messages so that all agents for $\vec{y}$ choose the values of $\vec{y}_0$. After that, let the agent for  $g_o$ consume all remaining messages, if any, and let the protocol simulate $\Gamma$ on $\vec{x}_0, \vec{y}_0$. Notice that the simulation can be carried out even if there are multiple agents for the same gate $g$. Indeed, in this case, for every argument $g'$ of $g$, we let at least one of the agents corresponding to $g'$ send the message with the correct value for $g'$ to all the agents for
$n$. Since  $\Gamma(\vec{x}_0, \vec{y}_0) = 0$ by assumption, the agents for $g_o$ eventually update their value to $0$, and eventually all agents change their opinion about the output of the circuit to $0$. Let $C'''$ be the configuration so reached. We claim that $C'''$ is a stable consensus of $0$. Indeed, the state of a gate cannot change without a change in the argument values or the output gate $g_o$. Therefore it is enough to prove that the input values cannot change.
Since no transition can change $\vec{x}_0$, this can only happen by changing the values $\vec{y}_0$ of the universal inputs. But these values can only change by the transitions of (7), which require the agent to receive a message $(g_o, 1)$. This is not possible because the current value of $g_o$ is $0$, and the claim is proved.

This concludes the reduction to the correctness problem for DO protocols. We can easily transform it into a reduction to the single-instance correctness problem. Indeed, it suffices to observe that the executions of the circuit $\Gamma$ correspond to the fair executions of $\PP_\Gamma$ from the unique initial configuration $C_0$ with exactly one agent in state $\iota(n)$ for every node $n$. So $\PP_\Gamma$ computes $0$ from $C_0$ if{}f $\exists \vec{x} \forall \vec{y} \Gamma(\vec{x}, \vec{y}) = 1$, and we are done. So we  have:

\begin{restatable}{theorem}{ThmDOHardness}\label{thm:do-hardness}
The single-instance correctness and correctness problems for DO protocols are $\Pi_2^p$-hard.
\end{restatable}

%% file: intro-reach-obs.tex
In the next three sections  we prove that the correctness problem is \PSPACE-complete for IO protocols and $\Pi_2^p$-complete for DO protocols.
These are the most involved results of this paper. They can only be obtained after a detailed study of the reachability problem of IO and DO protocols, which we believe to be of independent interest. The roadmap for the three sections is as follows. 

\medskip\noindent\textbf{Section \ref{sec:reach-obs}.} Section \ref{sec:mfdo} introduces \emph{message-free delayed-ob\-ser\-vation protocols} (MFDO), an auxiliary model very close to DO protocols, but technically more convenient. As its name indicates, agents of MFDO protocols do not communicate by messages. Instead, they directly observe the current \emph{or past} states of other agents. As a consequence, a configuration of an MFDO protocol is completely determined by the states of its agents, which has technical advantages. At the same time, MFDO and DO protocols are very close, in the following sense. We call a configuration of a DO protocol a \emph{zero-message} configuration if all messages sent by the agents have already been received.  Given a DO protocol $\PP$ we can construct an MFDO protocol $\widehat{\PP}$, with the same set of states, such that for any two zero-message configurations $Z, Z'$ of $\PP$, we have $Z \trans{*} Z'$ in $\PP$ if{}f $Z \trans{*} Z'$ in $\widehat{\PP}$. (Observe that, since $\PP$ and $\widehat{\PP}$ have the same set of states, a zero-message configuration of  $\PP$ is also a configuration of $\widehat{\PP}$.) So, any question about the reachability relation between zero-message configurations of $\PP$ can be ``transferred'' to $\widehat{\PP}$, and answered there.

The rest of the section is devoted to the Pruning and Shortening Theorems.
Say that  a configuration $C$ is coverable from $C'$ if there exists a configuration $C''$ such that $C' \trans{*} C'' \ge C$.  The Pruning Theorems state that if a configuration $C$ of a protocol with $n$ states is coverable from $C'$, then it is also coverable from a ``small" configuration $D \leq C'$, where small means $|D| \leq |C|+ f(n)$ for a low-degree polynomial $f$. The Shortening Theorem states that every execution $C \trans{*} C'$ can be ``shortened'' to an execution $C \trans{\xi} C'$, where $\xi=t_1^{k_1} t_2^{k_2} \ldots t_m^{k_m}$ and $m \leq f(n)$ for some low-degree polynomial $f$
that depends only on $n$, not on $C$ or $C'$.  Intuitively, if we assume that the $k_i$ occurrences of $t_i$ are executed synchronously in one step, then the execution only takes $m$ steps.

\medskip\noindent\textbf{Section \ref{sec:sym-reach-obs}.} This section applies the Pruning and Shortening Theorems to the reachability problem between \emph{counting sets} of configurations. Intuitively, a counting set of configurations is a union of \emph{cubes}, and a cube is the set of all configurations $C$ lying between a lower bound configuration $L$ and an upper bound configuration $U$ with possibly infinite components. Observe  that counting sets may be infinite, but always have a finite representation. The reachability problem for counting sets asks, given two counting sets 
$\mathcal{C}$ and $\mathcal{C}'$, whether some configuration of  $\mathcal{C}'$ is reachable from some configuration of $\mathcal{C}'$.  
The section proves two very powerful Closure Theorems for IO and DO. The Closure Theorems state that for every counting set $\mathcal{C}$,
 the set $\poststar(\mathcal{C})$ of all configurations reachable from $\mathcal{C}$ is also a counting set; further, the same holds for the set $\prestar(\mathcal{C})$ of all configurations from which $\mathcal{C}$ can be reached. So, loosely speaking, counting sets are closed under reachability. 
Furthermore, the section shows that if $\mathcal{C}$ has a representation with ``small'' cubes, in a sense to be determined, then so do $\prestar(\mathcal{C})$ and $\poststar(\mathcal{C})$.

\medskip\noindent\textbf{Section \ref{sec:coNP}.} This section applies the Pruning, Shortening, and Closure Theorems to prove the \PSPACE \  and $\Pi_2^p$ upper bounds for the correctness problems of IO and DO protocols, respectively. The section shows that this is also the complexity of the single-instance correctness problems. 

\medskip
\medskip\noindent\textbf{Notation.}Throughout these sections, the last three components of the tuples describing protocols (input symbol set $\Sigma$, initial set mapping $\iota$, and output mapping $o$) play no role.
Therefore we represent a DO protocol by the simplified tuple $(Q,M,\delta_s,\delta_r)$, and an IO protocol as just a pair $(Q,\delta)$.

\medskip
Section \ref{sec:pruning} proves the Pruning Theorems for IO and MFDO protocols. 
Section \ref{sec:shortening} proves the Shortening Theorem for MFDO protocols. Finally, 
making use of the tight connection between MFDO and DO protocols, Section \ref{sec:pruning-and-shortening-DO} proves the Pruning and Shortening Theorems for DO protocols.
\cwk{open to proposals of other places for this paragraph if this seems akward.}

%
%

%% file: mfdo-intro-JE.tex
Immediate observation and delayed observation protocols present similarities.
Essentially, in an immediate observation protocol an agent updates its state when it observes that another agent is \emph{currently} in a certain state $q$, while in a delayed observation protocol the agent observes that another agent \emph{was} in a certain state $q$, provided that agent \emph{emitted a message when it was in $q$}. In  a message-free delayed observation protocol we assume that a sufficient amount of such messages is always emitted by default; this allows us to dispense with the message, and directly postulate that an agent can observe whether another agent went through a given state in the past. So the model is message-free, and, since agents can observe events that happened in the past, we call it ``message-free delayed observation''.

\begin{definition}
A message-free delayed observation (MFDO) protocol is a pair $\PP = (Q, \delta)$, where $Q$ is a set of states and $\delta: Q^2 \rightarrow Q$ is a transition function.
Considering $\delta$ as a set of transitions, 
we write $q \trans{o} q'$ for
$((q, o), q') \in \delta$.  The set of finite executions of $\PP$ is the set of finite sequences of configurations defined inductively as follows. Every configuration $C_0$ is a finite execution. A finite execution $C_0, C_1, \ldots, C_i$ enables a transition $q \trans{o} q'$ if $C_i(q)\geq 1$ and there exists $j\leq i$ such that $C_j(o)\geq 1$.
(We say the agent of $C_i$ at state $q$ \emph{observes} that there \emph{was} an agent in state $o$ at $C_j$.)
If $C_i$ enables $q \trans{o} q'$, then
$C_0, C_1, \ldots, C_i, C_{i+1}$ is also a finite execution of $\PP$, where $C_{i+1} = C_i - \multiset{q} + \multiset{q'}$. An infinite sequence of configurations is an execution of $\PP$ if all its finite prefixes are finite executions.
\end{definition}

We assign to every DO protocol an MFDO protocol.

\begin{definition}
\label{def:corresponding-mfdo}
Let $\PP_\textit{DO}=(Q, M, \delta_r, \delta_s)$ be a DO protocol. The \emph{MFDO protocol
corresponding to $\PP_\textit{DO}$} is $\PP_\textit{MFDO}=(Q,\delta)$, where $\delta$ is  the
set of transitions $q \trans{o} q'$ such that $q' = \delta_r(q,m)$ for some message $m\in M$, and $o$ is a state satisfying $ \delta_s(o) = (m,o)$.
\end{definition}

Notice that if $Q$ has multiple states $o_1, \ldots, o_k$  such that $\delta_s(o_i)=(m,o_i)$ for every $1 \leq i \leq k$, then $\PP_\textit{MFDO}$ contains a transition $q \trans{o_i} q'$ for every $1 \leq i \leq k$.

\begin{example}\label{do-protocol-mfdo}
Consider the DO protocol $\PP_\textit{DO}=(Q, M, \delta_r, \delta_s)$
where  $Q=M=\set{a, b, ab}$ and  $\Sigma=\set{a,b}$.
The send transitions are given by $\delta_s(q)=(q, q)$ for all $q \in Q$, \ie, every state
can send a message with its own identity to itself, denoted $q \send{q} q$. The receive transitions are $\delta_r(a,b)=ab$ and $\delta_r(b,a)=ab$, denoted $a \trans{b-} ab$ and $b \trans{a-} ab$.

The corresponding MFDO protocol is $\PP_\textit{MFDO}=(Q, \delta)$, where $\delta$ contains the transitions $a \trans{b} ab$ and $b \trans{a} ab$.
\end{example}

Notice that an agent of a DO protocol can ``choose'' not to send a message
when it goes through a state, and thus not enable a future transition that consumes such a message.
This does not happen in MFDO protocols. In particular, if a configuration $C$ of an MFDO protocol enables a transition $q \trans{o} q$, then the transition remains enabled forever, and in particular $C^\omega$ is an execution. This is not the case for a transition $q \trans{o-} q'$ of a DO protocol, because each occurrence of the transition consumes one message, and eventually there are no messages left.

Despite this difference, a DO protocol and its corresponding MFDO protocol are equivalent with respect to reachability questions in the following sense. Observe that a configuration of $\PP_\textit{DO}$ with zero messages is also a configuration of $\PP_\textit{MFDO}$.
From now on, given a DO protocol, we denote by $\mathcal{Z}$ the set of its \emph{zero-message configurations}.
For every $Z \in \mathcal{Z}$, we overload the notation $Z$ by 	also using it to denote
the configuration of the corresponding MFDO protocol which is the restriction of $Z$ to a multiset over $Q$.
The following lemma shows that for any two configurations $Z$ and $Z'$ with zero messages, $Z'$ is reachable  from $Z$ in $\PP_\textit{DO}$ if{}f it is reachable in $\PP_\textit{MFDO}$.

\begin{restatable}{lemma}{LemmaDOtoMFDO}
\label{lm:do-to-mfdo}
Let $\PP_\textit{DO}=(Q, M, \delta_s, \delta_r)$ be a DO protocol, and let $\PP_\textit{MFDO}=(Q,\delta)$ be its corresponding MFDO protocol. Let $Z,Z' \in \mathcal{Z}$ be two zero-message configurations.
Then $Z\trans{*}Z'$ in $\PP_{DO}$ if and only if $Z \trans{*} Z'$ in $\PP_\textit{MFDO}$.
\end{restatable}
\begin{proof}
\noindent \textbf{DO to MFDO.}
Let $Z \trans{\xi} Z'$ be an execution of
$\PP_\textit{DO}$ with $Z,Z' \in \mathcal{Z}$.
Let $\xi = t_1 t_2 \cdots t_n$, and let $C_0, C_1,C_2,\ldots,C_n$ be the configurations describing the number of agents in each state along $\xi$.
In particular, $C_0=Z$ and $C_n=Z'$.
Define the sequence $\tau$ as follows. For every transition $t_i$:
\begin{itemize}
\item If $t_i$ is a send transition (\ie,  if $t_i = q \trans{m+} q$ for some $q$ and $ m$), then delete $t_i$.\\
Observe that, since the occurrence of $t_i$ does not change the state of any agent, we have $C_{i} = C_{i+1}$, and so in particular $C_{i} \trans{\epsilon} C_{i+1}$ in $\PP_\textit{MFDO}$.
\item If $t_i$ is a receive transition,  \ie,  if $t_i = q \trans{m-} q'$ for some $q$, $q'$, and $m$,
then replace it by the transition $q \trans{o} q'$, where $o$ is any state satisfying
$t_j = o \trans{m+} o$ for some index $j \leq i$.\\
Observe that the transition $t_j$ must exist, because every message received has been sent.
Further, since both $t_i$ and $q \trans{o} q'$ move an agent from $q$ to $q'$, we have $C_{i} \trans{u_i} C_{i+1}$ in $\PP_\textit{MFDO}$ for $u_i = q \trans{o} q'$.
\end{itemize}
The result follows from the fact that in both cases we have $C_{i} \trans{*} C_{i+1}$ in $\PP_\textit{MFDO}$.

\smallskip

\noindent \textbf{MFDO to DO.}
Let $Z \trans{\tau} Z'$ be an execution of $\PP_\textit{MFDO}$, and let $\tau = u_1 u_2 \cdots u_n$, where $u_i= q_i \trans{o_i} q_{i+1}$. We define the sequence $\xi$, such that $Z' \trans{\xi} Z$, in two steps as follows.
\begin{enumerate}
\item First replace every transition $u_i$ by $q_i \trans{m-} q_{i+1}$ for a message $m\in M$ such that $\delta_s(o_i)=(m,o_i)$.
Transition $q_i \trans{m-} q_{i+1}$ exists in $\PP_\textit{DO}$ by construction of $\PP_\textit{MFDO}$.
\item For each message $m \in M$ in $\xi$, denote by $q_m$ the state such that $\delta_s(q_m)=(m,q_m)$ and let $\#(m,\xi)$ denote the number of times $m$ is consumed along $\xi$.
If there are multiple states with such property, we choose the state that occurs earliest in the original execution.
Add $\#(m,\xi)$ iterations of $q_m \trans{m+} q_m$ at the first configuration along $\xi$ in which state $q_m$ is populated.
This ensures that the messages that the agents need to move from $q_i$ to $q_{i+1}$ are always available to be received and that all the messages will be consumed at the end of the execution.
\end{enumerate}
Thus $\xi$ is enabled and goes from $Z'$ to $Z$.
 \end{proof}



%% file: new-pruning-setting-JE.tex
The Pruning Theorems for IO and MFDO protocols are proved in the same way. Given an execution $C'' \trans{\xi} C' \geq C$, we examine the trajectories of the different agents during the execution of $\xi$. For this, we assign trajectories to the agents in an arbitrary way, but consistent with the configurations reached during the execution. For example, consider a protocol with states $q_1, q_2, q, q_1', q_2'$ in which two agents, initially in states $q_1$ and $q_2$, first move to $q$, after which one of them moves to $q_1'$ and the other to $q_2'$. Since the two agents are indistinguishable, we can choose to assume that their trajectories were $q_1, q, q_1'$ and $q_2, q , q_2'$, or that they were $q_1, q, q_2'$ and $q_2, q , q_1'$. After ``splitting'' the execution into a multiset of trajectories, one for each agent, we ``prune'' the multiset, keeping only those trajectories that are ``necessary" to cover $C$. This yields a smaller multiset, which we then ``transform back'' into an execution.

\subsubsection{Pruning Theorem for IO protocols}
\label{subsec:pruningIO}

\begin{definition}
A \emph{trajectory} of an IO protocol $\PP=(Q, \delta)$ is a sequence $\tau =q_1 \ldots q_n$ of states. We let $\tau(i)$ denote the $i$-th state of $\tau$. The \emph{$i$-th step} of $\tau$ is the pair $\tau(i)\tau(i+1)$ of adjacent states.

A \emph{history} is a multiset of trajectories of the same length. The length of a history is the common length of its trajectories. Given a history $H$ of length $n$ and index $1 \leq i \leq n$, the \emph{$i$-th configuration of $H$}, denoted $C_{H}^i$, is defined as follows: for every state $p$, $C_{H}^i(q)$ is the number of trajectories $\tau \in H$ such that $\tau(i)=q$.
The configurations $C_{H}^1$ and $C_{H}^n$ are called the \emph{initial} and \emph{final} configurations of $H$.
\end{definition}

\input{figure-history.tex}

\begin{example}
\label{history-ex}
Let $\PP=(Q,\delta)$  be the IO protocol with $Q=\{ q_1,q_2,q_3 \}$ and $\delta = \{t_1, t_2, t_3, t_4\}$, where
$$\begin{array}{lcl}
t_1 = q_1 \trans{q_1} q_2  & \quad  & t_3 = q_1 \trans{q_3} q_3\\
t_2 = q_2 \trans{q_2} q_3  &            & t_4 = q_2 \trans{q_3} q_3
\end{array}$$
We use this protocol as running example throuhout the section.
Histories of $\PP$ can be graphically represented. Figure \ref{figure-history} shows a history $H$ of length $7$. It consists of five trajectories: one trajectory from $q_3$ to $q_3$ passing only through $q_3$, and four trajectories from $q_1$ to $q_3$ which follow different state sequences.
The first configuration of $H$ is $C_{H}^1= (4,0,1)$ and the seventh and last configuration is $C_{H}^7=(0,0,5)$.
\end{example}

\begin{definition}
A history $H$ of length $n\geq 1$ is \emph{realizable} in an IO protocol $\PP$ if there exist transitions $t_1, \ldots, t_{n-1}$ of $\PP$ and numbers $k_1, \ldots, k_{n-1} \geq 0$ such that
$$C_{H}^1 \trans{t_1^{k_1}}C_{H}^2 \cdots  C_{H}^{n-1} \trans{t_{n-1}^{k_{n-1}}} C_{H}^n \ ,$$
\noindent where for every transition $t$ we define $C \trans{t^0} C'$ if{}f $C= C'$.
\end{definition}

\begin{remark}
Notice that histories of length $1$ are always realizable. Observe also that there may be more than one realizable history corresponding to a firing sequence, because the firing sequence does not keep track of which agent visits which states, while the history does.
\end{remark}

\begin{example}
\label{history-ex2}
The history $H$ of Figure \ref{figure-history} is realizable in $\PP$.
Indeed, we have  $C_{H}^1 \trans{t_3 \, t_1^2 \, t_3 \, t_2 \, t_4 } C_{H}^7$.
\end{example}

We introduce well structured histories. Intuitively, they are the histories in which at every step all agents that move execute the same transition, and so there are states $q, q'$ such that all the agents move from $q$ to $q'$.

\begin{definition}
\label{def:horizon}
A step $\tau(i)\tau(i+1)$ of a trajectory $\tau$ is \emph{horizontal} if $\tau(i) = \tau(i+1)$, and \emph{non-horizontal} otherwise.

A history $H$ of length $n$ is \emph{well structured} if for every $1 \leq i \leq n-1$ one of the two following conditions hold:
\begin{itemize}
\item[(i)] For every trajectory $\tau \in H$, the $i$-th step of $\tau$ is horizontal.
\item[(ii)] For every two trajectories $\tau_1, \tau_2 \in H$, if the $i$-th steps of $\tau_1$ and $\tau_2$ are non-horizontal, then they are equal.
\end{itemize}
\end{definition}

\begin{example}
The history of Figure \ref{figure-history} is well structured. The third step of all five trajectories is horizontal. The second step is horizontal for three trajectories, and non-horizontal for the other two; the two non-horizontal steps are equal, namely $q_1 \, q_2$.
\end{example}

\paragraph{Characterizing histories. }

We show that the set of executions of an IO protocol is completely determined by its well-structured and realizable histories. The proof is purely technical, and can be found in the Appendix.

\begin{restatable}{lemma}{LemmaRealizableIO}
\label{lem:realizable-io}
Let $\PP$ be an IO protocol.
For every configuration $C, C'$ the following holds: $C \trans{*} C'$ if{}f there exists a well-structured and realizable history in $\PP$ with $C$ and $C'$ as initial and final configurations.
\end{restatable}

We now proceed to give a syntactic characterization of the well-structured and realizable histories.

\begin{definition}
A history $H$ is \emph{compatible} with an IO protocol $\PP$
if for every trajectory $\tau$ of $H$ and for every non-horizontal step $\tau(i)\tau(i+1)$ of $\tau$,
the protocol $\PP$ contains a transition $\tau(i) \trans{o} \tau(i+1)$ for a state $o$ such that
$H$ contains a trajectory $\tau'$ with $\tau'(i)=\tau'(i+1)=o$.
\end{definition}

Intuitively, a history is compatible with a protocol if for every non-horizontal step from,
say, $q$ to $q'$, the protocol has a transition of the form $q \trans{o} q'$ for some observed state
$o$. Since the transition can only happen if an agent in $q$ observes $o$, there must be another agent in state $o$ (the
one with trajectory $\tau'$).

\begin{example}
The history of Figure \ref{figure-history} is compatible with the IO protocol of Example \ref{history-ex}.
Consider for example the trajectory $\tau = q_1 \, q_1 \, q_2 \,  q_2 \,  q_2 \,  q_3 \,  q_3$. It has two
non-horizontal steps, namely $\tau(2)\tau(3) = q_1 \, q_2$ and $\tau(5)\tau(6) = q_2 \, q_3$.
The corresponding transitions are $q_1 \trans{q_1} q_2$ and $q_2 \trans{q_2} q_3$.
\end{example}

\begin{restatable}{lemma}{LemmaHistoryIO}
\label{lem:compatible-io}
Let $\PP$ be an IO protocol. A well-structured history is realizable in $\PP$ if{}f it is compatible with $\PP$.
\end{restatable}

\paragraph{Pruning.}

We introduce \emph{bunches of trajectories}, and present a lemma about pruning bunches.
Then, we prove the Pruning Theorem for IO protocols.

\begin{definition}
 A \emph{bunch} is a multiset of trajectories of the same length and with the same initial and final states.
\end{definition}

\begin{example}
        The history of Figure \ref{figure-history} consists of a trajectory from $q_3$ to $q_3$ (which can be considered a bunch of size $1$),
        and a bunch of four trajectories with initial state $q_1$ and final state $q_3$.
\end{example}

We show that every well-structured and realizable history containing a bunch of more than $|Q|$ trajectories can be ``pruned'', meaning that the bunch can be replaced by a smaller one, while keeping the history well-structured and realizable.

\begin{lemma}
\label{lm:pruning}
Let $\PP=(Q,\delta)$ be an IO protocol. Let $H$ be a well-structured and realizable history of $\PP$ containing a bunch $B\subseteq H$ of size larger than $|Q|$. There exists a nonempty bunch $B'$ of size at most $|Q|$, of the same length and with the same initial and final states as $B$, such that the history $H' \defeq H - B + B'$ (where $+$ and  $-$ denote multiset addition and multiset subtraction, respectively) is also well-structured and realizable.
\end{lemma}
\begin{proof}
Let $Q_B$ be a set of all states visited by at least one trajectory in the bunch $B$.
        For every $q\in Q_B$, let $f(q)$ and $l(q)$ be the earliest and the latest
        moment in time at which $q$ is visited by any of the trajectories
        (the first and last occurrences can belong to different trajectories).

        For every $q \in Q_B$, let $\tau_q = \tau_{q,1} \tau_{q,2} \tau_{q,3}$, where $\tau_{q,1}$ is a prefix of length $f(q)-1$ of
        some trajectory of $B$ with $q$ at the moment $f(q)$; $\tau_{q,2} = q^{l(q)-f(q)}$; and $\tau_{q,3}$ is a suffix
        of some trajectory of $B$ with the state $q$ at the moment $l(q)$, starting at the moment $l(q)$. The prefix and the suffix
        exist by the definition of $f(q)$ and $l(q)$.

        Let $B'=\{\tau_q \mid q \in Q_B\}$, and let  $H' = H - B + B'$. We prove that $H'$
       is well structured and compatible with $\PP$. By Lemma \ref{lem:compatible-io}, this
       proves that $H'$ is well structured and realizable in $\PP$.

        Let us first show that $H'$ is well structured. Notice that every trajectory of $B'$ is the concatenation
        of a prefix of a trajectory of $B$, a sequence of horizontal steps, and a suffix of another trajectory of $B$.
        Hence, if $B'$ contains a trajectory whose $i$-th step is non-horizontal, then the same holds for $B$.
        It follows:
        \begin{itemize}
        \item If the $i$-th step of $H$ satisfies condition (i) of Definition \ref{def:horizon}, then so does
        the $i$-th step of $H'$.
        \item If the $i$-th step of $H$ satisfies condition (ii), then all its non-horizontal
        $i$-th steps are equal. So all non-horizontal $i$-th steps of $H'$ are also equal, which implies that the
        $i$-th step of $H'$ also satisfies condition (ii).
        \end{itemize}

        Let us now show that $H'$ is compatible with $\PP$. Let $\tau'$ be a trajectory of $H'$, and let
        $\tau'(i)\tau'(i+1)$ be a non-horizontal step of $\tau'$. We show that $\PP$ has a transition
        $\tau'(i) \trans{o'} \tau'(i+1)$, where the state $o'$ satisfies that some trajectory $\tau'' \in H'$ satisfies
        $\tau''(i)=\tau''(i+1)=o'$.

        Since $\tau'(i)\tau'(i+1)$ is a non-horizontal step, by the argument above $H$ contains a trajectory $\tau$ such that
        $\tau(i)\tau(i+1) = \tau'(i)\tau'(i+1)$. Further, $H$ is realizable in $\PP$ by assumption, and so by Lemma   \ref{lem:compatible-io} $H$ is compatible with $\PP$. So $\PP$ has a transition $\tau(i) \trans{o} \tau(i+1)$,
and $H$ has a trajectory $\tilde{\tau}$ such that $\tilde{\tau}(i) = \tilde{\tau}(i+1) = o$. Choose $o':=o$. Since
$\tau(i)\tau(i+1) = \tau'(i)\tau'(i+1)$, we have that $\tau'(i) \trans{o'} \tau'(i+1)$ is a transition of $\PP$.
It remains to show that some trajectory $\tau'' \in H'$ satisfies $\tau''(i)=\tau''(i+1)=o'$.
Consider two cases:
\begin{itemize}
\item $\tilde{\tau} \notin B$.  Then $\tilde{\tau} \in H'$. Since $\tilde{\tau}(i) = \tilde{\tau}(i+1) = o$, we can choose $\tau'' := \tilde{\tau}$.
\item  $\tilde{\tau} \in B$. Then, since  $\tilde{\tau}(i) = \tilde{\tau}(i+1) = o$, we have $o \in Q_B$. So $f(o)\leq i<i+1\leq l(o)$. By the definition of $B'$, the history $H'$ contains a  trajectory $\tau_o$ for the state $o$, which stays at state $o$ from time $f(o)$ to time $l(o)$. So we have $\tau_o(i) \tau_o(i+1)= o$, and we can choose $\tau'':=\tau_o$.
\end{itemize}
 \end{proof}

\input{figure-pruned.tex}
\begin{example}
Consider the well-structured and realizable history of Figure \ref{figure-history}. It leads from configuration $(4,0,1)$
to $(0,0,5)$. The bunch $B$ from $q_1$ to $q_3$ is of size four, and so bigger than $|Q|=3$.
The set $Q_B$ of states visited by trajectories of $B$ is equal to $Q$.

Figure \ref{figure-pruned} shows for every state $q\in Q_B$ the first and last moments $f(q)$ and $l(q)$.
Lemma \ref{lm:pruning} shows that we can replace $B$ in $H$ by the smaller bunch $B'$ consisting of the trajectories $\tau_{q_1},\tau_{q_2},\tau_{q_3}$, drawn in dashed lines in Figure \ref{figure-pruned}.
Notice that the non-horizontal $5$-th step in $H$ does not appear in the new
well-structured and realizable history $H' = H - B + B'$. The history $H'$ satisfies
$C_{H'}^1=(3,0,1) \trans{t_3 t_1 t_3 t_4 }(0,0,4)=C_{H'}^7$. 
\end{example}

Using Lemma \ref{lm:pruning} we can now prove the Pruning Theorem for IO protocols:

\begin{theorem} [IO Pruning]
\label{thm:pruning}
Let $\PP=(Q,\delta)$ be an IO protocol,
let $L'$ and $L$ be multisets of states of $\PP$,
and let
$C' \trans{*} C$ be an execution of $\PP$
such that $L' \leq C'$ and $C \geq L$.
There exist configurations $D'$ and $D$ such that
\begin{center}
\(
\begin{array}[b]{@{}c@{}c@{}c@{}c@{}c@{}}
C' &  \trans{\hspace{1em}*\hspace{1em}} & C  \\[0.1cm]
\geq &  & \geq \\[0.1cm]
D' & \trans{\hspace{1em}*\hspace{1em}} &D  \\[0.1cm]
\geq &  & \geq \\[0.1cm]
L' &  &L  \\[0.1cm]
\end{array}
\)
\end{center}
and $|D'| = |D| \leq |L| + |L'| + |Q|^3$.
\end{theorem}
\begin{remark}
We will often use the theorem when $L'$ or $L$ is empty,
which is why we call them multisets of states instead
of configurations.
\end{remark}
\begin{proof}
Let $L' \leq C' \trans{*}  C  \  \geq  \ L$.
By Lemma \ref{lem:realizable-io},
there is a well-structured realizable history $H$ with $C'$ and $C$ as initial and final configurations, respectively.
Let $H_L \subseteq H$ be an arbitrary sub(multi)set of $H$ with the multiset of final states $L$,
and $H_{L'}$ be a sub(multi)set of $H$ with multiset of final states $L'$.
Define $H_{0}$ as their union (maximum) $max(H_L,H_{L'})$, and let $H'=H-H_0$.
Further, for every $p, p' \in Q$, let $H'_{p,p'}$ be
the bunch of all trajectories of $H'$ with $p$ and $p'$ as initial and final states, respectively.
We have $$H' = \sum_{p,p' \in Q} H'_{p,p'}$$

So $H'$ is the union of $|Q|^2$ (possibly empty) bunches. Applying Lemma \ref{lm:pruning} to each bunch of $H'$ with more than $|Q|$ trajectories yields a new history

$$H'' = \sum_{p,p' \in p} H''_{p,p'}$$

\noindent where the sum represents multiset addition, such that $|H''_{p,p'}| \leq |Q|$ for every $p, p' \in Q$, and such that the history $H'' + H_0$ is well structured and realizable.

Let $D'$ and $D$ be the initial and final configurations of $H'' + H_0$.
We show that $D'$ and $D$ satisfy the required properties:
\begin{itemize}
\item $D' \trans{*} D$, because  $H'' + H_0$ is well structured and realizable.
\item $D' \geq L'$ and $D \geq L$, because $H_0 \leq H'' + H_0$.
\item $|D'| \leq |L'| + |L| + |Q|^3$ because $|H'' + H_0| = \sum_{p', p} |H''_{p,p'}| + |H_0| \leq |Q|^2 \cdot |Q| + |H_{L'}|+|H_{L}| = |L'| + |L| + |Q|^3$.
\end{itemize}
This concludes the proof.
 \end{proof}

\begin{remark}
A slight modification of our construction allows one to prove Theorem~\ref{thm:pruning} (but not Lemma \ref{lm:pruning}) with $2|Q|^2$ overhead instead of $|Q|^3$. We provide more details in the appendix.
However, since some results of Section \ref{sec:small-instance} explicitly rely on Lemma~\ref{lm:pruning}, we prove Theorem~\ref{thm:pruning} as a consequence of Lemma~\ref{lm:pruning} for simplicity.
\end{remark}

\subsubsection{Pruning Theorem for MFDO protocols}

The proof of the Pruning Theorem for MFDO protocols is similar to the one for IO protocols. It follows the same sequence of steps, but with some differences.

Trajectories and histories of MFDO protocols are defined as for DO protocols.
Well-structured and realizable histories also have the same definition, and Lemma \ref{lem:realizable-io}
holds, with the same proof. Let us see an example:

\input{figure-history-mfdo.tex}

\begin{example}
\label{history-ex-mfdo}
Recall the MFDO protocol $\PP_\textit{MFDO}=(Q, \delta)$ of Example \ref{do-protocol-mfdo},
with $Q=\{a, b, ab\}$ and $\delta = \{t_1, t_2\}$, where $t_1 = a \trans{b} ab$ and $t_2 = b \trans{a} ab$.
Figure \ref{figure-history-mfdo} shows a graphical representation of a history $H$ of $\PP_\textit{MFDO}$.  It consists of five trajectories: one trajectory from $a$ to $ab$, and four trajectories from $b$ to $ab$, following different state sequences. The first configuration of $H$ is $C_{H}^1= (1,4,0)$, and the fourth and last  configuration is $C_{H}^4=(0,0,5)$. The history is well structured and realizable. In particular, we have
$$C_{H}^1 \trans{t_2 \, t_1 \, t_2^3} C_{H}^4 \ .$$
\end{example}

For MFDO-protocols we also need the notion of the sets of states visited along a history.

\begin{definition}
Let $H$ be a history of an MFDO protocol of length $n$.
The set of states \emph{visited in the first $i$ steps of $H$} is  $\mathcal{S}_{H}^i:= \{\tau(j) \mid \tau \in H, j \leq i \}$.
The set of states \emph{visited} by $H$ is denoted $\mathcal{S}_{H}$ and defined by $\mathcal{S}_{H} := \mathcal{S}_{H}^n$.
\end{definition}

\begin{example}
Let $H$ be the history of Figure \ref{figure-history-mfdo}. We have $\mathcal{S}_{H}^1 = \set{a,b}$, $\mathcal{S}_{H}^i = \set{a,b,ab}$
for $i=2,3,4$, and $\mathcal{S}_{H} = \set{a,b,ab}$.
\end{example}

%% file: figure-history.tex
\begin{figure}
\centering
\begin{tikzpicture}
  \node[circle, draw]
    (Node11)
    {~~~}; 
    \node[right=1 of Node11, circle, draw]
    (Node12)
    {~~~};
    \node[right=1 of Node12, circle, draw]
    (Node13)
    {~~~};
    \node[right=1 of Node13, circle, draw]
    (Node14)
    {~~~};
    \node[right=1 of Node14, circle, draw]
    (Node15)
    {~~~};
    \node[right=1 of Node15, circle, draw]
    (Node16)
    {~~~};
    \node[right=1 of Node16, circle, draw]
    (Node17)
    {~~~};
    \node[below=0.7 of Node11, circle, draw]
    (Node21)
    {~~~};
    \node[right=1 of Node21, circle, draw]
    (Node22)
    {~~~};
    \node[right=1 of Node22, circle, draw]
    (Node23)
    {~~~};
    \node[right=1 of Node23, circle, draw]
    (Node24)
    {~~~};
    \node[right=1 of Node24, circle, draw]
    (Node25)
    {~~~};
    \node[right=1 of Node25, circle, draw]
    (Node26)
    {~~~};
    \node[right=1 of Node26, circle, draw]
    (Node27)
    {~~~};
    \node[below=0.7 of Node21, circle, draw]
    (Node31)
    {~~~};
    \node[right=1 of Node31, circle, draw]
    (Node32)
    {~~~};
    \node[right=1 of Node32, circle, draw]
    (Node33)
    {~~~};
    \node[right=1 of Node33, circle, draw]
    (Node34)
    {~~~};
    \node[right=1 of Node34, circle, draw]
    (Node35)
    {~~~};
    \node[right=1 of Node35, circle, draw]
    (Node36)
    {~~~};
    \node[right=1 of Node36, circle, draw]
    (Node37)
    {~~~};
        \node[left=0.3 of Node11]
        (Node11l)
        {$q_1$};
        \node[right=0.3 of Node17]
        (Node17r)
        {$q_1$};
        \node[left=0.3 of Node21]
        (Node21l)
        {$q_2$};
        \node[right=0.3 of Node27]
        (Node27r)
        {$q_2$};
        \node[left=0.3 of Node31]
        (Node31l)
        {$q_3$};
        \node[right=0.3 of Node37]
        (Node37r)
        {$q_3$};

  \draw
  ($(Node11.center)+(0,1.50mm)$)
    --
  ($(Node12.center)+(0,1.50mm)$)
  ;
  \draw
  ($(Node12.center)+(0,1.50mm)$)
    --
  ($(Node13.center)+(0,1.50mm)$)
  ;
  \draw($(Node13.center)+(0,1.50mm)$)
    --
  ($(Node14.center)+(0,1.50mm)$)
  ;
  \draw
  ($(Node14.center)+(0,1.50mm)$)
    --
  ($(Node35.center)+(0,1.50mm)$)
  ;
  \draw
  ($(Node35.center)+(0,1.50mm)$)
    --
  ($(Node36.center)+(0,1.50mm)$)
  ;
  \draw
  ($(Node36.center)+(0,1.50mm)$)
    --
  ($(Node37.center)+(0,1.50mm)$)
  ;
  \draw
  ($(Node11.center)+(0,0.0mm)$)
    --
  ($(Node12.center)+(0,0.0mm)$)
  ;
  \draw
  ($(Node12.center)+(0,0.0mm)$)
    --
  ($(Node23.center)+(0,0.0mm)$)
  ;
  \draw
  ($(Node23.center)+(0,0.0mm)$)
    --
  ($(Node24.center)+(0,0.0mm)$)
  ;
  \draw
  ($(Node24.center)+(0,0.0mm)$)
    --
  ($(Node25.center)+(0,0.0mm)$)
  ;
  \draw
  ($(Node25.center)+(0,0.0mm)$)
    --
  ($(Node36.center)+(0,0.0mm)$)
  ;
  \draw
  ($(Node36.center)+(0,0.0mm)$)
    --
  ($(Node37.center)+(0,0.0mm)$)
  ;
  \draw
  ($(Node11.center)+(0,0.75mm)$)
    --
  ($(Node12.center)+(0,0.75mm)$)
  ;
  \draw
  ($(Node12.center)+(0,0.75mm)$)
    --
  ($(Node23.center)+(0,0.75mm)$)
  ;
  \draw
  ($(Node23.center)+(0,0.75mm)$)
    --
  ($(Node24.center)+(0,0.75mm)$)
  ;
  \draw
  ($(Node24.center)+(0,0.75mm)$)
    --
  ($(Node25.center)+(0,0.75mm)$)
  ;
  \draw
  ($(Node25.center)+(0,0.75mm)$)
    --
  ($(Node26.center)+(0,0.75mm)$)
  ;
  \draw
  ($(Node26.center)+(0,0.75mm)$)
    --
  ($(Node37.center)+(0,0.75mm)$)
  ;
  \draw
  ($(Node11.center)+(0,-0.75mm)$)
    --
  ($(Node32.center)+(0,-0.75mm)$)
  ;
  \draw
  ($(Node32.center)+(0,-0.75mm)$)
    --
  ($(Node33.center)+(0,-0.75mm)$)
  ;
  \draw
  ($(Node33.center)+(0,-0.75mm)$)
    --
  ($(Node34.center)+(0,-0.75mm)$)
  ;
  \draw
  ($(Node34.center)+(0,-0.75mm)$)
    --
  ($(Node35.center)+(0,-0.75mm)$)
  ;
  \draw
  ($(Node35.center)+(0,-0.75mm)$)
    --
  ($(Node36.center)+(0,-0.75mm)$)
  ;
  \draw
  ($(Node36.center)+(0,-0.75mm)$)
    --
  ($(Node37.center)+(0,-0.75mm)$)
  ;
  \draw
  ($(Node31.center)+(0,-1.50mm)$)
    --
  ($(Node32.center)+(0,-1.50mm)$)
  ;
  \draw
  ($(Node32.center)+(0,-1.50mm)$)
    --
  ($(Node33.center)+(0,-1.50mm)$)
  ;
  \draw
  ($(Node33.center)+(0,-1.50mm)$)
    --
  ($(Node34.center)+(0,-1.50mm)$)
  ;
  \draw
  ($(Node34.center)+(0,-1.50mm)$)
    --
  ($(Node35.center)+(0,-1.50mm)$)
  ;
  \draw
  ($(Node35.center)+(0,-1.50mm)$)
    --
  ($(Node36.center)+(0,-1.50mm)$)
  ;
  \draw
  ($(Node36.center)+(0,-1.50mm)$)
    --
  ($(Node37.center)+(0,-1.50mm)$)
  ;

\end{tikzpicture}
\caption{ Realizable history in IO protocol with three states}
\label{figure-history}
\end{figure}
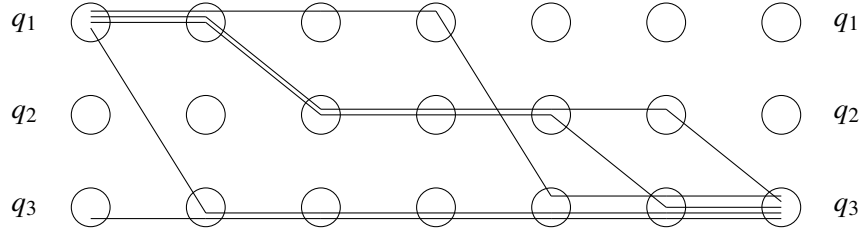

%% file: figure-pruned.tex
\begin{figure}
\centering
\begin{tikzpicture}

  \node[circle, draw]
    (Node11)
    {~~~}; 
    \node[right=1 of Node11, circle, draw]
    (Node12)
    {~~~};
    \node[right=1 of Node12, circle, draw]
    (Node13)
    {~~~};
    \node[right=1 of Node13, circle, draw]
    (Node14)
    {~~~};
    \node[right=1 of Node14, circle, draw]
    (Node15)
    {~~~};
    \node[right=1 of Node15, circle, draw]
    (Node16)
    {~~~};
    \node[right=1 of Node16, circle, draw]
    (Node17)
    {~~~};
    \node[below=0.7 of Node11, circle, draw]
    (Node21)
    {~~~};
    \node[right=1 of Node21, circle, draw]
    (Node22)
    {~~~};
    \node[right=1 of Node22, circle, draw]
    (Node23)
    {~~~};
    \node[right=1 of Node23, circle, draw]
    (Node24)
    {~~~};
    \node[right=1 of Node24, circle, draw]
    (Node25)
    {~~~};
    \node[right=1 of Node25, circle, draw]
    (Node26)
    {~~~};
    \node[right=1 of Node26, circle, draw]
    (Node27)
    {~~~};
    \node[below=0.7 of Node21, circle, draw]
    (Node31)
    {~~~};
    \node[right=1 of Node31, circle, draw]
    (Node32)
    {~~~};
    \node[right=1 of Node32, circle, draw]
    (Node33)
    {~~~}; 
    \node[right=1 of Node33, circle, draw]
    (Node34)
    {~~~};
    \node[right=1 of Node34, circle, draw]
    (Node35)
    {~~~};
    \node[right=1 of Node35, circle, draw]
    (Node36)
    {~~~};
    \node[right=1 of Node36, circle, draw]
    (Node37)
    {~~~};
        \node[left=0.3 of Node11]
        (Node11l)
        {$q_1$};
        \node[right=0.3 of Node17]
        (Node17r)
        {$q_1$};
        \node[left=0.3 of Node21]
        (Node21l)
        {$q_2$};
        \node[right=0.3 of Node27]
        (Node27r)
        {$q_2$};
        \node[left=0.3 of Node31]
        (Node31l)
        {$q_3$};
        \node[right=0.3 of Node37]
        (Node37r)
        {$q_3$};

        \node[above=0.0 of Node11](Node11u){$f$};
        \node[above=0.0 of Node14](Node14u){$l$};
        \node[above=0.0 of Node23](Node23u){$f$};
        \node[above=0.0 of Node26](Node26u){$l$};
        \node[above=0.0 of Node32](Node32u){$f$};
        \node[above=0.0 of Node37](Node37u){$l$};
        
  \draw[dashed]
  ($(Node11.center)+(0,1.50mm)$)
    --
  ($(Node12.center)+(0,1.50mm)$)
  ;
  \draw[dashed]
  ($(Node12.center)+(0,1.50mm)$)
    --
  ($(Node13.center)+(0,1.50mm)$)
  ;
  \draw[dashed]($(Node13.center)+(0,1.50mm)$)
    --
  ($(Node14.center)+(0,1.50mm)$)
  ;
  \draw[dashed]
  ($(Node14.center)+(0,1.50mm)$)
    --
  ($(Node35.center)+(0,1.50mm)$)
  ;
  \draw[dashed]
  ($(Node35.center)+(0,1.50mm)$)
    --
  ($(Node36.center)+(0,1.50mm)$)
  ;
  \draw[dashed]
  ($(Node36.center)+(0,1.50mm)$)
    --
  ($(Node37.center)+(0,1.50mm)$)
  ;
  \draw[dashed]
  ($(Node11.center)+(0,0.0mm)$)
    --
  ($(Node12.center)+(0,0.0mm)$)
  ;
  \draw[dashed]
  ($(Node12.center)+(0,0.0mm)$)
    --
  ($(Node23.center)+(0,0.0mm)$)
  ;
  \draw[dashed]
  ($(Node23.center)+(0,0.0mm)$)
    --
  ($(Node24.center)+(0,0.0mm)$)
  ;
  \draw[dashed]
  ($(Node24.center)+(0,0.0mm)$)
    --
  ($(Node25.center)+(0,0.0mm)$)
  ;
  \draw[dashed]
  ($(Node25.center)+(0,0.0mm)$)
    --
  ($(Node26.center)+(0,0.0mm)$)
  ;
  \draw[dashed]
  ($(Node26.center)+(0,0.0mm)$)
    --
  ($(Node37.center)+(0,0.0mm)$)
  ;
  \draw[dashed]
  ($(Node11.center)+(0,-0.75mm)$)
    --
  ($(Node32.center)+(0,-0.75mm)$)
  ;
  \draw[dashed]
  ($(Node32.center)+(0,-0.75mm)$)
    --
  ($(Node33.center)+(0,-0.75mm)$)
  ;
  \draw[dashed]
  ($(Node33.center)+(0,-0.75mm)$)
    --
  ($(Node34.center)+(0,-0.75mm)$)
  ;
  \draw[dashed]
  ($(Node34.center)+(0,-0.75mm)$)
    --
  ($(Node35.center)+(0,-0.75mm)$)
  ;
  \draw[dashed]
  ($(Node35.center)+(0,-0.75mm)$)
    --
  ($(Node36.center)+(0,-0.75mm)$)
  ;
  \draw[dashed]
  ($(Node36.center)+(0,-0.75mm)$)
    --
  ($(Node37.center)+(0,-0.75mm)$)
  ;
  \draw
  ($(Node31.center)+(0,-1.50mm)$)
    --
  ($(Node32.center)+(0,-1.50mm)$)
  ;
  \draw
  ($(Node32.center)+(0,-1.50mm)$)
    --
  ($(Node33.center)+(0,-1.50mm)$)
  ;
  \draw
  ($(Node33.center)+(0,-1.50mm)$)
    --
  ($(Node34.center)+(0,-1.50mm)$)
  ;
  \draw
  ($(Node34.center)+(0,-1.50mm)$)
    --
  ($(Node35.center)+(0,-1.50mm)$)
  ;
  \draw
  ($(Node35.center)+(0,-1.50mm)$)
    --
  ($(Node36.center)+(0,-1.50mm)$)
  ;
  \draw
  ($(Node36.center)+(0,-1.50mm)$)
    --
  ($(Node37.center)+(0,-1.50mm)$)
  ;

\end{tikzpicture}
\caption{ History $H$ of Figure \ref{figure-history} after pruning}
\label{figure-pruned}
\end{figure}

%% file: figure-history-mfdo.tex
\begin{figure}
\centering
\begin{tikzpicture}
  \node[circle, draw]
    (Node11)
    {~~~}; 
    \node[right=1 of Node11, circle, draw]
    (Node12)
    {~~~};
    \node[right=1 of Node12, circle, draw]
    (Node13)
    {~~~};
    \node[right=1 of Node13, circle, draw]
    (Node14)
    {~~~};
    \node[below=0.7 of Node11, circle, draw]
    (Node21)
    {~~~};
    \node[right=1 of Node21, circle, draw]
    (Node22)
    {~~~};
    \node[right=1 of Node22, circle, draw]
    (Node23)
    {~~~};
    \node[right=1 of Node23, circle, draw]
    (Node24)
    {~~~};
    \node[below=0.7 of Node21, circle, draw]
    (Node31)
    {~~~};
    \node[right=1 of Node31, circle, draw]
    (Node32)
    {~~~};
    \node[right=1 of Node32, circle, draw]
    (Node33)
    {~~~};
    \node[right=1 of Node33, circle, draw]
    (Node34)
    {~~~};
        \node[left=0.3 of Node11]
        (Node11l)
        {$a$};
        \node[right=0.3 of Node14]
        (Node14r)
        {$a$};
        \node[left=0.3 of Node21]
        (Node21l)
        {$b$};
        \node[right=0.3 of Node24]
        (Node24r)
        {$b$};
        \node[left=0.3 of Node31]
        (Node31l)
        {$ab$};
        \node[right=0.3 of Node34]
        (Node34r)
        {$ab$};

  \draw
  ($(Node11.center)+(0,1.50mm)$)
    --
  ($(Node12.center)+(0,1.50mm)$)
  ;
  \draw
  ($(Node12.center)+(0,1.50mm)$)
    --
  ($(Node33.center)+(0,1.50mm)$)
  ;
  \draw
  ($(Node33.center)+(0,1.50mm)$)
    --
  ($(Node34.center)+(0,1.50mm)$)
  ;
  \draw
  ($(Node21.center)+(0,0.0mm)$)
    --
  ($(Node22.center)+(0,0.0mm)$)
  ;
  \draw
  ($(Node22.center)+(0,0.0mm)$)
    --
  ($(Node23.center)+(0,0.0mm)$)
  ;
  \draw
  ($(Node23.center)+(0,0.0mm)$)
    --
  ($(Node34.center)+(0,0.0mm)$)
  ;
  \draw
  ($(Node21.center)+(0,0.75mm)$)
    --
  ($(Node22.center)+(0,0.75mm)$)
  ;
  \draw
  ($(Node22.center)+(0,0.75mm)$)
    --
  ($(Node23.center)+(0,0.75mm)$)
  ;
  \draw
  ($(Node23.center)+(0,0.75mm)$)
    --
  ($(Node34.center)+(0,0.75mm)$)
  ;
  \draw
  ($(Node21.center)+(0,-0.75mm)$)
    --
  ($(Node22.center)+(0,-0.75mm)$)
  ;
  \draw
  ($(Node22.center)+(0,-0.75mm)$)
    --
  ($(Node23.center)+(0,-0.75mm)$)
  ;
  \draw
  ($(Node23.center)+(0,-0.75mm)$)
    --
  ($(Node34.center)+(0,-0.75mm)$)
  ;
  \draw
  ($(Node21.center)+(0,-1.50mm)$)
    --
  ($(Node32.center)+(0,-1.50mm)$)
  ;
  \draw
  ($(Node32.center)+(0,-1.50mm)$)
    --
  ($(Node33.center)+(0,-1.50mm)$)
  ;
  \draw
  ($(Node33.center)+(0,-1.50mm)$)
    --
  ($(Node34.center)+(0,-1.50mm)$)
  ;

\end{tikzpicture}
\caption{ Realizable history in $\PP_\textit{MFDO}$ of Example \ref{do-protocol-mfdo}}
\label{figure-history-mfdo}
\end{figure}

%% file: pruning-theorems.tex
\paragraph{Characterizing Histories.}

As for IO protocols, we introduce a notion of compatibility.

\begin{definition}
\label{def:compatibleMFDO}
A history $H$ is \emph{compatible} with an MFDO protocol $\PP$
if for every trajectory $\tau$ of $H$ and for every non-horizontal step $\tau(i)\tau(i+1)$ of $\tau$, 
the protocol $\PP$ contains a transition $\tau(i) \trans{o} \tau(i+1)$ 
such that $o \in \mathcal{S}_H^i$, \ie, such that $o$ has been visited by time $i$.
\end{definition}

\begin{remark}
Notice the difference with IO protocols. In the IO case, compatibility requires  that some agent visits 
$o$ exactly at time $i$, a requirement captured by the condition $\tau'(i) = \tau'(i+1)=o$. In the MFDO case, compatibility requires that some agent visits state $o$ at time $i$ \emph{or earlier}, captured by the condition $o \in \mathcal{S}_H^i$.
\end{remark}

\begin{restatable}{lemma}{LemmaHistoryMFDO}
\label{lem:compatibleMFDO}
Let $\PP$ be an MFDO protocol.  A well-structured history is realizable in $\PP$ if{}f it is compatible with $\PP$.
\end{restatable}

\begin{example}
The history $H$ of Figure \ref{figure-history-mfdo} is well structured, realizable,
and compatible with the MFDO protocol of Example \ref{history-ex-mfdo}.
\end{example}

\paragraph{Pruning.}

We prove that the construction of the Pruning Theorem for IO protocols 
yields the same results for MFDO protocols.

\begin{theorem}[MFDO Pruning]
\label{thm:mfdopruning}
Let $\PP = (Q,\delta)$ be an MFDO protocol, 
let $L'$ and $L$ be multisets of states of $\PP$,
and let
$C' \trans{*} C$ be an execution of $\PP$
such that $L' \leq C'$ and $C \geq L$.
There exist configurations $D'$ and $D$ such that 
\begin{center}
\(
\begin{array}[b]{@{}c@{}c@{}c@{}c@{}c@{}}
C' &  \trans{\hspace{1em}*\hspace{1em}} & C  \\[0.1cm]
\geq &  & \geq \\[0.1cm]
D' & \trans{\hspace{1em}*\hspace{1em}} &D  \\[0.1cm]
\geq &  & \geq \\[0.1cm]
L' &  &L  \\[0.1cm]
\end{array}
\)
\end{center}
and $|D'| = |D| \leq |L| + |L'| + |Q|^3$.
\end{theorem}
\begin{proof}
Let $H$ be a well-structured and realizable history for the execution $L' \leq C' \trans{*} C \geq L$.
Let $H'$ be the result of pruning $H$ using the construction of theorem~\ref{thm:pruning}.
We already know that $H'$ is well-structured and covers $L'$ and $L$ by its initial and final configuration.
Let us show that it is compatible with $\PP$.  By the definition of compatibility (Definition \ref{def:compatibleMFDO}), and since $H' \subseteq H$, it suffices to show that $\mathcal{S}_H^i = \mathcal{S}_{H'}^i$ holds for every $i$. But this follows from the fact that, by the definition of $H'$, each state is \emph{first} visited in $H'$ at the same moment that it is \emph{first} visited in $H$.
 \end{proof}

\begin{remark}
For MFDO protocols we can also obtain a linear bound. 
Intuitively, the reason is that in order to construct the smaller history $H'$ from $H$
we no longer need to concatenate prefixes and suffixes of trajectories of $H$, but just pick an adequate subset of them.
We provide more details in the appendix.
One can apply the improved bound to the results of Section \ref{sec:small-instance},
but some technical special cases arise in the proofs,
therefore we use theorem~\ref{thm:mfdopruning} for simplicity and uniformity.
\end{remark}

%% file: shortening-q4-MR.tex
We introduce a new measure of the length of executions, the \emph{aggregated length} of an execution.

\begin{definition}
Let $\PP=(Q,\delta)$ be an MFDO protocol, and let $\xi$ be a nonempty sequence of transitions of $\PP$. Let $(k_1, \ldots, k_m)$ be the unique tuple of positive natural numbers such that $\xi=t_1^{k_1} t_2^{k_2} \ldots t_m^{k_m}$ and $t_i \neq t_{i+1}$ for every $i=1, \ldots, m-1$.
We say that $\xi$ has \emph{aggregated length} $m$, and let $|\xi|_a$ denote the aggregated length of $\xi$.
\end{definition}

The Shortening Theorem states that we can replace "long" executions of an MFDO protocol with shorter executions in terms of aggregated length.

\begin{theorem}[MFDO Shortening]
\label{thm:mfdoshortening}
Let $\PP = (Q,\delta)$ be an MFDO protocol, and let $C \trans{*} C'$ be an execution of $\PP$.
There exists a sequence $\xi$ such that $C \trans{\xi} C'$ and $|\xi|_a \leq |Q|^4$.
\end{theorem}
\begin{proof}
Let $H$ be a well-structured and realizable history
for the execution $C \trans{*} C'$, and let $n$ be the length of $H$.
We have $\mathcal{S}_H^1 \subseteq \mathcal{S}_H^2 \subseteq \cdots \subseteq \mathcal{S}_H^n$. Since $H$ is well structured, for every $1\leq i \leq n-1$ either
$\mathcal{S}_H^i = \mathcal{S}_H^{i+1}$, or $\mathcal{S}_H^{i+1}$ contains exactly one more state than $\mathcal{S}_H^{i}$.

Let $T_0=1$, let $T_1, T_2, \ldots, T_{k-1}$ be the time moments immediately before the set of visited states increases, that is, the set of indices satisfying $\mathcal{S}_H^{T_i} \subset \mathcal{S}_H^{T_{i+1}}$, and let $T_k = n$. Observe that $k \leq |Q|$.

For every $0 \leq j \leq k$, let $H_j$ be the initial segment of $H$ of length $T_j$.
We prove by induction over $j$ that there is a well-formed and realizable history $H'_j$
satisfying the following conditions:
\begin{itemize}
\item[(i)] $\mathcal{S}_{H_j} = \mathcal{S}_{H'_j}$, that is, $H_j$ and $H_j'$ visit the same states;
\item[(ii)] there exists a bijection $b$ between the trajectories of $H$ and $H_j'$ such that the $T_j$-th state of $\tau$ and the last state of $b(\tau)$  coincide; and
\item[(iii)] $H_j'$ has length at most $j(|Q|(|Q|-1)^2+1)$.
\end{itemize}
\noindent The theorem then follows from the fact that, since $H'_k$ has length at most $|Q|(|Q|(|Q|-1)^2+1)<|Q|^4$ and
        is realizable, it can be realized by an execution of aggregated length at most
$|Q|^4$.

\begin{figure}
\centerline{
\includegraphics{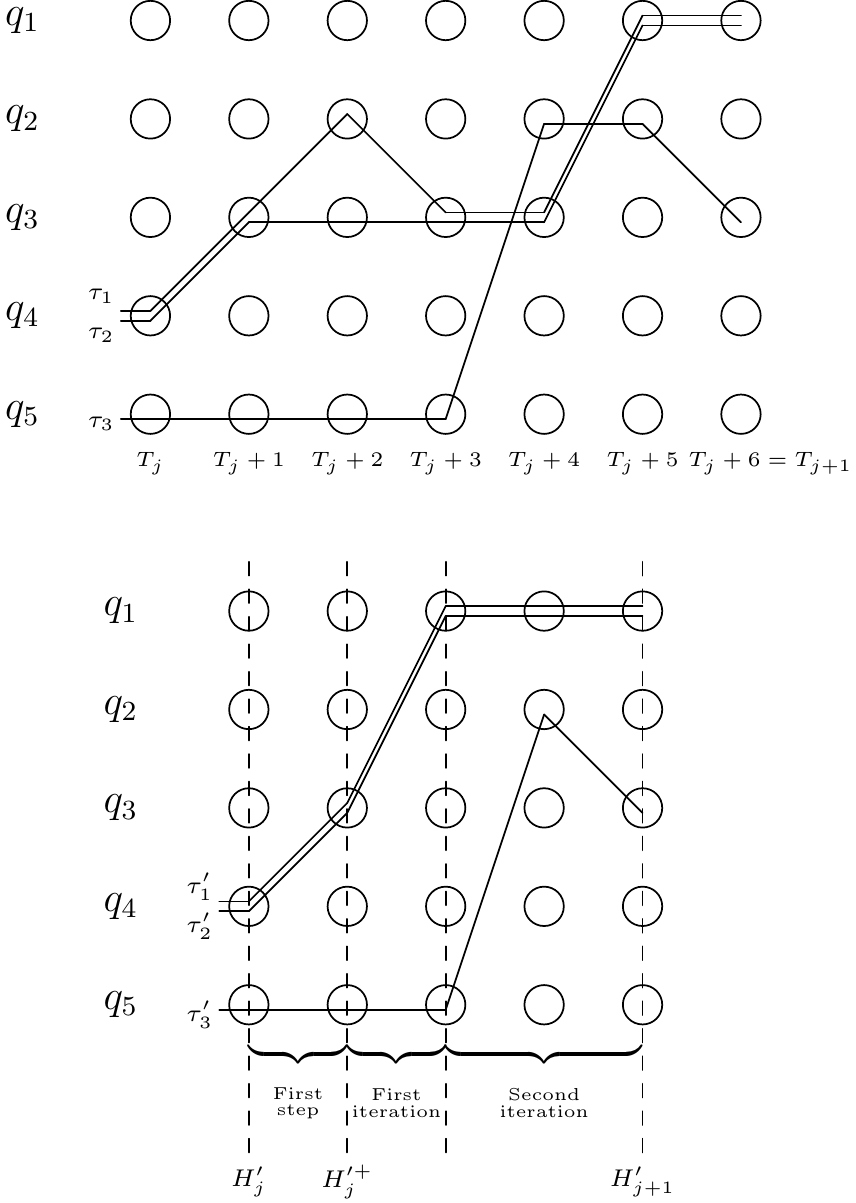}
}
\caption{Illustration of the proof of Theorem \ref{thm:mfdoshortening}}
\label{fig:shortening}
\end{figure}

The base case of the induction is $j=0$. Since $T_0=1$, we can set $H_0' \defeq H_0$. For the induction step, assume we have already constructed
$H_j'$ satisfying conditions (i)-(iii). We construct $H'_{j+1}$ by
extending each trajectory of $H_j'$. We
illustrate how to perform the extensions on the example of Figure \ref{fig:shortening}.

\begin{example}
Figure \ref{fig:shortening} shows at the top the fragment of a history $H$ between times $T_j$ and $T_{j+1}= T_j+6$. The history $H$ consists of three trajectories $\tau_1, \tau_2, \tau_3$. We assume that $\mathcal{S}_H^{T_j} = \{q_1, q_2, q_4,q_5\}$, i.e., up to time $T_j$ the three trajectories have visited all states but $q_3$. We then have $\mathcal{S}_H^{T_{j+1}} = \{q_1, \ldots, q_5\}$.
\end{example}

Let $\tau$ be an arbitrary trajectory of $H$, and for every $0 \leq i \leq j$ let $\tau_i$ be the
        prefix of $\tau$ of length $T_i$. By condition (ii), there exists a bijection $b$ that assigns to $\tau$ a trajectory $\tau'_j \defeq b(\tau)$ of $H_j'$. Further,  $\tau_j$ and $\tau'_j$ have the same initial and final states. We describe an algorithm that extends the history $H'_j$ to $H'_{j+1}$ with the same final configuration as $H_{j+1}$.

The algorithm initializes a variable $\tilde{\tau}:=\tau'_j$ for each trajectory $\tau \in H$.
In a first step, the algorithm sets $\tilde{\tau} := \tau_j' \tau(T_j + 1)$.  In our example, the three trajectories of $H_j'$ are extended as shown in the bottom part of Figure \ref{fig:shortening}.

Let $H'^+_j$ be the history obtained after applying this first step. It is
easy to see that, since $H_j$ and $H_j'$ satisfy conditions (i)-(iii), so do $H'^+_j$
and the prefix of $H$ of length $T_j+1$.

The algorithm now proceeds to execute a loop. Let $\mathcal{B}[q,\allowbreak q',\allowbreak j]$ be the bunch of trajectories $\tau \in H$ such that $\tau(T_j+1)=q$ and $\tau(T_{j+1})=q'$, and let $E$ be
be an arbitrary but fixed enumeration of the pairs $(q, q')$ of states such that
 $\mathcal{B}[q,q',j]$ is nonempty. The algorithm loops through every pair $(q, q') \in E$, extending each $\tilde{\tau}$ in a way to be described later. After the loop, the algorithm sets $\tau_{j+1}'$ to the final value of $\tilde{\tau}$. Observe that each variable $\tilde{\tau}$ gets extended $|Q|(|Q|-1)$ times.

\begin{example}
The history at the top of Figure \ref{fig:shortening} has two nonempty bunches, namely
$\mathcal{B}[q_3,q_1,j] = \{\tau_1, \tau_2\}$, and $\mathcal{B}[q_5,\allowbreak q_3,\allowbreak j] = \{ \tau_3\}$.
In what follows we assume that $E = (q_3,\allowbreak q_1) \, (q_5,\allowbreak q_3)$.
\end{example}

Before describing the body of the loop for a given pair $(q, q')$ of states, we need to state and prove a claim.

\smallskip
\noindent \textbf{Claim.}
For every $(q, q') \in E$ there exists a sequence $\mathit{sh}(q,q')$ (where $\mathit{sh}$ stands for ``short'') leading from $q$ to $q'$ and satisfying the following properties:
\begin{itemize}
\item each state in $\mathit{sh}(q,q')$ is in $\mathcal{S}_H^{T_j}$;
\item each step in $\mathit{sh}(q,q')$ corresponds to a protocol transition observing some state in $\mathcal{S}_H^{T_j}$;
\item $\mathit{sh}(q,q')$ has length $|Q|$.
\end{itemize}

\noindent To prove the claim, observe first that, by the definition of $E$, there exists at least one trajectory $\tau \in \mathcal{B}[q,q',j]$. Pick any such trajectory. The steps of $\tau$ between times $T_j$ and $T_{j+1}$ form a path in the oriented graph of transitions of the protocol enabled by the set $\mathcal{S}_H^{T_j}$ of visited states. Let $\mathit{sh}(q,q')$ be the result of removing all cycles from this path. By construction only states from $\mathcal{S}_H^{T_j}$ are used, and only transitions enabled by observing states from the same set are performed. Clearly, we have $|\mathit{sh}(q,q')| \leq |Q|$.

\begin{example} In Figure \ref{fig:shortening}, the segment of $\tau_1$ between times $T_{j}+1$ and $T_{j}+6=T_{j+1}$ is the sequence $q_3 q_2 q_3 q_3 q_1 q_1$ of states. The trajectory $\mathit{sh}(q_3,q_1)$ obtained from $\tau_1$ by ``cutting out the cycles'' is $q_3 q_1$.
\end{example}

\noindent  For each pair $(q, q') \in E$, the algorithm picks an arbitrary trajectory of $\mathcal{B}[q,q',j]$, constructs the shortened trajectory  $\mathit{sh}(q,q')$, and for every trajectory $\tau \in H$ it extends the current trajectory $\tilde{\tau}$ as follows:
\begin{itemize}
\item[(1)] If  $\tau \in \mathcal{B}[q,q',j]$, then the algorithm extends $\tilde{\tau}$ with $\mathit{sh}(q,q')$ (more precisely, with the result of dropping the first state in $\mathit{sh}(q,q')$).
\item[(2)] Otherwise, the algorithm extends $\tilde{\tau}$ by replicating its final state $|\mathit{sh}(q,q')|-1$ times. In other words, it extends $\tilde{\tau}$ with $|\mathit{sh}(q,q')|-1$ horizontal steps.
\end{itemize}
\noindent Observe that after each iteration of the loop all trajectories have the same length. The histories consisting of all the trajectories after the same iteration are well-formed (all added non-horizontal steps are copies of the same one) and realizable (because of the second condition in the claim). In particular, after the last iteration of the algorithm, we obtain a wellformed and realizable history.

\begin{example}
Recall that  $E = (q_3, q_1) \, (q_5, q_3)$. Assume that for $(q, q'):=(q_3, q_1)$ the algorithm
picks $\tau_1$ (it could also pick $\tau_2$). The algorithm sets $\mathit{sh}(q_3, q_1):= q_3 q_1$, and in the first iteration of the loop it extends $\tilde{\tau}_1$ and
$\tilde{\tau}_2$ with $q_1$, and $\tilde{\tau}_3$ with $q_5$ (see the bottom of Figure \ref{fig:shortening}).

For $(q, q'):=(q_5, q_3)$ the algorithm necessarily picks $\tau_3$ and sets
$\mathit{sh}(q_5, q_3):= q_5 \, q_2 \, q_3$. In the second iteration of the loop $\tilde{\tau}_1$ and
$\tilde{\tau}_2$ are extended with horizontal steps $q_1 q_1$, and $\tilde{\tau}_3$ with
$q_5 \, q_2 \, q_3$.
\end{example}


Let us show that the realizable history $H_{j+1}'$ constructed by the algorithm satisfies properties (i)-(iii). Property (i) follows directly from the fact that the algorithm only extends the trajectories of $H_{j}'$  with steps taken from the trajectories of $H_{j+1}$.
For property (ii), we observe that for every pair of states $(q, q')$, the bunches $\mathcal{B}[q,q',j+1]$ and  $\mathcal{B}'[q,q',j+1]$ (defined as $\mathcal{B}[q,q',j+1]$, but for the history $H'_{j+1}$) have the same size. So the bijection can be obtained as the union of bijections between these bunches. Finally, let us prove property (iii).  Since the sequences $\mathit{sh}(q,q')$ have length at most $|Q|$, they consist of at most $|Q|-1$ steps. Since $|E| \leq |Q| (|Q|-1)$, during the loop every trajectory gets  extended at most $|Q| (|Q|-1)$ times. So the trajectories of $H_{j+1}'$ have at most $|Q| (|Q|-1)^2 + 1$ more steps than the trajectories of $H_{j}'$, and at most $(j+1) (|Q| (|Q|-1)^2 + 1)$ steps. Since $H_{j+1}'$ is well structured, its aggregated length is bounded by the number of steps of its trajectories, and we are done.
 \end{proof}

\begin{remark}
An optimised version of the construction allows to obtain a \emph{quadratic} bound
for the aggregated length of the history after shortening.
We provide a rough outline in the appendix in case
the reader is interested in carrying out such optimisation.
\end{remark}

%% file: pruning-and-shortening-DO.tex
In Section \ref{sec:mfdo} we showed that reachability in MFDO and zero-message reachability in DO are essentially equivalent notions. Using this correspondence, we derive Pruning and Shortening Theorems for DO protocols from the corresponding results for MFDO protocols. 
Recall that we denote by $\mathcal{Z}$ the set of zero-message configurations of a DO protocol, and that a configuration $Z \in \mathcal{Z}$ can be seen both as a DO configuration and (by abuse of notation) as an MFDO configuration.

\begin{corollary}[DO Pruning]
\label{thm:dopruning}
let $Z, Z'\in \mathcal{Z}$ be zero-message configurations of $\PP$,
let $L'$ and $L$ be multisets of states of $\PP$,
and let
$Z' \trans{*} Z$ be an execution of $\PP$
such that $L' \leq Z'$ and $Z \geq L$.
There exist zero-message configurations $Y'$ and $Y$ such that
\begin{center}
\(
\begin{array}[b]{@{}c@{}c@{}c@{}c@{}c@{}}
Z' &  \trans{\hspace{1em}*\hspace{1em}} & Z  \\[0.1cm]
\geq &  & \geq \\[0.1cm]
Y' & \trans{\hspace{1em}*\hspace{1em}} &Y  \\[0.1cm]
\geq &  & \geq \\[0.1cm]
L' &  &L  \\[0.1cm]
\end{array}
\)
\end{center}
and $|Y'| = |Y| \leq |L| + |L'| + |Q|^3$.
\end{corollary}
\begin{proof}
By Lemma \ref{lm:do-to-mfdo}, if $Z'' \trans{*}  Z' \geq Z$ in DO protocol $\PP$, then $Z'' \trans{*}  Z' \geq Z$ in the corresponding MFDO protocol (see Definition \ref{def:corresponding-mfdo}).
By applying Theorem \ref{thm:mfdopruning} to $Z'' \trans{*} Z'\geq Z$ in the MFDO protocol, there exist $Y''$ and $Y'$ such that
\begin{center}
\(
\begin{array}[b]{@{}c@{}c@{}c@{}c@{}c@{}}
Z' &  \trans{\hspace{1em}*\hspace{1em}} & Z  \\[0.1cm]
\geq &  & \geq \\[0.1cm]
Y' & \trans{\hspace{1em}*\hspace{1em}} &Y  \\[0.1cm]
\geq &  & \geq \\[0.1cm]
L' &  &L  \\[0.1cm]
\end{array}
\)
\end{center}
and $|Y| \leq |L| + |L'| + |Q|^3$.
By Lemma \ref{lm:do-to-mfdo}, $Y' \trans{*}  Y$ is also valid in our DO protocol with $Y',Y \in \mathcal{Z}$.
 \end{proof}

\begin{corollary}[DO Shortening]
\label{thm:doshortening}
Let $\PP = (Q,M,\delta_s,\delta_r)$ be a DO protocol, let $Z$ and $Z'$ be zero-message configurations of $\PP$, and let $Z \trans{*} Z'$ be an execution of $\PP$.
There exists a sequence $\xi$ such that $Z \trans{\xi} Z'$ and $|\xi|_a \leq |Q|^4 + |Q|$.
\end{corollary}
\begin{proof}
By Lemma \ref{lm:do-to-mfdo}, if $Z \trans{*}  Z'$ in DO protocol $\PP$, then $Z \trans{*} Z'$ in the corresponding MFDO protocol.
By applying Theorem \ref{thm:mfdoshortening} to $Z \trans{*} Z'$, there exists $\xi$ such that $Z \trans{\xi} Z'$ in the corresponding MFDO protocol and $|\xi|_a \leq |Q|^4$.

Following the construction of a DO sequence from an MFDO sequence described in the proof of Lemma \ref{lm:do-to-mfdo}, we show that we can construct a sequence $\xi'$ in $\PP$ such that $Z \trans{\xi'} Z'$ and $|\xi'|_a \leq |\xi|_a + |M|$.
In the first step of the construction, we replace each transition of $\xi$ by a corresponding receive transition in $\PP$.
Then for each message $m\in M$ that appears in these receive transitions, we add a sequence of identical send transitions $q_m \send{m} q_m$ the first time that state $q_m$ that can send $m$ is reached.
Thus the constructed DO sequence $\xi'$ has an aggregated length of at most $|\xi|_a + |M|$, and since $|\xi|_a \leq |Q|^4$ and $|M|\leq |Q|$ we get our result.
 \end{proof}


%% file: small-instance-section.tex
We introduce counting sets, a class of possibly infinite sets of configurations with a finite representation in terms of so-called counting constraints. We then prove the Closure Theorems for IO and MFDO, stating that the sets of predecessors and successors of a counting set are also counting sets. Further, we show that if the original counting set has a representation with ``small'' cubes, then the sets of predecessors and successors also have succinct representations.

\paragraph{Counting constraints and counting sets.}
Let $\PP$ be an IO or MFDO protocol with set of states $Q$. A set $\mathcal{C}$ of configurations of $\PP$ is a \emph{cube} if there exist mappings $L \colon Q \rightarrow \N$ and $U \colon Q \rightarrow \N \cup \{\infty\}$ such that $C \in \mathcal{C}$ if{}f $L \leq C \leq U$. (Observe that the components of $U$ may be equal to $\infty$, and that both $L$ and $U$ are unique.) We call $L$ and $U$ the \emph{lower bound} and \emph{upper bound} of $\mathcal{C}$, respectively, and call the pair $(L, U)$ the \emph{representation} of $\mathcal{C}$. Given two mappings $L \colon Q \rightarrow \N$ and $U \colon Q \rightarrow \N \cup \{\infty\}$, the cube represented by $(L, U)$ is denoted $\sem{L,U}$.



A \emph{counting constraint} is a finite set $\Gamma = \{ (L_1, U_1),\allowbreak  \ldots,\allowbreak  (L_n, U_n)\}$ of representations of cubes. We say that $\Gamma$ \emph{represents} the set $\sem{\Gamma} \defeq\sem{L_1, U_1} \cup \cdots \cup  \sem{L_n, U_n} $. A set $\mathcal{S}$ is a \emph{counting set} if $\mathcal{S} = \sem{\Gamma}$ for some counting constraint $\Gamma$. \cwk{changed counting set notation from$\mathcal{C}$ to $\mathcal{S}$ to avoid confusion with cube notation.}

Observe that, while a cube has a unique representation, the same counting set may be represented by more than one counting constraint.  For example, consider a protocol with just one state. The counting constraints $\{ (1,3), (2,4) \}$, $\{ (1,2), (3,4) \}$, and $\{ (1,4) \}$ define the same counting set, namely the cube $\sem{1,4}$.

\paragraph{Measures of counting constraints.} We introduce two measures of the size of a counting constraint.
Let $\mathcal{C}$ be a cube with representation $(L, U)$. The \emph{$l$-norm} of
$\mathcal{C}$, denoted $ \lnorm{\mathcal{C}}$, is the sum of the components of $L$.
The \emph{$u$-norm} of $\mathcal{C}$, denoted $ \unorm{\mathcal{C}}$, is the sum of the components of $U$ that are not equal to $\infty$, if there are any, and $0$ otherwise.

The $l$-norm and $u$-norm of a counting constraint $\Gamma= \{\mathcal{C}_1, \ldots,\mathcal{C}_m\}$ are defined by

$$\begin{array}{lcl}
\lnorm{\Gamma} \defeq \displaystyle \max_{i\in [1,m]} \{ \lnorm{\mathcal{C}_i} \} &  & \unorm{\Gamma} \defeq \displaystyle \max_{i\in[1,m]} \{ \unorm{\mathcal{C}_i} \}.
\end{array}$$

The $l$-norm (respectively $u$-norm) of a counting set $\mathcal{S}$ is the smallest $l$-norm (respectively $u$-norm) of a counting constraint representing $\mathcal{S}$, that is
$$\begin{array}{lcl}
\lnorm{\mathcal{S}} \defeq \displaystyle \min_{\mathcal{S} = \sem{\Gamma}} \{ \lnorm{\Gamma} \} &  & \unorm{\mathcal{S}} \defeq \displaystyle  \min_{\mathcal{S} = \sem{\Gamma}} \{ \unorm{\Gamma} \}.
\end{array}$$

\begin{example}
Cube $\mathcal{C}$ with representation $(1,4)$ has $l$-norm $1$ and $u$-norm $4$.
The counting constraint $\Gamma= \set{ (2,4),(3,5) }$ has $l$-norm $3$ and $u$-norm $5$.
\end{example}

The following proposition, whose proof is given in the Appendix, shows that a Boolean combination of counting sets is still a counting set and bounds the size of the counting constraints representing such combinations.

\begin{restatable}[\cite{EsparzaGMW18}, Proposition 5]{proposition}{PropMeasureBooleanOp}
\label{prop:oponconf}
Let $\Gamma_1, \Gamma_2$ be counting constraints.
\begin{itemize}
\item There exists a counting constraint $\Gamma$ with $\sem{\Gamma} = \sem{\Gamma_1} \cup \sem{\Gamma_2}$ such that
$\unorm{\Gamma} \leq \max \{\unorm{\Gamma_1}, \unorm{\Gamma_2} \}$ and $\lnorm{\Gamma} \leq \max \{\lnorm{\Gamma_1}, \lnorm{\Gamma_2} \}$.
\item  There exists a counting constraint $\Gamma$ with $\sem{\Gamma} = \sem{\Gamma_1} \cap \sem{\Gamma_2}$ such that
$\unorm{\Gamma} \leq \unorm{\Gamma_1} + \unorm{\Gamma_2}$ and $\lnorm{\Gamma} \leq \lnorm{\Gamma_1} + \lnorm{\Gamma_2}$.
\item There exists a counting constraint $\Gamma$ with $\sem{\Gamma} = \N^n \setminus \sem{\Gamma_1}$ such that
$\unorm{\Gamma} \leq n\lnorm{\Gamma_1}$ and $\lnorm{\Gamma} \leq n\unorm{\Gamma_1} + n$.
\end{itemize}
 \end{restatable}

Loosely speaking, Proposition \ref{prop:oponconf} shows that applying boolean operations to counting sets does not increase much the
size of its representation. Now we prove the Closure Theorem, showing that this is also the case
for the operations of computing the set of successors or predecessors of a counting set.

\paragraph{Closure Theorem for IO protocols.} The Closure Theorem for IO protocols is an easy consequence of the following lemma:

\begin{lemma}
\label{lm:smallminterm}
Let $\PP$ be an IO protocol with state set $Q$ and let $\mathcal{C} \subseteq \pop{Q}$ be a cube.
For all $C' \in \prestar(\mathcal{C})$, there exists a cube $\mathcal{C}'$ such that
\begin{enumerate}
\item $C' \in \mathcal{C}' \subseteq  \prestar(\mathcal{C})$, and
\item $\lnorm{\mathcal{C}'} \leq \lnorm{\mathcal{C}}  + |Q|^3$ and $\unorm{\mathcal{C}'} \leq \unorm{\mathcal{C}}$.
\end{enumerate}
\end{lemma}
\begin{proof}
Let $L, U$ be mappings such that $\mathcal{C} = \sem{L,U}$.
Let $C'$ be a configuration of $\prestar(\mathcal{C})$.
There exists a configuration $C \in \mathcal{C}$ such that $C' \longrightarrow C$, and $C \geq L$.
By the Pruning Theorem  there exist configurations $D'$ and $D$ such that
\begin{center}
\(
\begin{array}[b]{@{}c@{}c@{}c@{}c@{}c@{}c@{}c@{}}
C' &  \trans{\hspace{1em}*\hspace{1em}} & C & \  \geq \ &L  \\[0.1cm]
\geq &  & \geq \\[0.1cm]
D' & \trans{\hspace{1em}*\hspace{1em}} &D  & \geq & L
\end{array}
\)
\end{center}
and $|D'| \leq |L| + |Q|^3$.
Since $C \in \mathcal{C}$, we have $U\geq C \geq D \geq L$. So $D\in \mathcal{C}$, and therefore $D' \in \prestar(\mathcal{C})$.

We  find a cube $\mathcal{C}'$ satisfying conditions (1) and (2). For this,
we choose appropriate lower and upper bounds $L', U'$, and set $\mathcal{C}'= \sem{L', U'}$. First, we set $L' \defeq D'$. For the definition of $U'$, we
use the tools of the Pruning Theorem section, in which the movements of the agents are de-anonymized into trajectories.
Let $H_{C'}$ be a well-structured realizable history of $\PP$ leading from $C'$ to $C$, and let $q$ be a state of $Q$. We define $U'(q)$ as follows:
\begin{itemize}
\item[(i)] If some trajectory of $H_{C'}$ starting at $q$ leads to a state $r$ such that $U(r) = \infty$, then set $U'(q) \defeq \infty$.
\item[(ii)] If every trajectory of $H_{C'}$ starting at $q$ leads to states $r$ such that $U(r) < \infty$, then set $U'(q)= C'(q)$.
\end{itemize}
We prove that  $\mathcal{C}' \defeq \sem{L', U'}$ satisfies the conditions of the lemma.

\medskip \noindent \textbf{Property 1:} $C' \in \mathcal{C}' \subseteq  \prestar(\mathcal{C})$.\\
Since $\mathcal{C}' \defeq \sem{L', U'}$, we first prove $L' \leq C' \leq U'$.
The inequality $L' \leq C'$ follows from $C' \geq D'$ (see the diagram above) and  $L' \defeq D'$.
Let us now show that $C'(q) \leq U'(q)$ holds for every state $q$.  If $U'(q) = \infty$ there is nothing to show. If $U'(q)$ is finite, \ie, if \emph{Case 2} above holds, then  $U'(q) = C'(q)$, and we are done.

It remains to prove $\sem{L', U'} \subseteq  \prestar(\mathcal{C})$, which requires more effort.
We show that for every configuration $R' \in \sem{L', U'}$ there exists a history $H_{R'}$ leading  from $R'$ to a configuration $R \in \mathcal{C}$, \ie, to a configuration $R$ satisfying $L \leq R \leq U$. Since $R' \in \sem{L', U'}$ and $L' \defeq D'$, we have $R' \geq D'$.
So we construct $H_{R'}$ by adding trajectories to $H_{D'}$: Since $H_{D'}$ leads to $D$, this guarantees that $H_{R'}$ leads to a configuration $R$ such that $R \geq D \geq L$ (see Figure \ref{figure-boost}). Further, to ensure that $H_{R'}$ starts at $R'$, for every $q \in Q$ we add
 to $H_{D'}$ exactly $(R'(q) - D'(q))$ trajectories starting at $q$.  It remains to choose these trajectories in such a way that $R \leq U$ holds.
 We add trajectories so that $R(q) \leq C(q)$ holds, which, since $C(q) \leq U(q)$ (see Figure \ref{figure-boost}), ensures $R(q) \leq U(q)$.

\input{figure-boost.tex}

We add trajectories to $H_{D'}$ by replication, \ie, we only add copies of trajectories already present in $H_{D'}$.  Recall that for every state $q \in Q$ we have to add $(R'(q) - D'(q))$ trajectories starting at $q$. We decide which trajectories to add according to two cases, very similar to the cases (i) and (ii) above:

\begin{itemize}
\item[(i$'$)] $H_{D'}$ contains a trajectory $\tau$ leading from $q$ to a state $r$ such that $U(r) = \infty$. \\
In this case we add $(R'(q) - D'(q))$ copies of $\tau$.
\item[(ii$'$)] Every trajectory of $H_{D'}$ leading from $q$ to some state $r$ satisfies $U(r) < \infty$. \\
In this case, by the definition of $U'$ (see (ii) above), we have $U'(q) = C'(q)$. Since $R' \leq U'$ by hypothesis, we get $D'(q) \leq R'(q) \leq C'(q)$, and so $(R'(q) - D'(q)) \leq (C'(q) - D'(q))$, \ie, we need to add at most $C'(q) - D'(q)$ trajectories.

For each state $r \in Q$, let $n_{C'}[q, r]$ and $n_{D'}[q, r]$ be the sizes of the bunches of trajectories of $H_{C'}$ and $H_{D'}$ leading from $q$ to $r$, respectively. By this definition, and the definition of the pruning operation, we have
\begin{itemize}
\item[(a)] $C'(q) - D'(q) = \sum_{r \in Q}  \; (n_{C'}[q, r] - n_{D'}[q, r])$.
\item[(b)]  For every $r \in Q$: $n_{C'}[q, r] \geq n_{D'}[q, r]$, and
\item[(c)] For every $r \in Q$: $n_{C'}[q, r] \geq 1$ implies $n_{D'}[q, r] \geq 1$.
\end{itemize}
We add trajectories as follows: we loop through the states $r$ such that $n_{C'}[q, r] \geq 1$. We take any trajectory of $H_{D'}$ leading from $q$ to $r$ (which exists by (c)), and replicate it
$n_{C'}[q, r] - n_{D'}[q, r]$ times or less, until the quota of $R'(q) - D'(q)$ trajectories has been reached.
The quota is eventually reached by (a).
\end{itemize}

We claim that this procedure produces a history $H_{R'}$ such that $n_{R'}[q, r] \leq n_{C'}[q, r]$ for every $q, r \in Q$ such that $U(r) < \infty$. Indeed, fix $r$ such that  $U(r) < \infty$. If $q$ satisfies (i$'$), then no trajectory from $q$ to $r$ is replicated, \ie, $n_{R'}[q, r] = n_{C'}[q, r]$. If $q$ satisfies (ii$'$), then $n_{R'}[q, r] \leq n_{C'}[q, r]$. By the claim, $R(r) \leq C(r)$ for every state $r$ such that
$U(r) < \infty$. Since $C \leq U$, we have $R \leq U$, and we are done.

\medskip \noindent \textbf{Property 2:} $\lnorm{\mathcal{C}'} \leq \lnorm{\mathcal{C}}  + |Q|^3$ and $\unorm{\mathcal{C}'} \leq \unorm{\mathcal{C}}$. \\
\noindent For the $l$-norm, recall that $L' \defeq D'$. Since $H_{D'}$ leads from $D'$ to $D$, we have $|L'|=|D'|= |D|$. By the Pruning Theorem
$$ \lnorm{(L',U')} \leq |L| + |Q|^3 = \lnorm{(L,U)} + |Q|^3 \ . $$

%
\noindent For the $u$-norm, notice that by (i) and (ii), every trajectory of $H_{C'}$ starting at a state $q$ satisfying $U'(q)<\infty$ leads to  a state $r$ satsfying $U(r) < \infty$. Using this observation, we get:

\newenvironment{myarray}{\def\arraystretch{1.2}\everymath={\displaystyle}\begin{array}{rccccr}}
{\end{array}}

\[
\begin{myarray}
&  & \unorm{(L',U')}   \\
& = & \sum_{q \mid U'(q) < \infty} & U'(q)   \\
& = & \sum_{q \in Q \mid U'(q) < \infty} & C'(q) &  & \mbox{\big(Def. of $U'$\big)} \\
& = & \sum_{q \in Q \mid U'(q) < \infty} & \sum_{r \in Q} & n_{C'}[q, r] & \mbox{\big(Def. of $n_{C'}[q,r]$\big)} \\
& \leq &  \sum_{q \in Q}  & \sum_{r \in Q \mid U(r) < \infty} & n_{C'}[q, r] & \qquad \mbox{(Observation)}\\
& = & \sum_{r \in Q \mid U(r)< \infty} & \sum_{q \in Q}    & n_{C'}[q, r]  & \mbox{(Algebra)} \\
& = & \sum_{r \in Q \mid U(r) < \infty} & C(r) &  & \mbox{($H_{C'}$ leads to $C$)} \\
& \leq & \sum_{r \in Q \mid U(r) < \infty} & U(r)  & & \mbox{($C \leq U$)} \\
& = & \unorm{(L,U)}
\end{myarray}
\]
 \end{proof}

\begin{restatable}[IO Closure]{theorem}{ThmCCReachabilityIO}
\label{thm:ccreach-io}
Let $\PP$ be an IO protocol with a set $Q$ of states, and let $\mathcal{S}$ be a counting set of configurations of $\PP$ represented by a counting constraint $\Gamma$. Then $\prestar(\mathcal{S})$ is also a counting set,  and there exists a counting constraint $\Gamma'$ satisfying $\sem{\Gamma'} = \prestar(\mathcal{S})$ and
$$
\unorm{\Gamma'} \leq \unorm{\Gamma} \text{ and }
\lnorm{\Gamma'} \leq  \lnorm{\Gamma} + |Q|^3
$$
The same holds for $\poststar$.
\end{restatable}
\cwk{now we've defined the norm of a counting set so this theorem could have a shorter text without the "there exists a counting constraint...". But then we would lose the fact that the same counting constraint has these bounds for l-norm and u-norm. Does it matter?}
\mr{I think it is pretty believeable that the upper and the lower norm can be obtained on the same constraint, and it doesn't really matter}
\begin{proof}
By the definition of a counting set,  there exist cubes $\mathcal{C}_1, \ldots,\mathcal{C}_k$ such that
$\mathcal{S} = \bigcup_{i=1}^k \mathcal{C}_i$, and therefore $\prestar(\mathcal{S}) = \bigcup_{i=1}^k \prestar(\mathcal{C}_i)$
By  Lemma \ref{lm:smallminterm}, for every configuration $C' \in \prestar(\mathcal{S})$ there is a cube $\mathcal{C}'$ such that $C' \in \mathcal{C}'$,  $\mathcal{C}' \subseteq \prestar(\mathcal{S})$, and
$\lnorm{\mathcal{C}'} \leq \lnorm{\mathcal{C}_i} + |Q|^3$, and $\unorm{\mathcal{C}'} \leq \unorm{\mathcal{C}_i}$ for some $1 \leq i \leq k$. So $\prestar(\mathcal{S}) = \bigcup_{C' \in \prestar(\mathcal{S})} \mathcal{C}'$. Since there are only finitely many cubes $\mathcal{C}'$ with a given bound on their lower and upper norms, $\prestar(\mathcal{S}) = \bigcup_{i=1}^{k'} \mathcal{C}'_i$ for some $k'$, and so a counting set.

Let $\Gamma$ and $\Gamma'$ be the counting constraint defined as the set of the representations of
$\{ \mathcal{C}_1, \ldots, \mathcal{C}_k \}$ and  $\{ \mathcal{C}'_1, \ldots, \mathcal{C}'_{k'} \}$, respectively. By the definition of the norm of a counting constraint, we have $\lnorm{\mathcal{C}'_i} \leq \lnorm{\Gamma} + |Q|^3$ and $\unorm{\mathcal{C}'_i} \leq \unorm{\Gamma}$ for every $1 \leq i \leq k'$.  So
$
\unorm{\Gamma'} \leq \unorm{\Gamma}
$
and
$
\lnorm{\Gamma'} \leq \lnorm{\Gamma} + |Q|^3.
$

The result for $\poststar(\mathcal{S})$ can be proven in the exact same way, as the pruning theorem is symmetric.
 \end{proof}

\paragraph{Closure Theorem for MFDO protocols.} The Closure Theorem for MFDO protocols can be proved in the same way as for IO protocols.

\begin{restatable}{lemma}{LemmaSmallCubeMFDO}
\label{lm:smallminterm-mfdo}
Let $\mathcal{C}$ be a cube of an MFDO protocol $\PP$ of with state set $Q$.
For all $C' \in \prestar(\mathcal{C})$, there exists a cube $\mathcal{C}'$ such that
\begin{enumerate}
\item $C' \in \mathcal{C}' \subseteq \prestar(\mathcal{C})$, and
\item $\lnorm{\mathcal{C}'} \leq \lnorm{\mathcal{C}} + |Q|^3$ and $\unorm{\mathcal{C}'} \leq \unorm{\mathcal{C}}$.
\end{enumerate}
\end{restatable}

\begin{restatable}[MFDO Closure]{theorem}{ThmCCReachability}
\label{thm:reachofcc}
Let $\PP$ be an MFDO protocol with a set $Q$ of states, and let $\mathcal{S}$ be a counting set defined by a counting constraint $\Gamma$.
Then $\prestar(\mathcal{S})$ is also a counting set and there exists a counting constraint $\Gamma'$ satisfying $\sem{\Gamma'} = \prestar(\mathcal{S})$, and
$$
\unorm{\Gamma'} \leq \unorm{\Gamma} \text{ and }
\lnorm{\Gamma'} \leq  \lnorm{\Gamma} + |Q|^3
$$
The same holds for $\poststar$.
\end{restatable}

The Closure Theorem for MFDO protocols yields a Closure Theorem for DO protocols.
In DO protocols, counting constraints are still defined as bounds associated to elements of $Q$, and thus they define counting sets which are sets of zero-message configurations.
To express the following result we need operators on zero-message configurations.

\paragraph{Zero-message predecessors and successors.}
Let $\PP$ be a DO protocol, and let $\mathcal{Z}$ be the set of its zero-message configurations.
For every set $\mathcal{M} \subseteq \mathcal{Z}$, we respectively define the set of zero-message predecessors and the set of zero-message successors as
$$\begin{array}{rcl}
\prezmstar(\mathcal{M}) & = & \prestar(\mathcal{M}) \cap \mathcal{Z} \\
\postzmstar(\mathcal{M})    & = & \poststar(\mathcal{M}) \cap \mathcal{Z}.
\end{array}$$

\begin{restatable}[DO Closure]{corollary}{ThmCCReachabilityDO}
\label{coro:reachofcc-do}
Let $\PP$ be a DO protocol with a set $Q$ of states, and let $\mathcal{S}$ be a counting set of zero-message configurations defined by a counting constraint $\Gamma$.
Then $\prezmstar(\mathcal{S})$ is also a counting set and there exists a counting constraint $\Gamma'$ satisfying $\sem{\Gamma'} = \prezmstar(\mathcal{S})$, and
$$
\unorm{\Gamma'} \leq \unorm{\Gamma} \text{ and }
\lnorm{\Gamma'} \leq  \lnorm{\Gamma} + |Q|^3
$$
The same holds for $\postzmstar$.
\end{restatable}

%% file: figure-boost.tex
\begin{figure}
\centering
\begin{tikzpicture}

\node[draw=none,fill=none] (Mp) {$C'$};%
\node[right=4 of Mp, draw=none,fill=none] (M) {$C$};%
\node[below=0.45 of Mp, draw=none,fill=none] (Np) {$R'$};%
\node[right=4 of Np, draw=none,fill=none] (N) {$R$};%
\node[below=0.45 of Np, draw=none,fill=none] (Sp) {$D'$};%
\node[right=4 of Sp, draw=none,fill=none] (S) {$D$};%
\node[below=0.0 of Mp, draw=none,fill=none] (small) {$\geq$};%
\node[below=0.0 of Np, draw=none,fill=none] (smal) {$\geq$};%
\node[below=0.0 of M, draw=none,fill=none] (sma) {$\geq$};%
\node[below=0.0 of N, draw=none,fill=none] (sm) {$\geq$};%
\node[right=0.1 of M, draw=none,fill=none] (U) {$\leq U$};%
\node[right=0.1 of S, draw=none,fill=none] (L) {$\geq L$};%
\node[left=0.1 of Mp, draw=none,fill=none] (Up) {$U'\geq$};%
\node[left=0.1 of Sp, draw=none,fill=none] (Lp) {$L'\leq$};%
%

\draw[->] (Mp) to node[below] {$H_C'$} (M);%
\draw[->] (Np) to node[below] {$H_R'$} (N);%
\draw[->] (Sp) to node[below] {$H_D'$} (S);%

\end{tikzpicture}
\caption{Construction of the proof of Lemma \ref{lm:smallminterm}}
\label{figure-boost}
\end{figure}

%% file: correctness-intro-CWK.tex
We use the Pruning, Shortening, and Closure Theorems proved in the past sections 
to prove that the correctness problem for IO protocols is in \PSPACE, and  that the correctness problem for DO protocols is in $\Pi_2^p$. These upper bounds match the lower bounds proved in  Theorem \ref{thm:io-correctness-hard} and Theorem 
\ref{thm:do-hardness}.

%

For the following results, we need the predicates $\varphi$ we consider to be describable by counting constraints.
A predicate $\varphi \colon \mathbb{N}^k \rightarrow \{0,1\}$ is \emph{describable by counting constraint} if there is a counting constraint $\Gamma$ such that  $\varphi(\vec{v})=1$ if{}f $\vec{v}$ satisfies $\Gamma$.
If $\varphi$ is a predicate over $\pop{\Sigma}$ that is describable by counting constraint, $k$ is the dimension of the symbol alphabet $\Sigma$, and populations $D \in \pop{\Sigma}$ are seen as vectors $\vec{v} \in \mathbb{N}^k$.
Fortunately, as mentioned in Section \ref{sec:expressivity}, Angluin \etal show in \cite{AAER07} that IO protocols compute exactly the predicates representable by counting constraints, and DO protocols compute a subset of these. 

\begin{lemma}
\label{lm:consensusandco-counting-sets}
Let $\PP$ be an IO or DO protocol with $Q$ its set of states, and let $\varphi$ be a predicate describable by a counting constraint $\Gamma$.
Then $I_b |_Q$ and $\Cons{b} |_Q$, the restrictions of $I_b$ and $\Cons{b}$ to their components over $Q$, are describable by counting constraints for  $b\in \set{0,1}$.
Moreover, the norms of these counting constraints are bounded in the norms of the counting constraint associated to $\varphi$ and in $n=|Q|$:
\begin{align*}
&\lnorm{I_0 |_Q}\leq n\unorm{\Gamma}+n \quad &\unorm{I_0|_Q}\leq n\lnorm{\Gamma} \\
&\lnorm{I_1|_Q}= \lnorm{\Gamma} \quad &\unorm{I_1|_Q}= \unorm{\Gamma} \\
&\lnorm{C_0|_Q}=\lnorm{C_1|_Q}= 0 \quad &\unorm{C_0|_Q}=\unorm{C_1|_Q}= 0 \\
\end{align*}
\end{lemma}
\begin{proof}
Let $\PP$ be an IO or DO protocol over an alphabet $\Sigma$ with initial state mapping $\iota$, and $Q$ its set of states.
Predicate $\varphi$ is a predicate describable by counting constraint $\Gamma$ which is over $\pop{\Sigma}$, \ie the bounds of the cubes of $\Gamma$ are mappings $\Sigma$ to $\N$.
We extend this to a counting constraint over agent configurations of $\PP$ by having the bounds of the cubes be mappings from $Q$ to $\N$: states of $\iota(\Sigma)$ map to $\N$ as before, and states to which no input symbols are mapped by $\iota$ have upper and lower bounds equal to $0$.
Without loss of generality we assume that each symbol of $\Sigma$ is mapped to one state, \ie $\iota$ is injective.
Notice that the norms of this extension of $\Gamma$ are still equal to $\lnorm{\Gamma}$ and $\unorm{\Gamma}$. 
We abusively also note this extension $\Gamma$.

Recall that $I_b=I(\varphi^{-1}(b))$ in the generalized protocol notation.
In the IO or DO notation, 
$${I}_b |_Q = \{ \iota(D) | \exists D\in \pop{\Sigma} \ . \ \varphi(D)=b \}$$
 where $\iota(D)$ is the agent configuration $\sum_{\sigma \in \Sigma} D (\sigma)\iota(\sigma)$.
Then ${I}_b |_Q$ is describable by the counting constraint $\Gamma$ for $b=1$ and by the counting constraint corresponding to $1-\varphi$ for $b=0$. 
The bounds on the norm of $I_0|_Q$ are a consequence of Proposition \ref{prop:oponconf}.

The set $\Cons{b} |_Q$ is given by the cube of upper bound equal to $0$ on all states $q$ with output $1-b$ and $\infty$ otherwise, and the lower bound equal to $0$ everywhere.
This cube is of upper and lower norm $0$.
 \end{proof}

\begin{remark}
Initial configurations are zero-message in all protocol models, so $I_b |_Q$ is exactly $I_b$.
For $\PP$ an IO protocol, $\Cons{b} |_Q$ is exactly $\Cons{b}$ for $b\in \set{0,1}$.
\end{remark}

%% file: io-correctness-sans-reachability-CWK.tex
Since IO protocols are well-behaved protocols (by Lemma \ref{lm:wellbehaved}), we can apply the reformulation of correctness as a reachability problem of Proposition \ref{prop:corr-2}.
An IO protocol $\PP$ is correct for a predicate $\varphi$ if and only if
\begin{equation}
\label{prop:well-specification}
\poststar({I}_{b}) \subseteq \prestar(\Stab{b} )
\end{equation}
for $b\in \set{0,1}$.
By Theorem \ref{thm:ccreach-io}, Propostion \ref{prop:oponconf} and Lemma \ref{lm:consensusandco-counting-sets} above, $\Stab{b}$ is a counting set of norms 
$\lnorm{\Stab{b}} \leq n \in O(n) , \ \unorm{\Stab{b}} \leq n^3 +n^2 \in O(n^3)$, with $n \defeq |Q|$.

Thus Equation (\ref{prop:well-specification}) formulates the problem of correctness of an IO protocol as a predicate with boolean and reachability operators over counting sets.
We use the results of Section \ref{sec:small-instance} to show that we only need to examine ``small" configurations to verify such predicates, thus yielding a \PSPACE \ algorithm for checking correctness.
We start by giving a lemma for general predicates with boolean and reachability operators over counting sets, then apply it to the predicate for correctness.

\begin{lemma}
\label{lm:pspace}
        Let $\mathcal{S}_1$ and $\mathcal{S}_2$ be two functions 
        that take as arguments an IO protocol $\PP$ and a counting constraint $X$, and return counting sets $\mathcal{S}_1(\PP,X)$ and $\mathcal{S}_2(\PP,X)$ respectively.

        Assume that $\mathcal{S}_1(\PP,X)$  and $\mathcal{S}_2(\PP,X)$ 
        have
        norms at most exponential in the size of the $(\PP,X)$,
        as well as
        \PSPACE{}-decidable membership
        (given input $(C,\PP,X)$, decide whether $C\in \mathcal{S}_i(\PP,X)$).
        
        Then the same is true about the counting sets
        $\mathcal{S}_1(\PP,X)\cap\mathcal{S}_2(\PP,X)$,
        $\mathcal{S}_1(\PP,X)\cup\mathcal{S}_2(\PP,X)$,
        $\compl{\mathcal{S}_1(\PP,X)}$,
        $\prestar(\mathcal{S}_1(\PP,X))$, and
        $\poststar(\mathcal{S}_1(\PP,X))$.
        Furthermore, given $\PP$ and $X$, the emptiness problem for these sets is in \PSPACE.
\end{lemma}
\begin{proof}
        The exponential bounds for the norms follow immediately from 
        Proposition~\ref{prop:oponconf} and Theorem~\ref{thm:ccreach-io}.
        The membership complexity for union, intersection and 
        complement is easy to see.
        Without loss of generality it suffices to prove that membership in
        $\poststar(\mathcal{S}_1(\PP,X))$
        is in \PSPACE.

        By Savitch's Theorem \NPSPACE=\PSPACE, so we provide a nondeterministic algorithm.
        Given $(C,\PP,X)$, we want to decide whether $C\in \poststar(\mathcal{S}_1(\PP,X))$.
        The algorithm
        first guesses a configuration $C_0 \in \mathcal{S}_1(\PP,X)$ of the same size as $C$,
        verifies that $C_0$ belongs to $\mathcal{S}_1(\PP,X)$, 
        and then guesses an execution starting at $C_0$, step by step,
        checking after each step if the reached configuration is $C$.
        Notice that all intermediate configurations of such an execution have the same size as $C$.
        At any moment in time the algorithm only stores three configurations,
        the current one, the next configuration in the execution, and the input one. 

        We can now observe that the  emptiness problem is in \PSPACE{} for any counting set
        with exponentially bounded norm and \PSPACE-decidable membership.
        We again use Savitch's Theorem.
        If the counting set is nonempty, it has an element
        of size equal to the $l$-norm of the set.
        Such an element can be described in polynomial space.
        Therefore we can guess it and verify the set membership.
 \end{proof}

\begin{theorem}
\label{thm:pspace-io-correctness}
The correctness problem for IO protocols is in \PSPACE.
\end{theorem}
%
\begin{proof}
Let $\PP=(Q,\delta,\Sigma, \iota, o)$ be an IO protocol, and $\varphi$ a predicate over $\pop{\Sigma}$.  According to Proposition \ref{prop:corr-2},  $\PP$ computes $\varphi$ if and only if
\begin{equation}
\label{correct}
\poststar({I}_{b}) \cap \compl{\prestar(\Stab{b} )} = \emptyset.
\end{equation}
for $b \in \{ 0,1 \}$. By Lemma \ref{lm:consensusandco-counting-sets}, ${I}_{b}$ and $\Cons{b}$ are counting sets of polynomial norm.

By repeated application of Lemma \ref{lm:pspace},
we observe that membership in
$\poststar({I}_{b})$,
${\prestar(\Stab{b})}$,
$\compl{\prestar(\Stab{b} )}$,
and finally
$\poststar({I}_{b}) \cap \compl{\prestar(\Stab{b} )}$
is in \PSPACE;
furthermore, emptiness of 
$\poststar({I}_{b}) \cap \compl{\prestar(\Stab{b} )}$
is in \PSPACE{} as a problem with input $\PP$ and $\varphi$.
 \end{proof}

%% file: mfdo-correctness-CWK.tex
We show that both the single-instance correctness and the correctness problem for DO protocols are in $\Pi_2^p$.

Throughout the section we use the symbol $Z$, possibly with accents or subscripts, to denote
zero-message configurations.
As before we denote the set of zero-message configurations by $\mathcal{Z}$.

We start with a characterization of non-correctness of a protocol for a given input.

\begin{lemma}
\label{lem:notcomputeDO}
Let $\PP$ be a DO protocol with input alphabet $\Sigma$, let $\varphi$ be a predicate over $\pop{\Sigma}$, and let $D \in \pop{\Sigma}$ be an input to $\PP$. $\PP$ does not compute $\varphi(D)$ on input $D$ if{}f there exist zero-message configurations $Z, Z_{nc}$ such that
 \begin{itemize}
 \item[(i)] $I(D) \trans{*} Z \trans{*} Z_{nc}$;
 \item[(ii)] $Z_{nc}$ is not a $\varphi(D)$-consensus; and
 \item[(iii)] for every $Z'$ reachable from $Z$ there exists $C$ such that  $Z' \trans{*} C$
 and $C|_Q = Z$.
 \end{itemize}
\end{lemma}
\begin{proof}
\noindent ($\Leftarrow$) Assume that there exist  $Z, Z_{nc}$ satisfying (i)-(iii). We show that no configuration reachable from $Z$ is a stable $\varphi(D)$-consensus, which implies that $\PP$ does not compute $\varphi(D)$. Let $\tilde{C}$ be an arbitrary configuration reachable from $Z$. By consuming all messages of $\tilde{C}$, the protocol can move from $\tilde{C}$ to some zero-message configuration $\tilde{Z}$ and, by (iii), to a configuration $C$ such that $C|_Q = Z$. By (i), there exists a transition sequence $\xi$ such that
$Z \trans{\xi}Z_{nc}$. Since $C|_Q = Z$, we have $C \trans{\xi} C_{nc}$ for some
configuration $C_{nc}$ such that $C_{nc}|_Q = Z_{nc}$ (the sequence just ``ignores'' the messages
of $C$).  Summarizing, we have

$$ Z \trans{*} \tilde{C} \trans{*} \tilde{Z} \trans{*} C \trans{*} C_{nc} $$

\noindent and so in particular $\tilde{C} \trans{*} C_{nc}$. By (ii) and $C_{nc}|_Q = Z_{nc}$, the configuration $C_{nc}$ is not a $\varphi(D)$-consensus, and so $\tilde{C}$ is not a stable $\varphi(D)$-consensus.

\noindent ($\Rightarrow$)
Assume that $\PP$ does not compute $\varphi(D)$ on input $D$.
Let $B$ be a bottom configuration reachable from $I(D)$ with no stable consensus reachable from it.
Let $Z$ be an arbitrary zero-message configuration reachable from $B$.
By the assumption that $B$ cannot reach a stable consensus,
there is a configuration $Z\trans{*}C_{nc}$ which contains an agent with the output $1-\varphi(D)$.
Recall that we always have at least two agents, because configurations of our protocol models are defined as the populations over $Q$ or $Q\cup M$, and populations are multisets with at least two elements.
Given a configuration $C \in \compl{\Cons{\varphi(D)}}$, we can keep one agent of $C$ ``aside" that has output $1-\varphi(D)$ and let the other agents of $C$ consume all the messages.
This method applied to $C=C_{nc}$ yields a zero-message configuration $Z_{nc}$ such that $Z\trans{*}C_{nc}\trans{*}Z_{nc}$
which is not a $\varphi(D)$-consensus.
        This proves properties (i) and (ii).
        To prove the property (iii) observe that $B$ was a bottom configuration and therefore
        for every $Z\trans{*}Z'$ we have $B\trans{*}Z\trans{*}Z'$ and therefore
        $Z'\trans{*}B\trans{*}Z$. We can now define $C=Z$.
 \end{proof}

\begin{theorem}
\label{thm:single-correctness-do}
Single-instance correctness of DO protocols is in $\Pi_2^p$.
\end{theorem}
 \begin{proof}
Let $\PP=(Q,M,\delta_r, \delta_s,\Sigma, \iota, o)$ be a DO protocol, let $\varphi$ a predicate over $\pop{\Sigma}$, and let $D \in \pop{\Sigma}$ be an input to $\PP$. We show that the problem of checking whether $\PP$ with input $D$ computes $\varphi(D)$ lies in $\Pi_2^p$.

It suffices to show that the  problem of checking the existence of $Z$ and $Z_{nc}$ satisfying conditions (i)-(iii) of Lemma \ref{lem:notcomputeDO} is in $\Sigma_2^p$. By the Shortening Theorem for DO protocols (Corollary \ref{thm:doshortening}), we can guess two configurations $Z$ and $Z_{nc}$ satisfying (i) and (ii) in polynomial time, by nondeterministically traversing a computation of polynomial length (recall that all configurations reachable from $I(D)$ have the same size as $I(D)$), and checking in linear time that $Z_{nc}$ is not a $\varphi(D)$-consensus. The rest of the proof shows that checking (iii) is in co-\NP. We proceed in three steps:

\begin{itemize}
\item We define the \emph{saturation} of a zero-message configuration.
\item We replace condition (iii) by an equivalent condition (iv) on the saturation of $Z'$ (see Claim 2 below)
\item We show that checking (iv) is in co-\NP.
\end{itemize}

\paragraph{Saturation.} Let $Z$ be an arbitrary zero-message configuration, and let $|Z|$ be the number of agents of $Z$.
For every state $q \in Q$ such that $Z(q)>0$, let one of the agents in $q$ send $|Z||Q|+|Q|^2$ messages $\delta_s(q)$.
As long as there are reachable states $q$ that have not yet sent $|Z||Q|+|Q|^2$ messages, let an agent go to $q$ by a shortest path (which is of length at most $|Q|-1$, see proof of Theorem \ref{thm:mfdoshortening}) and let the agent send $|Z||Q|+|Q|^2$ messages $\delta_s(q)$. The resulting configuration $\MS(Z)$, called the \emph{saturation} of $Z$, has the following properties:
\begin{itemize}
\item[(a)] $Z \trans{*} \MS(Z)$. \\
By definition.
\item[(b)] For every message-type $m$, either $\MS(Z)$ has no messages of type $m$, or it has at least $|Z||Q|$ of them. \\
Indeed at most $|Q|$ messages are consumed in the addition of a new message-type, as a shortest path has length at most $|Q|-1$ with at most one message consumed per step.
And at most $|Q|$ message-types can be added (as each state sends only one type of message), and therefore each message-type has at most $|Q|^2$ messages consumed in $Z \trans{*} \MS(Z)$.
\item[(c)] For every configuration $C$, if $\MS(Z) \trans{*} C$ then $\MS(Z) \trans{\xi} C'$ for some configuration $C'$ such that $C'|_Q=C|_Q$, and some sequence $\xi$ that does not send any messages.
        Indeed, no new message types can be added to $\MS(Z)$ because otherwise we would have added them during the saturation step.
There are enough messages of each type for $|Z|$ agents to move to new states by less than $|Q|$ steps (along the shortest paths), so no new messages are needed to reach $C$.
\end{itemize}

\paragraph{From condition (iii) to condition (iv).} We claim:

\medskip \noindent \textbf{Claim 1.} Let $Z$ be a zero-message configuration. Condition (iii) of Lemma \ref{lem:notcomputeDO} is equivalent to:
\begin{itemize}
\item[(iv)]  for every $Z'$ reachable from $Z$ there exists $C$ such that  $\MS(Z') \trans{*} C$
 and $C|_Q = Z$.
 \end{itemize}
To show that (iv) implies (iii), let $Z'$ be reachable from $Z$. By (iv), there exists $C$
 such that $\MS(Z') \trans{*} C$ and $C|_Q = Z$. Since  $Z' \trans{*} \MS(Z')$,
 we have $Z' \trans{*} C$. So (iii) holds. To prove that (iii) implies (iv), let $Z'$ be reachable from $Z$. Since $Z' \trans{*} \MS(Z')$, we have $Z \trans{*} \MS(Z')$. By (iii), there exists $C$ such that $\MS(Z') \trans{*} C $ and $C|_Q = Z$, and we are done.

\paragraph{Checking (iv) is in co-\NP.} Condition (iv) states that every $Z'$ reachable from $Z$ satisfies $P(Z,Z')$, where

$$P(Z,Z') \defeq \exists C . \; \MS(Z') \trans{*} C \; \wedge \; C|_Q = Z \ . $$

\noindent We prove that the negation of (iv), \ie, the existence of $Z'$ reachable from $Z$ satisfying $\neg P(Z,Z')$, is in \NP. By the Shortening Theorem (Corollary \ref{thm:doshortening}), $Z'$ can be guessed in polynomial time. So it suffices to prove the second and final claim:

\medskip \noindent \textbf{Claim 2.} For every zero-message configuration $Z'$, we can check in deterministic polynomial time whether $P(Z,Z')$ holds.

By property (c) of the saturation $\MS(Z')$ of $Z'$, checking $P(Z,Z')$ reduces to deciding if there is a history of length $|Q|-1$ whose trajectories transfer the agents from their states in $\MS(Z')$ to their states in $Z$, while sending no messages, and consuming only messages in $\MS(Z')$. We reduce this question to an integer max-flow problem, which can be solved in polynomial time by e.g. Edmonds-Karp algorithm. Consider the following directed graph $G_{Z,Z'}$ with capacities:

\begin{itemize}
\item The nodes of  $G_{Z,Z'}$ are $|Q|$ copies of $Q$, written $q^{(1)},\allowbreak q^{(2)},\allowbreak  \ldots\allowbreak  q^{(|Q|)}$ for each $q \in Q$,
plus a source node $s$, and a target node $t$.

\item $G_{Z,Z'}$ has edges from $s$ to each $q^{(1)}$ with capacity $\MS(Z')(q)$, and from each $q^{(|Q|)}$ to $t$ with capacity $Z(q)$.

\item For each $i=1,\ldots,|Q|-1$, $G_{Z,Z'}$ has an edge from $q^{(i)}$ to $q'^{(i+1)}$ whenever the protocol has
a receive transition from $q$ to $q'$ that consumes a message of $\MS(Z')$, or when $q=q'$.
These edges have infinite capacity.
\end{itemize}

A flow value in this graph cannot exceed $\sum_{q\in Q} \MS(Z')(q)= |Z|$.
Integer flows of value $|Z|$ naturally correspond to histories of length
$|Q|-1$ leading from $\MS(Z')$ to a configuration $C$ such that $C|_Q = Z$, and vice versa. The flow through an edge
$(q^{(i)}, q'^{(i+1)})$ gives the number of trajectories $\tau$ of $H$ such that $\tau(i) \tau(i+1) = q \; q'$.
So we have: $P(Z,Z')$ holds if{}f the maximum integer flow of $G_{Z,Z'}$ is equal to $|Z|$.
  \end{proof}

We formulate a new characterization of DO correctness, which considers only the reachability of zero-message configurations.


\begin{restatable}{proposition}{PropDOCharacterization}
\label{prop:do-correctness}
A DO protocol $\PP$ is correct for a predicate $\varphi$ if{}f the following holds for $b\in \set{0,1}$:
$$
\postzmstar({I}_{b}) \subseteq \prezmstar(\Stab{b}^{Z} )
$$
where $\Stab{b}^{Z}$ is the set of zero-message configurations $Z$ such that every zero-message configuration reachable from $Z$ has output $b$.
\end{restatable}
\begin{proof}
Notice that $\postzmstar({I}_{b})$ is well-defined because DO initial configurations are always zero-message.
By definition, $\Stab{b}^{Z}$ is the set of zero-message configurations described by $\compl{\prezmstar\left(\compl{ \Cons{b} } \cap \mathcal{Z} \right)}\cap \mathcal{Z}$.
We prove the following claim

\medskip \noindent \textbf{Claim.} The set equality $\Stab{b}^Z = \Stab{b} \cap \mathcal{Z}$ holds.

\noindent This can be rewritten as
$$\compl{\prezmstar\left(\compl{ \Cons{b} } \cap \mathcal{Z} \right)} \cap \mathcal{Z} = \compl{\prestar(\compl{ \Cons{b} })}  \cap \mathcal{Z}.$$
Consider a configuration $Z$ in $\compl{\prezmstar\left(\compl{ \Cons{b} } \cap \mathcal{Z} \right)} \cap \mathcal{Z}$. We assume  that $Z \notin \compl{\prestar(\compl{ \Cons{b} })}  \cap \mathcal{Z}$, \ie $Z \notin \compl{\prestar(\compl{ \Cons{b} })}$, and derive a contradiction.
Since $Z \notin \compl{\prestar(\compl{ \Cons{b} })}$, from $Z$ we can reach a configuration of $\compl{\Cons{b}}$.
We show that we can also reach a configuration of $\compl{\Cons{b}} \cap \mathcal{Z}$.
We again use that a configuration contains at least two agents.
Given a configuration $C \in \compl{\Cons{b}}$, we can keep one agent of $C$ ``aside" that has output $1-b$ and let the other agents of $C$ consume all the messages.
This is possible because $\delta_r$ is a total function, and thus every $C \in \compl{\Cons{b}}$ can reach a configuration of $\compl{ \Cons{b} } \cap \mathcal{Z}$, thus amounting to a contradiction for $Z$.

Conversely, let $Z \in \compl{\prestar(\compl{\Cons{b}})} \cap \mathcal{Z}$, and
assume $Z \notin \compl{\prezmstar\left(\compl{ \Cons{b} } \cap \mathcal{Z} \right)}$. Then $Z$ can reach a configuration of $\compl{ \Cons{b} } \cap \mathcal{Z}$, which is also a configuration of $\compl{ \Cons{b} }$. This is a contradiction, and so the claim is proved. \medskip

Recall the characterization of correctness in Proposition \ref{prop:corr-2} which states that a DO protocol $\PP$ is correct for a predicate $\varphi$ if and only if
\begin{equation}
\label{prop:wellspecification}
\poststar({I}_{b}) \subseteq \prestar(\Stab{b} )
\end{equation}
for $b\in \set{0,1}$.
We use the claim above to show that $\postzmstar({I}_{b}) \subseteq \prezmstar(\Stab{b}^{Z})$ holds if and only if (\ref{prop:wellspecification}) holds.

Suppose (\ref{prop:wellspecification}) holds.
Let $Z$ be a configuration of $\postzmstar({I}_{b})$.
Since $\postzmstar({I}_{b}) \subseteq \poststar({I}_{b})$, there exists some $C \in \Stab{b}$ such that $Z \trans{*} C$.
Because $\delta_r$ is a total function, we can let the agents of $C$ consume all the messages so that $C \trans{*} Z'$ for some zero-message configuration $Z'$.
All configurations reachable from $\Stab{b}$ are still in $\Stab{b}$ so $Z' \in \Stab{b} \cap \mathcal{Z} =  \Stab{b}^Z$ by the claim, and we are done.

Suppose $\postzmstar({I}_{b}) \subseteq \prezmstar(\Stab{b}^{Z})$ holds.
Let $C$ be a configuration of $\poststar({I}_{b})$.
As before we let the agents of $C$ consume all its messages so that $C \trans{*} Z$ for some zero-message configuration $Z$ that is thus in $\postzmstar({I}_{b})$.
By assumption, there exists some $Z' \in \Stab{b}^Z$ such that $Z \trans{*} Z'$.
Since $\Stab{b}^Z = \Stab{b} \cap Z \subseteq \Stab{b}$, we are done.
 \end{proof}

\begin{theorem}
        The correctness problem for DO
        protocols
        is in $\Pi_2^p$.
\end{theorem}
\begin{proof}
We prove that the non-correctness problem for DO protocols is in $\Sigma_2^p$.
Let $\PP$ be a DO protocol and let $\varphi$ be a predicate.
By definition, $\PP$ is not correct if there exists an input $D \in \pop{\Sigma}$ such that $\PP$ does not compute $\varphi(D)$ on input $D$. (Observe that, by the definition of DO protocols, the initial configuration $I(D)$ is a zero-message configuration.) We start with a claim:

\noindent \textbf{Claim}. If such an input $D$ exists, then it can be chosen \emph{of polynomial size} in $\PP$ and $\varphi$. \\
By Proposition \ref{prop:do-correctness}, $\PP$ computes $\varphi$ if and only if
\begin{equation}
\label{eq:do-correctness}
\postzmstar({I}_{b}) \cap \compl{\prezmstar(\Stab{b}^{Z})} = \emptyset.
\end{equation}
 We show that if (\ref{eq:do-correctness}) does not hold, then $\postzmstar({I}_{b}) \cap \compl{\prezmstar(\Stab{b}^{Z})}$ contains a configuration, say $Z$, with a polynomial number of agents in $\PP$ and $\varphi$. Since transitions do not change the number of agents of a configuration, there exist an input $D$ such that $I(D) \trans{*} Z$ and $|D|=|I(D)|=|Z|$,
 proving the claim.

\smallskip
By Lemma \ref{lm:consensusandco-counting-sets}, $\Cons{b}|_Q$ and ${I}_b|_Q$ are counting sets with norms of linear size in the size of $\PP$ and $\varphi$.
Sets $\Cons{b}|_Q$ and ${I}_b|_Q$ are the projections
onto $\N^Q$ of the sets $\Cons{b} \cap \mathcal{Z}$ and $I_b$, respectively.
Thus, by Proposition \ref{prop:oponconf} and Corollary \ref{coro:reachofcc-do}, the set
$\postzmstar({I}_{b}) \cap \compl{\prezmstar(\Stab{b}^{Z})}$ is represented by a counting constraint $\Gamma$ whose $l$-norm is polynomial in $\PP$ and $\varphi$. More precisely, we have
$$\lnorm{\Gamma} \leq |Q|^4+|Q|^3+|Q|^3 \in O(|Q|^4) \ .$$
So if (\ref{eq:do-correctness}) does not hold, then the set  $\postzmstar({I}_{b}) \cap \compl{\prezmstar(\Stab{b}^{Z})}$ contains a a zero-message configuration with $\lnorm{\Gamma}$ agents, and the claim is proved.

\smallskip\noindent
By Lemma \ref{lem:notcomputeDO} and the claim, $\PP$ does not compute $\varphi$ if{}f there exist an input $D$ \emph{of polynomial size in $\PP$ and $\varphi$}, such that there exist zero-message configurations $Z$, $Z_{nc}$ satisfying conditions (i)-(iii) of the lemma.

By Theorem~\ref{thm:single-correctness-do}, checking the existence of $Z$ and $Z_{nc}$ for a given input $D$ lies in $\Sigma_2^p$. Since the input $D$ and the boolean $b \in \{0,1\}$ can be guessed in polynomial time in $\PP$ and $\varphi$, checking that $\PP$ does not compute $\varphi$ also lies in $\Sigma_2^p$.
 \end{proof}

%% file: sec-tower-hardness-transition.tex
In this section we establish lower bounds for the complexity of the correctness problem of the different variants of transmission protocols. We show that deciding correctness for delayed and queued transmission protocols is TOWER-hard, even in the single-instance case, and that the general correctness problem is TOWER-hard for the three variants (immediate, delayed, queued) of transmission protocols.

In order to establish these lower bounds, we make use of the fact that the reachability problem for \emph{VASS} (vector addition systems with states) is TOWER-hard~\cite{CLLLM19}.
A VASS of some fixed dimension $k \in \N$ can be described as a pair $(Q, T)$ where $Q$ is a finite set of states, and $T \subseteq Q \times \Z^k \times Q$ is a transition relation.
We write $q \trans{\vec{v}} r$ whenever $(q, \vec{v}, r) \in T$. Furthermore, for two
vectors $\vec{w}, \vec{w}' \in \N^k$ and states $q, q' \in Q$, we write
$(q, \vec{w}) \trans{} (q', \vec{w}')$ whenever there exists a vector $\vec{v}$ such that $q \trans{\vec{v}} q'$ and $\vec{w}' = \vec{w} + \vec{v}$. As usual, by $\trans{*}$ we denote the reflexive-transitive closure of $\trans{}$. The reachability problem for VASS is the following problem: Given vectors $\vec{v}, \vec{w} \in \N^k$ in the dimension $k$ of a given VASS, and given states $q, r$, does $(q, \vec{v}) \trans{*} (r, \vec{w})$ hold?

We call a VASS $(Q, T)$ a $\pm1$-VASS, if every transition $q \trans{\vec{v}} q'$ in $T$ satisfy
that all components of $\vec{v}$ but one are equal to 0, and this component has value $1$ or $-1$.
For a given $\pm1$-VASS $\NN$ of some dimension $k$, and $1 \leq m \leq k$,
we write $q \trans{m++} q'$ whenever $q \trans{\vec{v}}_\NN q'$ holds for some $\vec{v}, q, q'$ such that $v_m = 1$.
Likewise, we write $q \trans{m--} q'$ whenever $q \trans{\vec{v}}_\NN q'$ holds for some $\vec{v}, q, q'$ such that $v_m = -1$. The following proposition holds:
\begin{restatable}{proposition}{propVassToPM}
	\label{prop:vass-to-pm-vass}
	For every unary-encoded VASS $\NN = (Q, T)$ and unary-encoded configurations $(q_0, \vec{v}_0)$,
	$(q, \vec{v})$, one can construct in polynomial time  a $\pm1$-VASS $\NN' = (Q', T')$ with distinct states $r_0, r \in Q'$ such that
	\begin{align*}
	(q_0, \vec{v}_0) \trans{*}_\NN (q, \vec{v}) & \Longleftrightarrow (r_0, \zero) \trans{*}_{\NN'} (r, \zero) \ .
	\end{align*}
\end{restatable}
\begin{proof}
The reduction is rather straightforward; details can be found in the appendix.
 \end{proof}

To simplify the coming proofs, we introduce nondeterministic delayed-transmission protocols. The definition of the nondeterministic version is identical to the deterministic version except
that $\delta_s$ now maps to sets of message/state pairs, $\delta_r$ maps to a non-empty set of states, and the scheduler must choose nondeterministically from these sets whenever a message is sent or received.

Nondeterminism adds no expressive power to delayed-transmission protocols, as the following proposition shows:

\begin{restatable}{proposition}{propDetToNonDet}
	\label{prop:deterministic-to-nondeterministic-dt}
	For every nondeterministic DT protocol $\PP$ there exists a deterministic DT protocol $\PP'$ that computes the same predicate as $\PP$.
	Moreover, $\PP'$ can be constructed in polynomial time.
\end{restatable}
\begin{proof}
	Let $\PP =(Q, M, \delta_s, \delta_r, \Sigma, \iota, o)$.
	In order to simulate the nondeterminism of $\PP$ in $\PP'$, each state $q \in Q$ is annotated with a round counter $i$ ranging from $1$ to $n$, where $n$ is the maximal number of nondeterministic choices per state. When an agent sends/receives a message from $M$, the counter $i$ determines the choice to be made.
	Additionally, agents may send and receive a special message $\increment$. Whenever an agent receives the message $\increment$, its round counter is incremented by one, that is, $i$ is set to $(i \text{ modulo } n) + 1$.
	To ensure full simulation of nondeterminism, we must ensure that there are always enough $\increment$ messages in circulation. We achieve this by letting every agent emit an $\increment$ message at the start of the computation, and enforcing re-emission of $\increment$ messages after receiving an $\increment$ message. Whether an agent must send an $\increment$ message is governed by an additional bit, which the agent stores in its state. We provide the full construction in the appendix.
 \end{proof}

We show:

\begin{restatable}{proposition}{propVassToProtocol}
\label{prop:pm-vass-to-protocol}
	Let $\NN = (Q^\NN, T^\NN)$ be a $\pm1$-VASS and let $r_0, r \in Q^\NN$. It is possible to construct in polynomial time a (nondeterministic) DT protocol $\PP$ and an initial configuration $C_0$ of $\PP$ such that
	$(r_0, \zero) \trans{*} (r, \zero)$ holds if and only if $\PP$ does not converge to $1$ for the initial configuration $C_0$.
\end{restatable}
\begin{proof}
        Intuitively, the protocol $\PP$ simulates the $\pm1$-VASS in a population of size $1$, with the current control state of $\NN$ being stored in the state of the single agent, and the current counting vector represented in the message pool by messages denoted $1, \ldots, k$. For example, if the configuration of the machine is $q, (6, 4)$, then the agent is in state $q$, and the message pool contains $6$ messages denoted by $1$, and $4$ messages denoted by $2$. Decrementing/incrementing a counter is implemented by sending/receiving messages.

When the agent reaches state $r$, it can nondeterministically guess that the current vector is $\zero$, and then alternate indefinitely between a false and a true state, say $r_\bot$ and $r_\top$, which constitutes  a non-stabilizing fair execution in the case where $r_0, \zero \trans{*} r, \zero$ holds. If the agent makes a wrong guess, then the message pool is non-empty at that time, and by fairness the agent eventually receives a message which lets the the agent turn to a permanent true state, say, $\top$. This ensures that every fair execution converges to $1$ in the case where $r_0, \zero \not \trans{*} r, \zero$.

Let us now define $\PP$ formally.
Given the $\pm1$-VASS $\NN$ of some dimension $k$ and the states $r_0, r$, the protocol $\PP= (Q, M, \delta_s, \delta_r, \Sigma, \iota, o)$ is constructed as follows:
\begin{itemize}
	\item $Q \defeq Q^\NN \cup \{r_\top, r_\bot, \top \}$
	\item $M \defeq \{1, \ldots, k\} \cup \{\epsilon\}$
	\item $\delta_s $ is given by:
	      \begin{align*}
	      	\delta_s(r) & \defeq \{(q', m) \mid r \trans{m++} q' \} \cup \{(r_\bot, \epsilon) \} \\
	      	\delta_s(r_\bot) & \defeq \{(r_\top, \epsilon) \} \\
	      	\delta_s(r_\top) & \defeq \{(r_\bot, \epsilon) \} \\
	      	\delta_s(q) & \defeq \{(q', m) \mid q \trans{m++} q' \} \quad \text{ for every } q \in Q^\NN \setminus \{r \}.
	      \end{align*}
	\item $\delta_r$ is given by:
		\begin{align*}
			\delta_r(r_\top, \epsilon) & = \delta_r(r_\bot, \epsilon)  \defeq \{r_\top \} \\
			\delta_r(q, m) & \defeq \{q' \mid q \trans{m--} q' \} & \text{ if } q \trans{m--} q'
			                                                       \text{ for some } q' \\
			\delta_r(q, m) & \defeq \{ \top \} & \text{ in all remaining cases. }
		\end{align*}
	\item $\Sigma \defeq \{r_0\}$
	\item $\iota = \text{id}$
	\item $o(r_\bot) \defeq 0$ and $o(q') \defeq 1$ for every $q' \neq r_\bot$.
\end{itemize}
We define the initial configuration by setting $C_0 \defeq \multiset{r_0}$.

We associate a configuration $C \in \pop{\{1, \ldots, k\}}$ with its corresponding vector in $\N^k$ via the bijection
$\varphi \colon \pop{Q} \rightarrow \N^k$ given by $\varphi(C) \defeq (C(1), \ldots, C(k))$.
By construction, for every sequence of states $q_1, \ldots, q_m \in Q^\NN$, and
every sequence of vectors $\vec{v}_1, \ldots, \vec{v}_m \in \N^k$ we have:
\begin{align*}
	& (r_0, \zero) \trans{} (q_1, \vec{v}_1) \trans{} (q_2, \vec{v}_2) \trans{} \ldots \trans{} (q_m, \vec{v}_m) \\
	& \Longleftrightarrow  \\
	& \multiset{r_0} \trans{} \left(\multiset{q_1} + \varphi^{-1}(\vec{v}_1)\right) \trans{} \ldots \trans{} \left(\multiset{q_m} + \varphi^{-1}(\vec{v}_m)\right).
 \end{align*}

It remains to prove that $\PP$ does not converge to $1$ for $C_0 = \multiset{r_0}$ if and only if $r_0, \zero \trans{*} r, \zero$. We only prove the direction $(\Leftarrow)$; the converse direction is similar.
Assume $r_0, \zero \trans{*} r, \zero$ holds. Then by the previous consideration we have:
$
C_0 \trans{*} \multiset{r}
$.
Thus we obtain:
$$
C_0 \trans{*} \multiset{r} \trans{} \multiset{r_\bot, \epsilon} \trans{} \multiset{r_\top} \trans{} \multiset{r_\bot, \epsilon} \trans{} \multiset{r_\top} \trans{} \ldots
$$
The above execution is fair, but does not converge to a consensus, as $o(r_\top) \neq o(r_\bot)$. Hence $\PP$ does not converge to $1$ for $C_0$, which concludes the proof for this direction.

Formally, the population should have at least two agents. One of the ways to resolve this problem is to say that we have an extra state $\bot$ with output $0$,
and an extra agent starting in the state $\bot$. It never sends messages, and if it ever receives a message, it switches to $\top$.
We can let $\top$ send a special message $m_\top$ turning the other agent into $\top$.
If there is a finite execution producing $r_\bot$ and leaving no messages, it can happen despite existence of the extra $\bot$ agent;
otherwise we reach $\top$ like we did before.
 \end{proof}

Combining the previously established propositions, we obtain:

\begin{theorem}\label{thm:one-instance-well-specification}
The single-instance correctness problem is TOWER-hard for DT and QT protocols.
\end{theorem}
\begin{proof}
Since delayed-transmission protocols are a subclass of queued-transmission protocols, it suffices to show the claim for delayed-transmission protocols.

By propositions ~\ref{prop:vass-to-pm-vass} and~\ref{prop:pm-vass-to-protocol}, the TOWER-hard reachability problem for VASS
is polynomially Turing-reducible to 1-instance correctness of delayed-transmission protocols. This shows the theorem.
 \end{proof}

We establish the same hardness result for the general correctness problem:

\begin{theorem}\label{thm:well-specification-transmission}
	The correctness problem for DT and QT protocols is TOWER-hard.
\end{theorem}
\begin{proof}
	Since delayed-transmission protocols are a subclass of queued-transmission protocols, we only need to prove the theorem for delayed-transmission protocols. In the appendix, we prove the following claim: For every delayed-transmission protocol $\PP=(Q, M, \delta_r, \delta_s, \Sigma, \iota, o)$ and every initial configuration $C \in \pop{I}$, one can construct in polynomial time a delayed-transmission protocol $\PP'=(Q', M', \delta'_r, \delta'_s, \Sigma, \iota', o')$ such that $\PP'$ computes constant $1$ if and only if $\PP$ converges to $1$ for the single instance $C$. By Theorem~\ref{thm:one-instance-well-specification}, the claim entails Theorem~\ref{thm:well-specification-transmission}, and we are done.
 \end{proof}

Perhaps surprisingly, even in the restricted setting of immediate-transmission protocols,  the general correctness problem remains TOWER-hard:

\begin{theorem}\label{thm:well-specification-immediate-transmission}
The correctness problem for IT protocols is TOWER-hard.
\end{theorem}
\begin{proof}
Let $\NN = (Q, T)$ be a $\pm1$-VASS and let $q, r \in Q$. We claim that we can construct in polynomial time an immediate-transmission protocol $\PP$ that computes constant $1$ if and only if  $q, \mathbf{0} \trans{*} r, \mathbf{0}$ does \emph{not} hold.
The claim entails the theorem by Proposition~\ref{prop:vass-to-pm-vass} and TOWER-hardness of VASS-reachability. In the appendix we provide a construction that shows the claim.
 \end{proof}

On the other hand, the single-instance correctness problem for immediate transmission protocols is not TOWER-hard.
It is in fact \PSPACE-complete.

\begin{theorem}
The single-instance correctness problem for IT protocols is \PSPACE-complete.
\end{theorem}
\begin{proof}
Let $\PP=(Q,\delta,\Sigma, \iota, o)$ be an IT protocol, $\varphi$ a predicate over $\pop{\Sigma}$ and $C_0$ a configuration.
We reuse the notation of section \ref{sec:coNP}, and let $C_0$ be a configuration in $I_b$ for $b\in \set{0,1}$, \ie a fair execution starting in $C_0$ must converge to  $b$ if the protocol is correct.
The proof is the same as for single-instance correctness of IO protocols in Theorem \ref{thm:pspace-io-correctness}: using the correctness characterization of Proposition \ref{prop:corr-2}, we guess a configuration $C$ of size $|C_0|$ and check that it is in the intersection $\poststar({I}_{b}) \cap \compl{\prestar(\Stab{b} )}$ using \NPSPACE \ procedures.
The only difference with the IO proof lies in the step relation, which remains checkable in polynomial time.

\PSPACE-hardness follows from the fact that IO protocols are IT protocols, and the hardness result of Theorem \ref{thm:io-correctness-hard}.
 \end{proof}

%% file: section-decidability-CWK.tex
We present a generic result showing that the correctness problem is decidable for a class of protocols satisfying certain properties. All protocol models considered in the paper, with the exception of QT, satisfy the properties. The proof follows closely the one of \cite{EsparzaGLM17} for standard population protocols. 
However, the presentation emphasizes the role played by each of the properties, allowing us to pinpoint why the proof of \cite{EsparzaGLM17} can be generalized to DT protocols, but not to QT protocols. While we leave the decidabililty of correctness for QT open, we also argue that the notion of fairness chosen in \cite{AAER07}, and also used in our paper, is questionable for QT, making the correctness problem for QT less interesting than for the other five models.

Recall the property defined in Section \ref{sec:formalizing-correctness}: a protocol is \emph{well-behaved} if every fair execution contains a bottom configuration.
We introduce some further properties of protocols:

\begin{definition} \label{def:wbP}
	A protocol $\PP=(\Conf, \Sigma, \Step, I, O)$ is 
	\begin{itemize}
		\item \emph{finitely generated} if  $\Conf \subseteq \N^k$ for some $k \geq 0$, and
			there is a finite set $\Delta \subseteq \Z^k$ such that 
			$(C, C') \in \Step$ if{}f $C' - C \in \Delta$; we say that $\Step$ is \emph{generated} by $\Delta$.
		\item \emph{input-Presburger} if for every effectively Presburger set $L \subseteq \pop{\Sigma}$ of inputs the set $I(L) \subseteq \pop{Q}$ is an effectively computable Presburger set of configurations.
		\item \emph{output-Presburger} if $O^{-1}(0)$ and $O^{-1}(1)$ are effectively Presburger sets of configurations.
	\end{itemize}

We call a protocol that is well-behaved, finitely generated, and input/output-Presburger a \emph{WFP-protocol}.	
\end{definition}

Recall the characterization of correctness for well-behaved protocols that we obtained in Proposition \ref{prop:corr}. 
\PropCorrectness*
We show that this reachability condition is decidable for WFP-protocols. Observe that a finitely generated protocol $\PP=(\Conf, \Sigma, \Step, I, O)$ can be easily represented as a VAS. Indeed, if $\Conf \subseteq \N^k$ and $\Step$ is generated by $\Delta$, then the VAS has dimension $k$ and has $\Delta$ as set of transitions. Using this fact, and the powerful result stating the decidability of the reachability problem in a VAS between effectively Presburger sets of configurations, we obtain:

\begin{proposition}[\cite{EsparzaGLM17}]\label{prop:effectPres}
Let $\mathcal{C}, \mathcal{C}'$ be two effectively Presburger sets of configurations of a finitely generated protocol. It is decidable if some of configuration of $\mathcal{C}'$ is reachable from
some configuration of  $\mathcal{C}$. 
\end{proposition}

By Proposition \ref{prop:effectPres}, in order to prove the decidability of correctness it suffices to show that the sets $I(\varphi^{-1}(b))$ and $\B\setminus\B_b$ of a WFP-protocol are effectively Presburger sets. $I(\varphi^{-1}(b))$ holds by the definition of WFP-protocols (recall that $\varphi^{-1}(b)$ is always a Presburger set). It remains to show that $\B \setminus \B_b$ is effectively Presburger. Since effectively Presburger sets are closed under boolean operations, it suffices to show that $\B$ and $\B_b$ are effectively Presburger. This is a nontrivial result, but already proved in  \cite{EsparzaGLM17}:

\begin{proposition}[\cite{EsparzaGLM17}, Proposition 14]
\label{prop:homesemi} 
There is an algorithm that takes as input a finitely generated, output-Presburger protocol, and returns Presburger 
predicates denoting the sets \(\B\), \(\B_0\), and \(\B_1\). 
\end{proposition} 

So we finally obtain:

\begin{theorem}\label{prop:maincorr}
The correctness problem is decidable for WFP-protocols.
\end{theorem}

Applying Theorem \ref{prop:maincorr} we can easily prove that the correctness problem is decidable for PP and DT. 
Indeed, PP protocols and DT protocols are WFP as they are well-behaved by Lemma \ref{lm:wellbehaved}, and finitely generated and input/output Presburger by hypothesis.
Since IT and IO are subclasses of PP and DO is a subclass of DT, the proof is valid for them as well.


\begin{corollary}\label{cor:main}
The correctness problem is decidable for PP, DT, and their subclasses.
\end{corollary}

However, queued-transmission protocols are not necessarily well-behaved (as shown in Example \ref{ex:fair}), and so not necessarily WFP. 
Currently, to the best of our knowledge the decidability of the well-specification and correctness problems for queued-transmission protocols is open. At the same time, Example \ref{ex:fair} shows that our fairness condition is questionable for queued-transmission models: An execution $C_0, C_1, \ldots$ in which only one agent acts, even if other agents have enabled actions in $C_i$ for every $i \geq 0$, can still be fair. Is the fairness notion of \cite{AAER07} adequate for queued-transmission protocols?

%% file: decidability-probabilistic.tex
In \cite{AAER07}, Angluin \etal state that the fairness condition ``may be viewed as an attempt to capture useful probability 1 properties in a probability-free model''. Indeed, population protocols are often introduced in a probabilistic setting, which assigns a probability to the set of executions that converge to a value. Once a probabilistic model is fixed, we have two different definitions of when a  protocol $\PP$ computes a predicate $\varphi$:
\begin{itemize}
	\item $\PP$ \emph{f-computes} $\varphi$ if for every input $\sigma \in \pop{\Sigma}$, every fair execution starting at $I(\sigma)$ converges to $\varphi(\sigma)$.
	\item $\PP$ \emph{p-computes} $\varphi$ if for every input $\sigma \in \pop{\Sigma}$, the set of all executions starting at $I(\sigma)$ that converge to $\varphi(\sigma)$ has probability 1.
\end{itemize}

The question whether the fairness condition is adequate for a class of protocols can now be rephrased as: Do f-computation and p-computation coincide for the class? 
In this section we examine this question in some detail.

In order to formalize a probabilistic protocol model we must specify the random experiment that determines the next step carried out by the protocol. For standard population protocols there is agreement in the literature on the experiment: At each step two agents of the population are chosen uniformly at random, and they interact. However, for the delayed and queued-transmission models there is no canonical experiment. We consider the following family of random experiments parameterized by a probability $p$.

\begin{definition}
	Let $\PP= (Q,M,\delta_s,\delta_r, I, O)$ be a queued-transmission protocol, and let $0 < p < 1$. For every state $q \in Q$, let $R(q)$ denote the set of messages that an agent can receive in state $q$.
	The \emph{s:p/r:(1-p) probabilistic model} \footnote{Short for ``send with probability $p$, receive with probability $(1-p)$".} is described by the following random experiment. Assume the current configuration is $C$. First, choose an agent uniformly at random, and let $q$ be its current state; then:
	\begin{itemize}
		\item with probability $p$, let the agent send the message specified by the send function;
		\item with probability $1-p$: if $R(q)\neq\emptyset$, choose a message from the multiset $\bigcup_{m \in R(q)} C(m)$ uniformly at random, and let the agent receive it; otherwise, the agent does nothing.
	\end{itemize}
\end{definition}

\noindent Recall that in the delayed-transmission model we have $R(q)=M$ for every state $q$, i.e., agents can never refuse receiving a message.

In the rest of the section we examine the relation between f-computation and p-computation for our protocol models, and obtain the following results:
\begin{itemize}
	\item For standard population protocols and their subclasses, f-computation and p-computation coincide.
	\item For delayed-transmission protocols and s:p/r:(1-p) models,
                f-computation and p-computation coincide
                if{}f $p \leq 1/2$.
	\item For queued-transmission protocols, f-computation and p-computation are incomparable notions
                under fairly general conditions on probabilistic models. In particular, there are protocols that f-compute a predicate but do not p-compute any predicate in any s:p/r:(1-p) model, and vice-versa.
\end{itemize}

\paragraph{Standard population protocols.}
Recall that in the probabilistic model at each step two agents are chosen uniformly at random. We have:

\begin{proposition}
	Let $\PP$ be a standard population protocol, and let $\varphi$ be a predicate. $\PP$ f-computes $\varphi$ if{}f $\PP$ p-computes $\varphi$.
\end{proposition}
\begin{proof}
	By Proposition \ref{prop:corr}, $\PP$ f-computes $\varphi$ if{}f for every input $a$ the set $\B \setminus \B_{\varphi(a)}$ is not reachable from $I(a)$. We show that this is the case if{}f $\PP$ p-computes $\varphi$.

	Since every configuration of a standard population protocol has a finite number of successors, an execution starting at $I(a)$ almost surely visits a bottom configuration. So $\PP$ p-computes $\varphi$ if the set of executions visiting
	$\B_{\varphi(a)}$ has probability $1$.
        Since every finite execution
        leading from $I(a)$ to a configuration of $\B$ has positive probability, this
	is the case if{}f $\B \setminus \B_{\varphi(a)}$ is not reachable from $I(a)$.
 \end{proof}

\paragraph{Delayed-transmission protocols.}

We show that for delayed-transmission protocols and s:p/r:(1-p)-models f-computation and p-computation coincide if{}f $p \leq 1/2$.

\begin{lemma}
	\label{lem:zero}
	Let $\PP$ be a delayed-transmission protocol
	in the s:p/r:(1-p) model with $p \leq 1/2$.
	With probability $1$, an execution of $\PP$ visits infinitely often configurations with no messages in transit.
\end{lemma}
\begin{proof}
We prove that the number $k$ of messages in transit behaves similarly to
a random walk in which the probability of reducing $k$ is at least as high as the probability of increasing it.


	For a configuration $C$, let $\text{Pr}(C)$
	denote the probability that an execution
	starting from $C$ only visits
	configurations with at least one message in transit.
	Further, let $\text{Pr}(n,k)$ be the maximum value of $\text{Pr}(C)$
	among all configurations with $n$ agents and $k$ messages in transit.
	Observe that $\text{Pr}(n,0)=0$, because in this case $C$ itself has no messages in transit.
	We prove that $\text{Pr}(n,k)=0$ for every $k \geq 0$, which is equivalent to the statement of the lemma.

	Let $n$ and $k>0$, and let
	$C_{max}$ be a configuration with $n$ agents and $k$ messages
	satisfying $\text{Pr}(C_{max})=\text{Pr}(n,k)$.
	A step from configuration $C_{max}$
	consumes a message with probability at least $\frac{1}{2}$
	(in a delayed transmission protocol
	an agent can always receive any message), and produces a message
	with probability $0 \leq p \leq \frac{1}{2}$.
	So we have
     \begin{align*}
      \text{Pr}(n,k) & = \text{Pr}(C_{max}) \\
                           & \leq  \frac{1}{2} \, \text{Pr}(n,k-1) + p \, \text{Pr}(n,k+1) + \left( \frac{1}{2}-p  \right) \, \text{Pr}(n,k)
     \end{align*}
     \noindent which can be rewritten as
     \begin{equation*}
             \text{Pr}(n,k) \leq \frac{ \frac{1}{2} \, \text{Pr}(n,k-1) + p \, \text{Pr}(n,k+1)}{\frac{1}{2}+p}
     \end{equation*}
        The right side is the weighted average of $\text{Pr}(n,k-1)$ and $\text{Pr}(n,k+1)$, with
        weight $p$ between $0$ and $\frac{1}{2}$.
        It can be bounded by the weighted average for one of the extremal
        values of $p$, and so we have $\text{Pr}(n,k)<\text{Pr}(n,k-1)$ or $\text{Pr}(n,k)\leq \frac{1}{2}\text{Pr}(n,k-1)+\frac{1}{2}\text{Pr}(n,k+1)$. Rewriting the second case, we finally obtain  that the following disjunction holds for all $n, k > 0$:
        \begin{align*}
        & \text{Pr}(n,k)<\text{Pr}(n,k-1)  \quad \text{or}  \\
        & \text{Pr}(n,k+1) - \text{Pr}(n, k) \geq \text{Pr}(n,k)-\text{Pr}(n,k-1) \ .
        \end{align*}
        Assume there is a smallest number $z$ such that $\text{Pr}(n,z) > 0$ and $\text{Pr}(n,z-1) = 0$.
	  Then, by the disjunction above and $\text{Pr}(n,z) - \text{Pr}(n,z-1) = \text{Pr}(n, z)$, we have $\text{Pr}(n,z+i) \geq (i+1) \text{Pr}(n,z)$ for every $i \geq 0$ (easy induction on $i$). This contradicts that $1  \geq \text{Pr}(n,z+i)$ holds for every $i \geq 0$, and so $z$ does not exist. Since $\text{Pr}(n, 0)=0$ by definition, we have $\text{Pr}(n,k)=0$ for every $k \geq 0$.
 \end{proof}

\begin{proposition}
	Let $\PP$ be a delayed-transmission protocol in a s:p/r:(1-p) model with $p \leq 1/2$, and let $\varphi$ be a predicate. $\PP$ f-computes $\varphi$ if{}f $\PP$ p-computes $\varphi$.
\end{proposition}
\begin{proof}
	Assume $\PP$ f-computes $\varphi$. We show that it p-computes $\varphi$.
	For this it suffices to show that for every initial configuration $C_0$ the set of fair executions starting at $C_0$ has probability 1,
	or, in other words, that an execution is fair with probability 1.

	Fix an initial configuration $C_0$, and let $C$ be an arbitrary configuration. Let ${\cal Z}$ be the set of configurations reachable from $C_0$ with zero messages in transit. Since the number of agents remains constant, ${\cal Z}$ is finite. For each $Z \in {\cal Z}$, either $C$ is unreachable from $Z$, or there is a shortest sequence of transitions leading from $Z$ to $C$ (possibly not unique). Such a sequence has a positive probability of occurring from $Z$. Let $p_{min}$ be the minimal probability of all the probabilities of shortest paths from any $Z\in \mathcal{Z}$ to $C$, and $\ell$ be the maximum length of a shortest path.

	By Lemma \ref{lem:zero}, an execution starting at $C_0$ reaches a configuration $Z_1 \in {\cal Z}$ with probability $1$.
	Either $C$ is unreachable from $Z_1$, or the probability of
	reaching $C$ in at most $\ell$ steps is at least $p_{min}$.
	If $C$ is not reached in $\ell$ steps but remains reachable,
	with probability $1$ we reach a configuration $Z_2 \in {\cal Z}$ from $Z_1$.
	Iterating this reasoning, we observe that the execution visits a sequence of configurations $Z_1, Z_2, \ldots \in {\cal Z}$ such that for every $Z_i$, the probability that in the next $\ell$ steps $C$ is reached or becomes unreachable is at least $p_{min}$. Therefore, the event ``$C$ becomes unreachable or it is reached infinitely often'' has probability 1. So an execution is fair with probability $1$.

	Assume $\PP$ does not f-compute $\varphi$. We show that it does not p-compute $\varphi$.
Since $\PP$ does not f-compute $\varphi$, there is a fair execution $\pi$ that does not converge to the value specified by $\varphi$, call it $b$. Let $C_0$ be the initial configuration of $\pi$ and, as above, let ${\cal Z}$ be the finite set of configurations reachable from $C_0$ with zero messages in transit. Further, let $\textit{Rec}(\pi)$ be the set of configurations of ${\cal Z}$ that occur in $\pi$ infinitely often.

        Since $\PP$ is a delayed-transmission protocol, every configuration of $\pi$ can reach some configuration of ${\cal Z}$. Therefore, by fairness and finiteness of $\cal{Z}$, $\textit{Rec}(\pi) \neq \emptyset$, and $\textit{Rec}(\pi)$ is closed under reachability.
	We claim that an execution that reaches $\textit{Rec}(\pi)$ converges to $b$ with probability 0. Since there is a positive probability that a execution reaches $\textit{Rec}(\pi)$, it follows that $\PP$ does not p-compute $\varphi$. To prove the claim, observe that, since $\pi$ does not converge to $b$, some configuration $C$ reachable from $\textit{Rec}(\pi)$ is not a $b$-consensus. Since $\textit{Rec}(\pi)$ is finite, there exists $p > 0$ such that $C$ is reachable from every configuration of $\textit{Rec}(\pi)$ with probability at least $p$. Therefore, an execution that reaches $\textit{Rec}(\pi)$ visits $C$ infinitely many times with probability 1, and so it converges to $b$ with probability 0.
 \end{proof}

\begin{restatable}{proposition}{PropositionMoreThanHalf}
	There is a delayed-transmission protocol $\PP$
        that p-computes the value $0$ on a certain input
	in every s:p/r:(1-p) model with $p > 1/2$,
        but that does not f-compute any value on the same input.
\end{restatable}

\begin{proof}
        Consider the protocol with states $\{q_0,q_1,q_2\}$;
        \cwk{pass 2}\mr{first pass}
        output function given by  $O(q_0)=O(q_1)=0$ and $O(q_2) = 1$;
        messages $\{a,b\}$;
        and
        transitions
        $$\begin{array}{lllll}
        q_0\trans{a+}q_0 & \quad &  q_1\trans{b+}q_0 & \quad &  q_2\trans{b+}q_1 \\
        q_0\trans{a-}q_0   & &  q_1\trans{a-}q_1 & &  q_2\trans{a-}q_2 \\
        q_0\trans{b-}q_1 & &  q_1\trans{b-}q_2 & &  q_2\trans{b-}q_2
        \end{array}$$
        Consider the input configuration
        $\multiset{q_2}$.

        For the sake of simplicity we allow configurations
        with a single agent.
        The behaviour is qualitatively the same for
        multiple agents (as required by the definition of population), up to
        some technicalities in probablility calculations. \cwk{rewording}

        In each configuration of each execution
        the sum of the index of the state and
        the number of messages of type $b$
        is equal to $2$. This protocol does not f-compute any value on $\multiset{q_2}$
        because the configuration $\multiset{q_2}$ with no messages
        is reachable from each configuration in the execution, as well as
        the configuration $\multiset{q_0,b,b}$ with $2$ messages of type $b$.
        These two configurations occur infinitely often
        in each fair execution and have different output values.

        The proof that an execution from $\multiset{q_2}$
        converges with probability $1$ if $p>\frac{1}{2}$ is based on the following
        observations.
        \begin{itemize}
        \item The number of messages changes independently of the configuration
        change, so it is a biased random walk with linear growth.
        \item The state $q_0$ can always be reached with probability at least $1/4$, and
        so it is reached infinitely many times.
        \item Going from $q_0$ to $q_2$ requires receiving two $b$s
        without sending in-between.
        \item The probability to receive two $b$s is proportional to $1/n^2$, where
        $n$ is the number of messages. Since the series $\sum_{i=1}^\infty 1/n^2$
        converges, so with probability $1$ the state $q_2$
        is only observed a finite number of times.
        \end{itemize}
        We show that therefore $q_2$ occurs only a finite
        number of times with probability $1$, and that
        the protocol p-computes value $0$.
        The rest of the proof presents this argument in detail; it is purely technical and can be found in the appendix.
         \end{proof}

\paragraph{Queued-transmission protocols.}

Unfortunately, in queued-transmission protocols there is no useful relation between f-computation and p-computation. We show this with the help of two examples.
The first one computes a predicate in every model from a general class, but does not f-compute any predicate. The second f-computes a predicate, but does not compute a predicate in any probabilistic model from the same general class.

\begin{definition}
        A probabilistic model of execution for queued-transmission protocols is
        \begin{itemize}
        \item \emph{positive} if for every configuration $C$ every step $C \rightarrow C'$ has positive probability.
        \item \emph{markovian} if for every configuration $C$ the probability of a step $C \rightarrow C'$ is independent of the previous history.
        \item \emph{bounded} if for every $n \geq 1$ and $\alpha>0$ there is $c(n,\alpha) > 0$ with the following property. Consider any configuration with $n$ agents and at least one message in transit. If the fraction of messages receivable by at least one agent is larger than $\alpha$, the probability of receiving a message is at least $c(n, \alpha)$.
        \item \emph{uniform} if for every configuration $C$ and agent $a$, every message in transit that can be received by $a$ at $C$ is received with the same probability.
\end{itemize}
\end{definition}

\begin{remark}
Each s:p/r:(1-p) model is positive, markovian, bounded, and uniform.
\end{remark}

In the following constructions we again use single-agent configurations.
We implicitly assume that an agent in a special state that can neither send
nor receive is always added to the configuration to obtain a valid population.

\begin{proposition}
There is a queued-transmission protocol $\PP$ that p-computes the value $1$ on a certain input
 in all positive, bounded, and markovian models, but that does not f-compute any value on this input.
\end{proposition}
\begin{proof}
	Consider the protocol with states $\{q_0, q_1\}$; messages $M = \{a\}$; transitions $q_0 \trans{a+} q_0$ and $q_0 \trans{a-} q_1$; and
        output function given by  $O(q_0)=O$ and $O(q_1) = 1$.
         Consider the input configuration $\multiset{q_0}$.
                \cwk{pass 2}\mr{first pass}

	In this protocol, the unique agent sends messages until it receives a message and moves to $q_1$.
Note that all the messages are receivable by the agent in state $q_0$.
In any positive, bounded markovian model the agent eventually reaches $q_1$ with probability 1 and stays there.
So the protocol p-computes the value $1$ on input $\multiset{q_0}$.  We show that the protocol does not f-compute any value on this input, because it has a fair execution converging to $0$ and fair executions converging to $1$. The fair executions converging to $1$ are those in which the agent reaches $q_1$. The unique fair execution converging to $0$ is the one in which the agent stays in $q_0$ forever. To prove that this execution is fair observe that (a) along the execution the number of messages grows continuously, and (b) every configuration reachable from a configuration of the execution with $m$ messages in transit has at least $m-1$ messages in transit. So no configuration of the protocol is reachable from infinitely many configurations of the execution.
 \end{proof}

\begin{proposition}
There is a queued-transmission protocol $\PP$ that f-computes the value $1$ on a certain input,
but that does not p-compute any value on this input in any positive, markovian and uniform model.
\end{proposition}
\begin{proof}
	Consider the protocol with
        states $\{q_0, q_1, q_2,\allowbreak  q^+_0,\allowbreak  q^+_1,\allowbreak  q^+_2,\allowbreak  q^+_3,\allowbreak  q^-,\allowbreak  q\}$,
        messages $M = \{p, m, c\}$,
        and transitions
        $$
        \begin{array}{lll}
                q_0\trans{p+}q_1   & \quad  q_1\trans{m+}q_2 \\
                q_2\trans{p-}q^+_0 & \quad  q_2\trans{m-}q^- \\
                q^+_0\trans{c+}q^+_1
                & \quad
                q^+_1\trans{c+}q^+_2
                & \quad
                q^+_2\trans{c+}q^+_3
                \\
                q^+_3\trans{m-}q   & \quad  q^-\trans{p-}q \\
                q\trans{c-}q_0
	\end{array}
        $$
	The output function maps $q$ to $1$ and all other states to $0$.
        Consider the input configuration $\multiset{q_0}$.
        \cwk{pass 2}\mr{first pass}

	In this protocol, starting from $\multiset{q_0}$, every configuration can reach the configuration $\multiset{q}$ in which the unique agent is in state $q$, and there are no messages in transit. So every fair execution eventually reaches $\multiset{q}$ and, since no message can be sent from $q$, stays in it forever. Therefore, the protocol f-computes the value $1$ on input $\multiset{q_0}$. We now show that the protocol does not p-compute any value on the same input in any positive, markovian, uniform model. Indeed, after reaching the state $q_0$ the execution must proceed to reach the state $q_2$ creating two messages of types $p$ and $m$. The only way to proceed is to receive either $p$ or $m$, which in uniform models is equally likely. Afterwards, both $p$ and $m$ are consumed, and either three messages of type $c$ or none are created. To proceed, the agent needs to receive a message of type $c$. The number of messages of type $c$ follows a random walk with possible changes $+2$ and $-1$ until it tries to go below zero. There is a positive probability that it will never return to zero and grow linearly. In this case all the states will be observed infinitely many times, so the protocol does not compute any value.
 \end{proof}

These propositions show that the correctness problem for probabilistic queued-transmission protocols cannot be reduced to the same problem for the fairness model. So in the queued-transmission model fairness does not capture useful probability 1 properties, which questions the interest of the fairness-based model in a probability-free model. At the same time, it opens the question of the decidability of correctness for probabilistic  queued-transmission protocols. Cummings, Doty and Soloveichik have recently proved that Chemical Reaction Networks can compute with probability 1 a superset of the Turing-computable functions \cite{CummingsDS16}, and using this result we can easily prove that correctness is undecidable.

\begin{theorem}
	In any positive, markovian, and uniform probabilistic model, the single-instance correctness problem for queued-transmission protocols is undecidable.
	\cwk{updated to single-instance}
\end{theorem}

\begin{proof}
	We only sketch the argument. According to \cite{CummingsDS16}, binary chemical reaction networks
	with uniform rates can p-compute all recursively enumerable predicates (in fact even more, see \cite{CummingsDS16}).
	In such a network we are initially given set of chemical reactions, like e.g. $A + B \rightarrow 2 C + D + E$, a multiset
	of molecules of different species (A, B, C, \ldots). At every step,  two molecules are picked uniformly at random and allowed to interact according to one of the reactions, which results in an arbitrary number of product molecules.
	A binary chemical reaction network can be modelled by a queued-transmision protocol with a single agent. Molecules are modeled by messages.
	The agent sends an initial set of messages, which corresponds to the initial multiset of molecules, and moves to a new state,
	from which it repeatedly receives two randomly chosen messages, and sends the results of the reaction.
        At each stop the agent can either only send or only receive,
        and if it can receive it can receive any message.
        Uniformity and Markov property
        guarantee that each pair of messages is selected with
        equal probability regardless of the details of the model,
        and positivity ensures that the protocol will make progress
        in modelling the chemical reaction network.
        As every binary reaction network can be modeled in such a way,
        and the problem of checking whether a Turing machine
        computes the constant true function is undecidable, the result follows.
 \end{proof}

%% file: related.tex
We have studied the correctness problem for the population protocol models introduced
by Angluin \etal in \cite{AAER07}.  Section 2 of \cite{AAER07} presents a detailed comparison 
with other models, focusing on expressivity questions. In this section we discuss work
on models that are related to those of \cite{AAER07}, and moreover address verification questions.

The IO and DO observation models of \cite{AAER07} are closely related to 
Reconfigurable Broadcast Networks (RBN), introduced by Delzanno \etal in \cite{DelzannoSTZ12}, 
and further studied in \cite{DelzannoST16,BertrandBM19}.
\footnote{We thank an anonymous reviewer for pointing this out.}
In RBNs, networks of finite-state agents communicate through broadcast. The network is 
modeled as an undirected graph $G=(V,E)$, with an agent at each node of $V$. An agent in 
state $q$ can execute a transition $q \trans{a!!} q'$, which broadcasts the
message $a$ to all neighbours, and updates the state of the agent to $q'$. All neighbours of the agent 
must react to the message according to transitions of the form $r \trans{a??} r'$
for every state $r$.  
The crucial feature of RBNs is that between any two broadcasts the network can 
nondeterministically reconfigure itself into \emph{any} other network with the same
set of nodes.  This makes RBNs equivalent to symmetric, fully connected networks
in which agents nondeterministically choose whether to react to a broadcast or not. 
Symmetry makes the agents indistinguishable, and so the configuration of an RBN
is completely determined by the number of agents in each state. As a consequence, 
given an instance of an IO protocol with $n$ agents, one can construct an equivalent RBN as follows. 
The network has $n$ nodes. For every transition $q_1 \trans{q_2} q_3$ of the IO protocol, we add to the network 
 transitions $q_2 \trans{a!!} q_2$, $q_1 \trans{a??} q_3$, and $q \trans{a??} q$ for every $q \neq q_1$. 
So IO protocols are a special case of RBNs. However, the analysis problems we study are more general than the ones studied in 
\cite{DelzannoSTZ12,DelzannoST16,BertrandBM19}.  The parameterized reachability problem
studied in \cite{DelzannoSTZ12} corresponds to the problem whether a given counting set is reachable from a cube $[L, U]$ such that $
L(q) = 0$ and $U(q) \in \{0, \infty\}$ for every state $q$ (i.e., from configurations that
can put arbitrarily many agents in some states, and no agent in others). We solve the more general problem 
of reachability between two arbitrary counting sets. Further, our solution allows us to prove 
that counting sets are closed under reachability, a question not considered in \cite{DelzannoSTZ12}.  
The results of \cite{BertrandBM19} on minimal length of covering 
executions have the same flavour as our Shortening Theorem for IO, but only consider the case in
which the configuration $C$ to be covered satisfies $C(q) \in \{0,1\}$ for every state $q$. We conjecture that
at least some of our results extend to RBNs, and leave this question for future research.

The standard population protocol model is closely related to Petri nets 
and Vector Addition Systems. The decidability of correctness for PP is proved in \cite{EsparzaGLM17}
using results of Leroux and others on reversible and cyclic Petri nets \cite{Leroux13LMCS,Leroux13LICS}.
The TOWER-hard lower bound is also proved in \cite{EsparzaGLM17} by reduction to the reachability
problem for Petri nets, which is shown to be TOWER-hard in \cite{DBLP:conf/stoc/CzerwinskiLLLM19}.
Practical verification algorithms for PP have been given in \cite{BlondinEJM17,BlondinEJ18,BlondinEH0M20}.
The complexity of other verification problems beyond correctness is studied in \cite{EsparzaGLM16}.
 
Population protocols are also closely related to Chemical Reaction Networks \cite{SoloveichikCWB08}.
Our result on the undecidability of correctness of queued-transmission protocols in positive, markovian and
uniform probabilistic models is based on the results on the computational
power of Chemical Reaction Networks by Cummings \etal \cite{CummingsDS16}.

After Angluin \etal proved in \cite{AAER07} that population protocols can only compute Presburger 
predicates, several models have been proposed that increase the expressive power. These include
community protocols \cite{Guerraoui07abstract}, passively mobile logarithmic space machines 
(PALOMA)\cite{journals/corr/abs-1004-3395}, mediated protocols \cite{DBLP:journals/tcs/MichailCS11},
clocked population protocols \cite{Aspnes17} and broadcast population protocols \cite{BlondinEJ19}. All these models 
can compute all predicates $\N^k \rightarrow \{0,1\}$ in $\NSPACE(\log n)$ or more, where $n$ is
the number of agents.  This makes the correctness problem for all these models undecidable. 
To prove this we can for example reduce from the
halting problem for Turing machines started on empty tape. Indeed, given a machine $T$, 
the predicate $\varphi_T(n)$ that holds for $n$ if the computation of $T$ on empty tape terminates
and visits at most $\log n$ cells is a symmetric predicate in $\NSPACE(\log n)$, and so it can be 
computed by a protocol. So $T$ fails to terminate if{}f the protocol computes the false predicate.

From a verification point of view, the correctness problem for population protocols is a 
so-called parameterized verification problem, in which one has to show that a system of identical agents
satisfies a property independently of the number of agents. Parameterized verification problems 
have been intensely studied, and we refer the reader to \cite{2015Bloem,Esparza16,AbdullaST18} for 
survey articles. Most work, however, concerns the verification of safety or liveness under adversarial schedulers; in other words, the property must hold even if the scheduler that selects which agents interact at each step tries to break it. Correctness of population protocols is however a liveness property under stochastic schedulers, which choose the agents at random. This distinguishes our work from recent  contributions to parameterized verification \cite{LinR16,LengalLMR17}.

%% file: conclusion.tex
We have determined the computational complexity of the correctness problem
for population protocols with different communication mechanisms, completing a research program initiated in \cite{EsparzaGLM17}. We have followed the classification used by Angluin \etal in \cite{AAER07} to study the expressive power of the models.

Our main results concern the observation-based models IO and DO. A first surprise is the fact that checking correctness of a protocol \emph{for all inputs} is not harder than checking it \emph{for one input}. Further, both problems have the same complexity as many standard verification problems for concurrent systems, which are typically \PSPACE-complete \cite{PitermanP18}. Moreover, our upper bounds are obtained by means of algorithms that suggest clean verification procedures. In particular, they show that the verification of properties of IO and DO protocols can be achieved by conducting symbolic state space exploration with counting sets represented by counting constraints. This opens the door to efficient implementations using SMT-solving technology \cite{BarrettT18}.

From a more theoretical point of view, we have derived our upper bounds from a number of fundamental results about the dynamics of the IO and DO models. We have encapsulated them in the Pruning, Shortening, and Closure Theorems, which could be of independent interest. In particular, the connection between IO protocols and models for enzymatic reactions is intriguing \cite{MarwanWW11}.

The second surprise is the huge complexity gap between observation-based and transmission-based models. Thanks to the recent result by Czerwinski \etal \cite{CLLLM19}, we can show that the correctness problem is TOWER-hard for all transmission-based models. This is in contrast with the limited computational power of the model, and raises the question whether there exists a natural model of computation by indistinguishable agents which is able to compute all Presburger predicates, and has a more manageable correctness problem. 
Another important insight
 is the fact that for 
all delayed-transmission models the problem is already TOWER-hard in the single-instance case. 
This already makes the application of model-checking technology to checking correctness for a few instances very difficult, and suggests a number of questions for further research.

Our investigation leaves one question open, namely whether the correctness problem is decidable for queued-transmission problems. We have explained that for this model the fairness assumption used by Angluin \etal in \cite{AAER07} is questionable, since it can no longer be seen as an ``over-approximation'' of the probabilistic behavior of the system.  However, settling the question
can be relevant for stochastic models with assumptions concerning the size of the pool of messages.
%
%
%
%

%% file: appendix-obs-3-1.tex
\section{Appendix for Section \ref{sec:io-hardness}}

The following definition and lemma are introduced to help  prove that our IO protocol implementation of a Turing machine does indeed simulate its functioning.
To recall the notation, let us start with an illustration of transitions modelling a single step of the Turing machine into the protocol.
The fragment of the protocol is represented as a diagram with some of the states and transitions of the IO protocol.

\input figure-turing-io.tex

Figure~\ref{figure-turing-io} illustrates transitions involved in modelling a single step of a Turing machine
that reads $0$, writes $1$, moves head to the right and switches the control state from $q$ to $q'$.

\begin{definition}
 A configuration of $\PP_M$ is a \emph{modelling configuration} if the following conditions hold.
\begin{enumerate}
        \item For every $1\leq n\leq K$ exactly one of the $2|\Sigma|$ states 
        $\act{\sigma}{n}, \pass{\sigma}{n}$ is populated, and it is populated with a single agent. \\
        (Intuitively: every cell is either \textit{on} or \textit{off} and contains exactly one symbol.)
        \item Exactly one of all the head states is populated (again, with a single agent).
        \item If a cell state $\act{\sigma}{n}$ is populated, then a head state $\stable{q}{n}$ or $\switch{q}{\sigma'}{n}{d}$ is populated for some $\sigma'$ and $d$.
        \item If a head state $\switch{q}{\sigma}{n}{d}$ is populated,
                either $\act{\sigma'}{n}$ is populated for some $\sigma'$,
                of $\pass{\sigma}{n}$ is populated.
\end{enumerate}
\end{definition}

\begin{remark}
Note that for every configuration $c$ of $M$ the configuration $C_c$ described in Definition \ref{def:turing-configuration} is a modelling configuration.
\end{remark}

\begin{lemma}
        \label{lemma:modellingmarkingevolution}
For every modelling configuration $C$ of $\PP_M$:
\begin{itemize}
\item[(1)] $C$ enables at most one transition.
\item[(2)] If $C$ enables no transitions, then it populates states $\act{\sigma}{n}$ and $\stable{q}{n}$ for some $q \in Q$, $\sigma \in \Sigma$, and $1 \leq n \leq K$.
\item[(3)] If $C \trans{} C'$, then $C'$ is also a modelling configuration. 
\end{itemize}
\end{lemma}
\begin{proof}
\noindent (1) All possible transitions require agents at two states,
        one of type $\act{\cdot}{n}$ or $\pass{\cdot}{n}$
        and one of type $\stable{\cdot}{n}$ or $\switch{\cdot}{n}{\cdot}{\cdot}$, with the same $n$.
        But the modelling condition requires that there can be at most one such pair.

\noindent (2) If a $\switch{\cdot}{\cdot}{\cdot}{\cdot}$ state is populated,
        a transition is always possible by definition
        of the list of $\switch{\cdot}{\cdot}{\cdot}{\cdot}$ states.
        The same for the case where a $\stable{\cdot}{\cdot}$ state
        is populated but no $\act{\cdot}{\cdot}$ case is populated.
        If there are populated states of types $\act{\cdot}{n}$ and $\stable{\cdot}{n}$,
        the transition may fail
        to exist if either the Turing machine halts or if it goes outside the allocated space.

\noindent (3) Every transition consumes and produces one agent 
        at $\pass{\cdot}{n}$ or $\act{\cdot}{n}$ state,
        and the new state has the same $n$.
        Every transition consumes and produces one agent
        at $\switch{\cdot}{\cdot}{\cdot}{\cdot}$ or $\stable{\cdot}{\cdot}$ state.
        If an $\act{\cdot}{n}$ state becomes populated after a transition,
        it has the same $n$ as the populated $\stable{\cdot}{n}$ state
        of both configurations (before and after);
        if an $\act{\cdot}{n}$ state stays populated,
        the agent is moved from a $\stable{\cdot}{n}$ to a $\switch{\cdot}{n}{\cdot}{\cdot}$
        state with the same $n$.
When $\switch{q}{\sigma}{n}{d}$ becomes populated,
        the transition needs a populated $\act{\cdot}{n}$ state.
        When $\switch{q}{\sigma}{n}{d}$ stays populated,
        the transition populates a $\pass{\sigma}{n}$ state.
        
 \end{proof}

%% file: figure-turing-io.tex

\newcounter{realfigure}
\setcounter{realfigure}{\thefigure}
\setcounter{figure}{1}
\addtocounter{figure}{-1}
\FigureTuringIO
\setcounter{figure}{\therealfigure}

%% file: appendix-obs-3-2.tex
\section{Appendix for Section \ref{sec:mfdo-do-hardness}}

\LemmaCircuitEvaluation*
\begin{proof}
Every input node can change its own current value if it is $\square$
and cannot otherwise; by fairness of the execution and definition of how an input node can update its $\square$ value, the input nodes will eventually all change their values
from $\square$ and keep these afterwards. 

By induction over the depth of an operation node,
we can see that the value of each operation node eventually converges
        to the value of this node in circuit $C$, without ever holding the opposite value; 
        moreover,
once a node adopts a value, the value stays stable.
Once the output value converges, each node will eventually 
learn it.

Notice that since the transitions of the circuit evaluation protocol always depend only on current values of nodes, the DO protocol cannot have a problem with lack of old messages.
 \end{proof}

%% file: appendix-obs-4-2.tex
\section{Appendix for Section \ref{sec:pruning}}

\LemmaRealizableIO*
\begin{proof}
One direction is obvious by definition: if we have a realizable history,
it also describes an execution.
Let us prove the other direction.

Informally, we just implement the ``de-anonymisation" of the agents, that is the assignment of a trajectory to each agent (in an arbitrary but consistent way).
A formal proof can be given by induction in the number of transitions in the execution.

\noindent
\emph{Base case}. If there are no transitions, we create a multiset of trajectories of length one such that the initial states of the trajectories are exactly the states (with multiplicity) of the initial marking of the execution.
This is well-structured because there are no steps.

\noindent
\emph{Induction step}. Consider a sequence of transitions and a corresponding well-structured history.
Now let us add a single enabled transition. 
To build the new history, we choose an arbitrary trajectory of the existing history such that this trajectory ends in the state corresponding to the source state of the added transition.
Such a trajectory exists because the transition is enabled and therefore its source state must be populated (one agent at least must be in the source state).
We extend the chosen trajectory with a step from the source state to the destination state of the added transition,
and we extend the rest of the trajectories with one horizontal step each.
We obtain a multiset of trajectories of same length, thus constituting a history.
It is realizable using the considered sequence of transitions followed by the new enabled transition.
As we add only a single non-horizontal step at that moment of time, we do not
break the well-structuring condition.
 \end{proof}

\LemmaHistoryIO*
\begin{proof} 
Let $H$ be a well-structured history of $\PP$.

Assume that $H$ is realizable. Let $\tau \in H$, and let $\tau(i) \tau(i+1)=q q'$ be an arbitrary non-horizontal step of $\tau$.  Since $H$ is well-structured, for every trajectory $\tau'$, if  $\tau'(i) \tau'(i+1)$ is non-horizontal then $\tau'(i) \tau'(i+1)=q q'$. Since $H$ is realizable, $C_H^i$ enables a transition $q \trans{o} q'$ of $\PP$. So $C_H^i(o) \geq 1$, and therefore there is a trajectory $\tau' \in H$ such that $\tau'(i) = o$. By the definition of step in IO protocols we have $\tau' \neq \tau$. Since $H$ is well-structured, the $i$-th step of $\tau'$ is horizontal, and so $\tau'(i+1) = o$.

Now assume that $H$ is compatible with $\PP$. We prove that $H$ is realizable by induction
on the length $n$ of $H$. If $n=1$, there is nothing to show. If $n > 1$, let $H'$ be the result of removing the last state from every trajectory of $H$. It follows immediately from the 
definitions that $H'$ is compatible with $\PP$. So there exist transitions $t_1, \ldots, t_{n-2}$ of $\PP$ and numbers $k_1, \ldots, k_{n-2} \geq 0$ such that 
$$C_{H}^1 \trans{t_1^{k_1}}C_{H}^2 \cdots  C_{H}^{n-2} \trans{t_{n-2}^{k_{n-2}}} C_{H}^{n-1} \ .$$
We show that there is a transition $t_n$ and $k_n \geq 0$ such that $C_{H}^{n-1} \trans{t_{n-1}^{k_{n-1}}} C_{H}^{n}$.
Consider the last steps of all trajectories of $H$. If they are all horizontal, then
$C_H^{n-1} = C_n$. So we can choose $t_n$ as any transition of $\PP$, and $k_n :=0$. 
If at least one of them is non-horizontal, let $s \geq 1$ be the number of non-horizontal steps. Since $H$ is well-structured, all non-horizontal steps are equal, say $q \, q'$. Further, $\PP$ has a transition $t = q \trans{o} q'$ and a trajectory $\tau' \neq \tau$ such that $\tau'(i) = \tau'(i+1) =o$. So we can choose $t_{n-1} := t$ and $k_{n-1}:= s$. 
 \end{proof}

\begin{theorem} [Quadratic Pruning Theorem]
Let $\PP$ be an IO protocol, let $L'$ and $L$ be multisets of states of $\PP$, and let
$C' \trans{*} C$ be an execution of $N$ such that $C' \geq L'$ and $C \geq L$.
There exist configurations $D'$ and $D$ such that 
\begin{center}
\(
\begin{array}[b]{@{}c@{}c@{}c@{}c@{}c@{}}
C' &  \trans{\hspace{1em}*\hspace{1em}} & C  \\[0.1cm]
\geq &  & \geq \\[0.1cm]
D' & \trans{\hspace{1em}*\hspace{1em}} &D  \\[0.1cm]
\geq &  & \geq \\[0.1cm]
L' &  &L  \\[0.1cm]
\end{array}
\)
\end{center}
and $|D'| = |D| \leq |L| + |L'| + 2|Q|^2$.
\end{theorem}
\begin{proof}
The proof is similar to the proofs of Lemma~\ref{lm:pruning} and Theorem~\ref{thm:pruning}.
The main difference is the following.
In Lemma~\ref{lm:pruning} we keep trajectories that belong to small bunches,
and prune each large bunch separately.
To prove the quadratic lower bound we keep trajectories from and to small states,
then prune all the remaining trajectories together.
The state is called small if it has less than $|Q|$ incoming or outgoing trajectories.

Let $L' \ \leq \ C' \trans{*}  C'  \  \geq  \ C$. 
By Lemma \ref{lem:realizable-io}, there is a well-structured realizable history $H$ with $C'$ and $C$ as initial and final configurations, respectively.
Let $H_0 \subset H$ be an arbitrary minimal sub(multi)set of $H$
with initial multiset of states at least $L'$ and final multiset of states at least $L$.
Let also $H'=H-H_0$.
We further reduce $H'$ by repeatedly removing all the trajectories with initial or final state
having less than $|Q|$  trajectories still in $H'$. 
We can perform at most $2|Q|$ steps like that, removing at most $|Q|-1$ trajectories per step.
At the end, we will add back these trajectories as well as those of $H_0$.

Now we can define $Q$ as the set of all states reached by the remaining trajectories in $H'$,
and $f(q)$ and $l(q)$ for $q\in Q$ 
as the earliest and the latest moment in time when this state has been used by any of the trajectories
(possibly on different trajectories, and possibly on trajectories with different initial and final state).

We now build a trajectory for every $q\in Q$ by reaching it by the moment $f(q)$ and leaving it after $l(q)$.
As all the trajectories in $H'$ have initial and final state with at least $|Q|$ trajectories in $H'$,
the set of trajectories that we build will have the initial and final configurations covered
by the corresponding configurations of $H'$.

The rest of the proof is identical to the proofs
of Lemma~\ref{lm:pruning} and Theorem~\ref{thm:pruning}.
 \end{proof}

\LemmaHistoryMFDO*
\begin{proof} 
Let $H$ be a well-structured history of $\PP$.

Assume that $H$ is realizable. Let $\tau \in H$, and let $\tau(i) \tau(i+1)=q q'$ be an arbitrary non-horizontal step of $\tau$.  Since $H$ is well-structured, for every trajectory $\tau'$, if  $\tau'(i) \tau'(i+1)$ is non-horizontal then $\tau'(i) \tau'(i+1)=q q'$. Since $H$ is realizable, $C_H^i$ enables a transition $q \trans{o} q'$ of $\PP$, and so $o \in \mathcal{S}_H^i$. So $H$ is compatible with $\PP$.

Now assume that $H$ is compatible with $\PP$. We prove that $H$ is realizable by induction
on the length $n$ of $H$. If $n=1$, there is nothing to show. If $n > 1$, let $H'$ be the result of removing the last state from every trajectory of $H$. It follows immediately from the 
definitions that $H'$ is compatible with $\PP$. So there exist transitions $t_1, \ldots, t_{n-2}$ of $\PP$ and numbers $k_1, \ldots, k_{n-2} \geq 0$ such that 
$$C_{H}^1 \trans{t_1^{k_1}}C_{H}^2 \cdots  C_{H}^{n-2} \trans{t_{n-2}^{k_{n-2}}} C_{H}^{n-1} \ .$$
We show that there is a transition $t_n$ and $k_n \geq 0$ such that $C_{H}^{n-1} \trans{t_{n-1}^{k_{n-1}}} C_{H}^{n}$.
Consider the last steps of all trajectories of $H$. If they are all horizontal, then
$C_H^{n-1} = C_n$. So we can choose $t_n$ as any transition of $\PP$, and $k_n :=0$. 
If at least one of them is non-horizontal, let $s \geq 1$ be the number of non-horizontal steps. Since $H$ is well-structured, all non-horizontal steps are equal, say $q \, q'$, and $\PP$ has a transition $t = q \trans{o} q'$ such that $o  \in \mathcal{S}_H^{n-1}$. So we can choose $t_{n-1} := t$ and $k_{n-1}:= s$. 
 \end{proof}

\begin{theorem}[Linear MFDO Pruning]
\label{thm:mfdopruninglinear}
Let $\PP = (Q,\delta)$ be an MFDO protocol,
let $L'$ and $L$ be multisets of states of $\PP$,
and let
$C' \trans{*} C$ be an execution of $\PP$
such that $C' \geq L'$ and $C \geq L$.
\begin{center}
\(
\begin{array}[b]{@{}c@{}c@{}c@{}c@{}c@{}}
C' &  \trans{\hspace{1em}*\hspace{1em}} & C  \\[0.1cm]
\geq &  & \geq \\[0.1cm]
D' & \trans{\hspace{1em}*\hspace{1em}} &D  \\[0.1cm]
\geq &  & \geq \\[0.1cm]
L' &  &L  \\[0.1cm]
\end{array}
\)
\end{center}
and $|D'| = |D| \leq |L| + |L'| + |Q|$.
\end{theorem}
\begin{proof}
Let $H$ be a well-structured and realizable history for the execution $L'\leq C' \trans{*} C \geq L$.
For every $q \in \mathcal{S}_H$, let $f(q)$ be the smallest index $i$ such that $q \in \mathcal{S}_H^i$, that is, $f(q)$ is the earliest moment at which $q$ is visited. Pick a trajectory $\tau_q$ of $H$ such that $q$ is reached by $\tau_q$ at moment $f(q)$ (we may pick the same trajectory for two different states).

Let $H'$ be union of the set $\{ \tau_q \mid q \in \mathcal{S}_H \}$ of trajectories, and an arbitrary sub(multi)set of trajectories of $H$ such that the initial configuration of $H'$ covers $L'$ and the final configuration of $H'$ covers $L$.
We need at most $|L|+|L'|$ of these additional trajectories.

It follows immediately from the definition that $H'$ is a history, covers $L'$ and $L$ by its initial and final configuration, and has at most $|Q|+|L|+|L'|$ trajectories. Let us show $H'$ is well structured and realizable. By Lemma \ref{lem:compatibleMFDO} it suffices to show that $H'$ is well structured and compatible with $\PP$. 
It is well-structured because $H' \subseteq H$, which is a well-structured history. Let us show that it is compatible with $\PP$.  By the definition of compatibility (Definition \ref{def:compatibleMFDO}), and since $H' \subseteq H$, it suffices to show that $\mathcal{S}_H^i = \mathcal{S}_{H'}^i$ holds for every $i$. But this follows from the fact that, by the definition of $H'$, each state is \emph{first} visited in $H'$ at the same moment that it is \emph{first} visited in $H$.

 \end{proof}

\input{figure-pruned-mfdo.tex}

\begin{example}
Consider the well-structured and realizable extended history of Figure \ref{figure-history-mfdo}, leading from $(1,4,0)$ to $(0,0,5)$, which covers configuration $(0,0,2)$.
The set of states visited by the trajectories is equal to $Q$.
Figure \ref{figure-pruned-mfdo} is annotated with the first moment $f(q)$ for every $q \in Q$.
We pick the trajectories $\tau_{a}$ and $\tau_{b}$ drawn in dashed lines in Figure \ref{figure-pruned-mfdo}, and choose $H'=\{\tau_a, \tau_b\}$. 
We have $C_{H'}^1=(1,1,0) \trans{t_2 t_1}(0,0,2)=C_{H'}^4$ and $C_{H'}^4 \geq (0,0,2)$. 
\end{example}

\begin{theorem}[Quadratic MFDO Shortening]
\label{thm:mfdoshorteningquadratic}
Let $\PP = (Q,\delta)$ be an MFDO protocol, and let $C \trans{*} C'$ be an execution of $\PP$.
There exists a sequence $\xi$ such that $C \trans{\xi} C'$ and $|\xi|_a \leq |Q|^2$.
\end{theorem}

\begin{proof}[Outline]
We optimise separately the construction for the last segment and for the other ones.

        We require that: 
        \\(i) the trajectories of $H'_j$ are in bijective correspondence
with trajectories of $H$ (and therefore $H_j$),
        \\(ii) the set of visited states $\mathcal{S}_{H'_j}$ is the same
as the set $\mathcal{S}_{H_j} = \mathcal{S}_H^{T_{j}}$ of states visited by $H_j$,
and 
        \\(iii) for each trajectory $\tau'\in H'_j$ corresponding to $\tau\in H_j$,
there is a path from final state of $\tau'$ to final state of $\tau$ 
visiting and observing only states in $\mathcal{S}_{H'_j}$.

In other words, instead of saying that 
the corresponding trajectories reach the same states, 
we say that the corresponding trajectories \emph{could}
reach the same states.

This can be maintained in the same way as in the proof of theorem~\ref{thm:mfdoshortening}
with two changes.
We extend only one trajectory with the shortest path to the newly reachable state
(this is possible because if state $q'$ is reached by $\tau\in H_{j+1}$
starting from $q$ at the moment $T_j$,
the corresponding $\tau'\in H'_j$ can reach $q$ and then $q'$ from $q$).
The remaining trajectories are extended with horizontal steps, and the new
reachability requirement is also satisfied by transitivity of reachability.

There is a linear number of non-last segments, and each will correspond 
to a linear-length replacement segment.
        Therefore the aggregated length of all the segments (except the last)
        together is quadratic.

In the last segment we need to make all the trajectories to reach the
final configuration of $H$  from the final configuration in $H'_n$.
Note that there is some feasible multiset of such trajectories because of the
condition (iii).
Also observe that as we don't change the set of visited states, the steps of
the trajectories do not depend on each other.

        Consider the multiset of trajectories leading from the final configuration of $H'_n$
to the final configuration of $H$ (maybe violating the initial trajectory 
correspondence) with the shortest total number of steps across the trajectories.
In such a multiset a union of all trajectories doesn't contain any cycles,
as otherwise we would be able to cut and reconnect the trajectories to remove
the steps along the cycle. Therefore we can consider topological ordering
corresponding to the union of these trajectories.
As each trajectory traverses the states in the ascending order according 
to the topological ordering, running the steps in the lexicographic order
        of the source and target states correctly traverses each trajectory.
        As all the equal steps are ran at the same time, we obtain
        quadratic aggregated length for the final segment.
 \end{proof}

%% file: figure-pruned-mfdo.tex
\begin{figure}
\centering
\begin{tikzpicture}
  \node[circle, draw]
    (Node11)
    {~~~}; 
    \node[right=1 of Node11, circle, draw]
    (Node12)
    {~~~};
    \node[right=1 of Node12, circle, draw]
    (Node13)
    {~~~};
    \node[right=1 of Node13, circle, draw]
    (Node14)
    {~~~};
    \node[below=0.7 of Node11, circle, draw]
    (Node21)
    {~~~};
    \node[right=1 of Node21, circle, draw]
    (Node22)
    {~~~};
    \node[right=1 of Node22, circle, draw]
    (Node23)
    {~~~};
    \node[right=1 of Node23, circle, draw]
    (Node24)
    {~~~};
    \node[below=0.7 of Node21, circle, draw]
    (Node31)
    {~~~};
    \node[right=1 of Node31, circle, draw]
    (Node32)
    {~~~};
    \node[right=1 of Node32, circle, draw]
    (Node33)
    {~~~};
    \node[right=1 of Node33, circle, draw]
    (Node34)
    {~~~};
        \node[left=0.3 of Node11]
        (Node11l)
        {$a$};
        \node[right=0.3 of Node14]
        (Node14r)
        {$a$};
        \node[left=0.3 of Node21]
        (Node21l)
        {$b$};
        \node[right=0.3 of Node24]
        (Node24r)
        {$b$};
        \node[left=0.3 of Node31]
        (Node31l)
        {$ab$};
        \node[right=0.3 of Node34]
        (Node34r)
        {$ab$};

        \node[above=0.0 of Node11](Node11u){$f$};
        \node[above=0.0 of Node21](Node21u){$f$};
        \node[above=0.0 of Node32](Node32u){$f$};

  \draw[dashed]
  ($(Node11.center)+(0,1.50mm)$)
    --
  ($(Node12.center)+(0,1.50mm)$)
  ;
  \draw[dashed]
  ($(Node12.center)+(0,1.50mm)$)
    --
  ($(Node33.center)+(0,1.50mm)$)
  ;
  \draw[dashed]
  ($(Node33.center)+(0,1.50mm)$)
    --
  ($(Node34.center)+(0,1.50mm)$)
  ;
  \draw[dashed]
  ($(Node21.center)+(0,-1.50mm)$)
    --
  ($(Node32.center)+(0,-1.50mm)$)
  ;
  \draw[dashed]
  ($(Node32.center)+(0,-1.50mm)$)
    --
  ($(Node33.center)+(0,-1.50mm)$)
  ;
  \draw[dashed]
  ($(Node33.center)+(0,-1.50mm)$)
    --
  ($(Node34.center)+(0,-1.50mm)$)
  ;

\end{tikzpicture}
\caption{ History $H$ of Figure \ref{figure-history-mfdo} after pruning}
\label{figure-pruned-mfdo}
\end{figure}
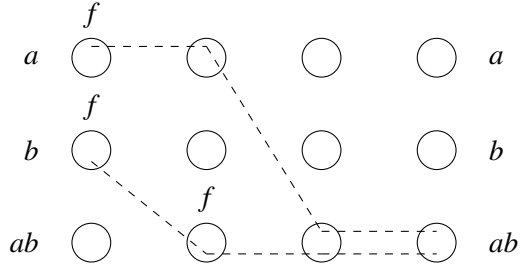

%% file: appendix-obs-4-3.tex
\section{Appendix for Section \ref{sec:small-instance}}

\PropMeasureBooleanOp*
\begin{proof}[Adapted from Proposition 5 of~\cite{EsparzaGMW18}.]

\noindent \emph{Union.}
Let $\Gamma$ be the union of the two counting constraints $\Gamma_1, \Gamma_2$, \ie the set of the cube representations of $\Gamma_1$ and of $\Gamma_2$. 
It is still a counting constraint as a set of cube representations, and the result follows from the definition of representation and norms.
\medbreak

\noindent \emph{Intersection.}
For this proof, we consider a cube representation $(L,U)$ as a collection of constraints over $n=|Q|$ variables $x_1, \ldots, x_n$ of the form $[l_i\leq x_i ]$ or $[x_i \leq u_i]$ with $l_i \in \N$ and $u_i \in \N \cup \infty$. 
Each variable $x_i$ is associated to a state $q_i \in Q$, for an arbitrary ordering of $Q$, and it intuitively denotes the number of agents in $q_i$.
A cube representation can now be seen as a conjunction of such constraints, one lower bound and one upper bound for each $x_i$ for $i \in \set{1,\ldots, n}$.
We call such a $2n$-conjunction a \emph{minterm}.
Counting constraints $\Gamma_1, \Gamma_2$ are thus disjunctions of minterms, noted $\gamma_1,\gamma_2$ respectively.
The intersection of $\Gamma_1, \Gamma_2$ is the conjunction $\gamma_1 \land \gamma_2$.

We rearrange this conjunction into a disjunction of  minterms by using the following steps: Put $\gamma_1 \land \gamma_2$ in disjunctive normal form. 
Remove conjunctions containing the unsatisfiable constraints $l \leq x_i \wedge x_i \leq u$ with $l > u$. 
Remove redundant constraints inside conjunctions, e.g., replace $(l_1 \leq x \wedge l_2 \leq x)$ by $\max\{l_1, l_2\} \leq x$. 
If a conjunction does not contain a lower bound (upper bound) for $x_i$, add $0 \leq x_i$ ($x_i \leq \infty$), thus making it a minterm.
The disjunction of these minterms is the counting constraint $\Gamma$ we are looking for, and the norm bounds follow from the fact that the new bounds are a mix of the old bounds.
\medbreak

\noindent \emph{Complement.}
We reuse the constraint and minterm formalism above.
The complement is represented by the negation of the disjunction of minterms.
We rearrange it into a disjunction of minterms using the rules above as well as $\sem{\neg (x_i \leq c)}=\sem{x_i \geq c+1}$; and $\sem{\neg (x_i \geq c)}=\sem{x_i \leq c-1}$ if \(c\in\N\setminus \{ 0\}\) and remove the enclosing conjunction otherwise.
We obtain $n$-conjunctions with lower bounds of the form $u+1$, with $u \leq \unorm{\Gamma_1}$ an upper bound in a minterm of the original constraint. This yields $\lnorm{\Gamma} \leq n\unorm{\Gamma_1} + n$ and the reasoning is similar for the $u$-norm.
 \end{proof}

\ThmCCReachabilityIO*
\begin{proof}
Consider a finite decomposition into cubes $\mathcal{S} = \cup_{i=1}^k \mathcal{C}_i$ of counting set $\mathcal{S}$, which exists by definition of a counting set.

Lemma \ref{lm:smallminterm} states that for every cube $\mathcal{C}$ of this decomposition, for every configuration $C'$ in $\prestar(\mathcal{C})$, there is a ``small" cube $\mathcal{C}'$ such that $C' \in \mathcal{C}'$ and $\mathcal{C}' \subseteq \prestar(\mathcal{C})$.
So $\prestar(\mathcal{C}) = \cup_{C' \in \prestar(\mathcal{C})} \mathcal{C}'$.
By the norm restrictions on the representation of $\mathcal{C}'$, there are only a finite number of such ``small" cubes.
So $\prestar(\mathcal{C})$ is a finite union of  cubes, and by extension $\prestar(\mathcal{S}) = \cup_{i=1}^k \prestar(\mathcal{C}_i)$ is too.
Thus by definition, $\prestar(\mathcal{S})$ is a counting set.

Let $\Gamma$ be the counting constraint defined as the set of the representations of the $\mathcal{C}_i$.
Let $\Gamma'$ be the counting constraint defined as the set of the representations of the ``small" cubes whose unions equal the $\prestar(\mathcal{C}_i)$.
Then by the bounds in Lemma \ref{lm:smallminterm} and by definition of the norms,
$
\unorm{\Gamma'} \leq \unorm{\Gamma}
$ 
and 
$
\lnorm{\Gamma'} \leq \lnorm{\Gamma} + |Q|^3.
$

The results also hold for $\poststar(\mathcal{S})$, 
as the pruning theorem and our use of it are symmetric.
 \end{proof}

\LemmaSmallCubeMFDO*
\begin{proof}
        The proof is exactly the same as for Lemma~\ref{lm:smallminterm},
        as adding a copy of a trajectory into a well-structured realizable
        history produces a realizable history for MFDO protocols as well.
 \end{proof}

\ThmCCReachability*
\begin{proof}
The proof is the same as for Theorem \ref{thm:ccreach-io}, except that Lemma \ref{lm:smallminterm-mfdo} is used instead of Lemma \ref{lm:smallminterm}.
 \end{proof}

\ThmCCReachabilityDO*
\begin{proof}
The set $\prezmstar(\mathcal{S})$ is the set of zero-message configurations $Z' \in \mathcal{Z}$ such that there exists $Z\in \mathcal{S}$ and $Z'\trans{*}Z$ in DO protocol $\PP$.
By Lemma \ref{lm:do-to-mfdo}, $Z'\trans{*}Z$ in DO protocol $\PP$ if and only if $Z'\trans{*}Z$ in the corresponding MFDO protocol.
Since $\mathcal{S}$ can also be seen as a counting set of MFDO configurations (by  our usual overloading of configurations of $\mathcal{Z}$),  $\prezmstar(\mathcal{S})$ in $\PP$ is equal to $\prestar(\mathcal{S})$ in the corresponding MFDO protocol.
Thus we obtain the result by application of Theorem \ref{thm:reachofcc}.

The result for $\postzmstar(\mathcal{S})$ is proved in the same way. 
 \end{proof}

%% file: app-lower-bounds.tex
\section{Appendix for Section~\ref{sec:transmission}}
\propVassToPM*
\begin{proof}
Let $(Q, T)$ have dimension $k \in \N$.
The $\pm1$-VASS $(Q', T')$ of the same dimension is constructed as follows: add to $Q$ new states $r$ and $r_0$, and add to $T$ the transitions $r_0 \trans{\vec{v}_0} q_0$ and $q \trans{-\vec{v}} r$. Then replace every transition of the form
$q \trans{(w_1, \ldots, w_m)} q'$ by
$$\begin{array}{lll}
   q & \trans{(w_1, 0, \ldots, 0)} & q_{w_1} \\
    & \trans{(0, w_2, 0, \ldots, 0)} & q_{w_2} \\
   & \ldots & \\
  & \trans{(0, \ldots, 0, w_{m-1}, 0)} & q_{w_{m-1}}\\
    & \trans{(0, \ldots, 0, w_m)} & q'.
\end{array}$$
where $q_{w_1}, \ldots, q_{w_{m-1}}$ are newly added states.
Then replace every transition of the form $q \trans{\zero} q'$ by $q \trans{1++} q'' \trans{1--} q'$, where $q''$ is a newly added state.
Finally, replace every transition of the form $q \trans{\vec{w}} q'$ where $w_m \neq 0$ for a fixed $m \in \{1, \ldots, k\}$ by the following:
\begin{align*}
  & q \trans{m++} q_{1} \trans{m++} q_{2} \trans{m++} \ldots \trans{m++} q_{w_m-1} \trans{m++} q' & \text{ if } w_i > 0,\\
  & q \trans{m--} q_{1} \trans{m--} q_{2} \trans{m--} \ldots \trans{m--} q_{w_m-1} \trans{m--} q' & \text{ otherwise,}
\end{align*}
where $q_{1}, \ldots, q_{w_m - 1}$ are newly added states.
The resulting construction is a $\pm1$-VASS and clearly satisfies the properties in our claim.
 \end{proof}

\propDetToNonDet*
\begin{proof}
  Fix some linear order $<$ on $Q \cup (M \times Q)$,  and let $n \defeq |Q\times M|$.
  We define the deterministic protocol $\PP'=(Q', M', \delta'_s, \delta'_r, \Sigma \iota', o')$ as follows:

  \begin{itemize}
    \item $Q' \defeq Q \times \{1, \ldots, n\} \times \{0, 1\}$. An agent in state $(q, i, b) \in Q'$
    simulates an agent from $\PP$ in state $q$, picks choice $i$ to resolve nondeterminism, and may send an $\increment$ message if and only if $b = 1$.

    \item $M' \defeq M \cup \{\increment\}$
    \item $\delta'_s$ is defined as follows:
          \begin{itemize}
            \item For every $q \in Q$ such that $\delta_s(q) = \{(m_1, q_1), \ldots, (m_k, q_k)\}$ for some $(m_1, q_1) < \ldots < (m_k, q_k)$, define:
            $$\delta'_s((q, i, 0) \defeq (m_j, (q_j, i, 0)) \text{ with } j = (i \text{ mod } k) + 1 \text{ and } j > 0.$$

            This implements the resolution of nondeterminism for outgoing messages from $M$.

            \item For every $(q, i, 1) \in Q'$, define:
                  $$\delta'_s((q, i, 1)) \defeq (\increment, (q, i, 0))$$

            This enforces that whenever the last bit is set to $1$, an agent will send an $\increment$ message exactly once.

          \end{itemize}
    \item $\delta'_r$ is defined as follows:
          \begin{itemize}
            \item For every $(q, m) \in Q \times M$ such that $\delta_s((q, m)) = \{q_1, \ldots, q_k\}$ for some
            $q_1 < \ldots < q_k$, and every $i, b$, define:
            $$\delta'_r((q,i,b), m) \defeq (q_{j}, i, b) \text{ with } j = (i \text{ mod } k) + 1$$

            This resolves the nondeterminism for incoming messages from $M$.

            \item Define for every $(q, i, b) \in Q'$:
            $$\delta'_r((q, i, b), \increment) \defeq \left(q, (i \text{ mod } n) + 1, 1\right)$$

            This implements the incrementation of the round counter after receiving an $\increment$ message. Moreover, $b$ is set to $1$, so that at least one $\increment$ will eventually be put back into the message pool.
          \end{itemize}
    \item $\iota'(a) \defeq (\iota(a),1, 1)$ for every $a \in \Sigma$.
    \item $o'((q, i, b)) = o(q)$ for every $(q, i, b) \in Q'$.
  \end{itemize}

$\PP'$ can be constructed in polynomial time. It remains to prove that $\PP$ and $\PP'$ compute identical predicates.
To this end, fix some input $X \in \pop{\Sigma}$ and let $b \in \{0, 1\}$. We must show that every fair execution of $\PP$ starting in $I(X)$ stabilizes to $b$ if and only if every fair of $\PP'$ starting in $I'(X)$ stabilizes to $b$.
Before we prove this equivalence, let us introduce some notation. For every $C \in \pop{Q'}$, we define the projection $\pi \left(C\right) \in \pop{Q}$ through $$\pi \left(C\right)(q) \defeq \sum_{(i, b) \in \{1, \ldots, n\} \times \{0, 1\}} C\left((q, i , b) \right).$$
For a given $C \in \pop{Q'}$, we  write $C(b = 1)$ as shorthand for
$$\sum_{(q, i, 1) \in Q'} C\left((q, i, b) \right).$$
We make the following observations that easily follow from the construction of $\PP'$:
\begin{enumerate}
  \item For every $C, C' \in \pop{Q'}$ such that $C(b=1) > 0 \lor C(\increment ) > 0$ and $C \trans{*} C'$, it must hold that $C'(b=1) > 0 \lor C'(\increment ) > 0$. \label{obs_one}
  \item For every $C \in \pop{Q'}$ and every $C' \in \pop{Q}$, we have:
        If $\pi(C) \trans{*} C'$ and $C(b=1) > 0 \lor C(\increment ) > 0$, then there exists some $C'' \in \pop{Q'}$ satisfying $\pi(C'') = C'$ and $C \trans{*} C'$. \label{obs_two}
  \item For every $C, C' \in \pop{Q'}$ such that $C \trans{} C'$, we have $\pi(C) \trans{*} \pi(C')$. \label{obs_three}
  \item For every $C' \in \pop{Q'}$ and every $C \in \pop{Q}$, $\{C'' \in \pop{Q'} \mid C' \trans{*} C'' \land \pi\left(C''\right) = C \}$ is finite.
  \label{obs_four}
\end{enumerate}

Let us now prove the equivalence.

$(\Leftarrow)$ Let $C'_0, C'_1,  C'_2, \ldots$ be a fair execution of $\PP'$ starting in $C'_0 = I'(X)$. By definition of $\iota'$, we have $C'_0(b=1) > 0$. By Observation~\ref{obs_one}, this gives
\begin{align}
\label{obs_five}
C'_i(b=1) > 0 \lor C'_i(\increment ) > 0 \text{ for every } i.
\end{align}
Now consider the sequence of configurations $C_0, C_1, C_2, \ldots = \pi(C'_0),\allowbreak  \pi(C'_1),\allowbreak  \pi(C'_2),\allowbreak  \ldots$, and let $\hat{i}_1 < \hat{i}_2 < \hat{i}_3 < \ldots$ be the maximal sequence of indices such that $C_{\hat{i}_1} = C_0$ and $C_{\hat{i}_1} \trans{} C_{\hat{i}_2} \trans{} \ldots$. By Observation~\ref{obs_three}, such a sequence of indices exists.
Moreover $\rho = C_{\hat{i}_1} C_{\hat{i}_2} C_{\hat{i}_3} \ldots$ is a fair execution of $\PP$:
By (\ref{obs_five}),  Observation~\ref{obs_two}, and Observation~\ref{obs_four}, for every $C$ that can be reached from  infinitely many configurations in $\rho$, there exists a configuration $C' \in \pi^{-1}\left(C\right)$ that can be reached from $C'_i$ for infinitely many $i$.
By fairness of $C'_0, C'_1, C'_2, \ldots$, we thus obtain that every configuration which can be reached infinitely often in $\rho$ is reached infinitely often. Hence $\rho$ is fair. Moreover, by definition of $o'$, $C'_0, C'_1,  C'_2, \ldots$ and $\rho$ converge to the same consensus. This proves the direction $(\Leftarrow)$.

$(\Rightarrow).$ The converse direction can be proven analogously.

 \end{proof}


We prove the claim made in the proof of Theorem \ref{thm:well-specification-transmission}.

\begin{proposition}
For every delayed-transmission protocol $\PP$ and every initial configuration $C$ of $\PP$, one can construct in polynomial time a delayed-transmission protocol $\PP'$ 
such that $\PP'$ computes the constant predicate $1$ if and only if $\PP$ stabilizes to $1$ for the single instance $C$.
\end{proposition}
\begin{proof}
  Fix $\PP=(Q, M, \delta_r, \delta_s, \Sigma, \iota, o)$ and $C$. Let $\C \defeq \{C' \in \pop{Q} \mid C' \leq C \}$.

  Each agent in $\PP'$ carries a state of $\PP$ and simulates interactions from $\PP$. Moreover, each agent carries a boolean flag $b \in \{0, 1\}$. The flag $b$ indicates that the initial configuration is $\geq C$. Initially, $b$ is set to $0$ for every agent. If $b$ is equal to $1$ and the agent carries some state $q \in Q$, its opinion is equal to $o(q)$. If $b=0$, the agent has opinion $1$. This ensures that the computation stabilizes to $1$ if the initial configuration is strictly smaller than $C$.

  In order to be able to detect whether the initial configuration is $\geq C$, the agents additionally store configuration from $\C$. Initially, if an agent carries the state $q$ from $Q$,  it stores the configuration $\multiset{q}$. Agents can transfer states from one stored configuration to another agent through message passing. If the initial configuration is equal to $C$ in $\PP$, by fairness a single agent will eventually store $C$ in the corresponding execution of $\PP'$, while all other agents store an empty configuration. When an agent stores $C$, it knows that the initial configuration must be $\geq C$, and in this case it is allowed to send a message
  that flips the flag $b$ of any receiving agent to $1$, and by fairness, eventually all agents have their flag $b$ set to $1$. If the initial configuration is $> C$, then at some point a state is transferred to an agent that already stores $C$. Such an agent assumes an error state, say $\top$, that maps to opinion $1$, and eats up all other states via message passing. This ensures that every execution starting in a configuration $>C$ stabilizes to $1$.

  Formally, the delayed-transmission protocol $\PP'=(Q', M', \delta'_r,\allowbreak  \delta'_s,\allowbreak  \Sigma,\allowbreak  \iota', o')$ is constructed as follows:
  \begin{itemize}
    \item $Q' \defeq \left(Q \times \C  \times \{0, 1 \} \right) \cup \{\top \}$
    \item $M' \defeq M \cup \{\top, \one \} \cup Q$
    \item $\delta'_r$ is defined as follows:
        \begin{itemize}
          \item For every transition $q \trans{m} r$ in $\delta$ and every $C \in \C$ and every $b \in \{0, 1\}$, add:
                $$(q, C, b) \trans{m} (r, C, b)$$

          \item For every $(q, C, b) \in Q \times \C \times \{0, 1\}$, add:
                $$(q, C, b) \trans{\one} (q, C, 1) $$

          \item For every $m \in Q$ and every $(q, C', b) \in Q \times \C \times \{0, 1\}$ s.t. $\left(C' + \multiset{q}\right) \in \C$, add:
                $$(q, C', b) \trans{m}  (q, C' + \multiset{q}, b)$$

          \item All remaining transitions to be defined transition to $\top$.
        \end{itemize}
    \item $\delta'_s$ is defined as follows:
      \begin{itemize}
        \item For every $(q, C', b) \in Q \times \C \times \{0, 1\}$ and every $q' \in Q$ such that $\multiset{q'} \leq C'$, add:
              $$\left(q, C', b\right) \trans{q'} \left(q, C' - \multiset{q'}, b \right)$$
        \item For every $q \in Q$, and every $b \in \{0,1 \}$, add:
            $$\left(q, C, b \right) \trans{\one} (q, C, b)$$
        \item For every transition $q \trans{m} r$ in $\delta_s$, and every $C' \in \C$, add:
              $$\left(q, C', 1 \right) \trans{m} \left(r, C', 1 \right)$$
          \item Further add:
                $$ \top \trans{\top} \top $$
          \item $\iota'(a) \defeq (\iota(a), \multiset{\iota(a)}, 0) $ for every $a \in \Sigma$.
          \item Set $o'(q, C, 1) \defeq O(q)$ for every $(q, C) \in Q \times \C$, and $o'(\vec{q}) \defeq 1$ for every other $\vec{q}$.
      \end{itemize}
  \end{itemize}
   \end{proof}


We prove the claim made in the proof of Theorem \ref{thm:well-specification-immediate-transmission}.

\begin{proposition}
Let $\NN = (Q, T)$ be a $\pm1$-VASS and let $r_0, r \in Q$.
It is possible to construct in polynomial time an immediate-transmission protocol $\PP$ that computes constant $1$ if and only if  $(r_0, \mathbf{0}) \trans{*} (r, \mathbf{0})$ does \emph{not} hold.
\end{proposition}
\begin{proof}
Let the dimension of $\NN=(Q, T)$ be $k$.
Like in the proof of Proposition~\ref{prop:pm-vass-to-protocol}, we represent the control state of $\NN$ in a single agent. The remaining agents either represent a reservoir of tokens by assuming states of the form $\free_i$ or $\token_i$ for every vector component $1 \leq i \leq k$, or they are of the form $t, \vect{t}$ for any given $t \in T$, or in some additional helper state.
A configuration $q, \vec{v}$ of $\NN$ is represented in a configuration $C$ of $\PP$ satisfying
\begin{align*}
  C(q) & = 1, \\
  C(q') & = 0 && \text{ for every } q' \in Q \setminus \{q\}, \\
  C(\token_i) & = v_i && \text{ for every } 1 \leq i \leq k.
\end{align*}

The states $\free_i$, $\vect{t}, t$ for every $t \in T$, and other helper states, may be populated by arbitrarily many agents. When the control agent interacts with an agent of the form $\vect{t}$, a transition of $\NN$ is simulated.
For example, a transition $t \in T$ of the form $q \trans{i++} q'$ is implemented in $\PP$ by a sequence of two transitions, namely  
$(t, q)  \trans{} (\vect{t}, t)$ followed by  $(t, \free_i)  \trans{} (q', \token_i)$
Similarly, a transition $t \in T$ of the form $q \trans{i--} q'$ is implemented in $\PP$ by the sequence consisting of
$(t, q)  \trans{} (\vect{t}, t)$ followed by $(t, \token_i) \trans{} (q', \free_i)$.
Thus, incrementation at position $i$  is implemented by turning $\free_i$ into $\token_i$, and symmetrically, decrementation is implemented by transforming $\token_i$ into $\free_i$.
Initially, no agent is in a state of the form $\token_i$, which reflects the fact that the initial vector of $\NN$ in the reachability query equals $\zero$.

Moreover, there are states $\final$ and $\compl{\final}$. When the agent representing the control state of $\NN$ assumes $r$, it can non-deterministically guess that the current vector is $\zero$, and signal this guess via transitioning to state $\final$ through the step $\compl{\final}, r \trans{} \compl{\final}, \final$. The state $\final$ is the only state that maps to false. If the guess was right, then the agent permanently remains in state $\final$, and thus the protocol does not compute constant $1$. If the guess was wrong, then by fairness the agent eventually meets some agent in state $\token_i$ for some $i$, and then turns into some error state, say $\top$, that maps to true and that converts all other states to $\top$, thus ensuring that the protocol eventually stabilizes to $1$.

Formally, we define $\PP = (Q^\PP, \delta^\PP, I^\PP, O^\PP)$ as follows:

\begin{itemize}
  \item We add the following states to $Q^\PP$:
        \begin{itemize}
          \item For every $1 \leq i \leq k$, add states $\free_i$ and $\token_i$.

          \item Add an ``error'' state $\top$.

          \item Add ``final'' states $\final, \compl{\final}$.

          \item For every state $q \in Q$, add $q$ to $Q^\PP$.

          \item For every transition $t \in T$, add
              $t$ and $\vect{t}$.

        \end{itemize}
  \item We define $\delta^\PP(x, y) = (\delta_1(x), \delta_2(x, y))$ by adding the following transitions:
      \begin{itemize}
        \item For every $t \colon q_1 \trans{} q_2 \in T$, add:
              $$\vect{t}, q_1 \trans{} \vect{t}, t.$$

        \item For every $t \colon q_1 \trans{i++} q_2$,
         add:
                  $$t, \free_i \trans{} q_2, \token_i.$$
            \item For every $t \colon q_1 \trans{i--} q_2$, add:
                  $$t, \token_i \trans{} q_2, \free_i .$$

            \item Add:
                $$\compl{\final}, r \trans{} \compl{\final}, \final$$

            \item For every $x \in Q^\PP$, add:
                  $$\top, x \trans{} \top, \top.$$
                  This ensures that the $\top$ eats up every other state.

            \item For every $t \colon q_1 \trans{i++} q_2$, and every $x \neq \free_i$, add:
                  $$t, x \trans{} q_2, \top.$$

            \item For every $t \colon q_1 \trans{i--} q_2$, and every $x \neq \token_i$, add:
                  $$t, x \trans{} q_2, \top.$$

            \item For every $1 \leq i \leq k$, add:
                  $$\token_i, \final \trans{} \token_i, \top.$$
                  This ensures that an agent only remains in $\final$ if the current marking is $\mathbf{0}$, otherwise everyone is sent to $\top$.

            \item For every $x, y \in Q \cup T \cup \{\final\}$, set:
                  $$\delta_2(x, y) = \top.$$
                  This ensures that at most one agent is in a control state of $\N$, otherwise everyone is sent to $\top$.

            \item In all remaining cases, set $\delta_1(x) = x$ and $\delta_2(x, y) = y$
        \end{itemize}
    \item $I \defeq \{r_0 \} \cup \{\free_i \mid 1 \leq i \leq k\}$

    \item $O(\final) \defeq 0$ and $O(x) = 1$ for every $x \neq \final$.
\end{itemize}
 \end{proof}

%% file: app-decidability.tex
\section{Appendix for Section~\ref{sec:decidability-probabilistic}}

\PropositionMoreThanHalf*
\begin{proof}
        Recall the given protocol with states $\{q_0,q_1,q_2\}$; 
        output function given by  $O(q_0)=O(q_1)=0$ and $O(q_2) = 1$;
        messages $\{a,b\}$; 
        transitions
        $$\begin{array}{lllll}
        q_0\trans{a+}q_0 & \quad &  q_1\trans{b+}q_0 & \quad &  q_2\trans{b+}q_1 \\
        q_0\trans{a-}q_0   & &  q_1\trans{a-}q_1 & &  q_2\trans{a-}q_2 \\
        q_0\trans{b-}q_1 & &  q_1\trans{b-}q_2 & &  q_2\trans{b-}q_2
        \end{array}$$
        and the input configuration $\multiset{q_0}$.

        In each configuration of each execution 
        the sum of the index of the state and 
        the number of messages of type $b$
        is equal to $2$. This protocol does not f-compute any value on $\multiset{q_0}$
        because the configuration with no messages and the agent in the state
        $q_2$ is reachable from each configuration in the execution, as well as 
        the configuration with $2$ messages of type $b$ and the agent in the 
        state $q_0$. These two configurations occur infinitely often
        in each fair execution and have different output values.

        The proof that an execution from $\multiset{q_0}$ 
        converges with probability $1$ if $p>\frac{1}{2}$ is based on the following 
        observations.
        \begin{itemize}
        \item The number of messages changes independently of the configuration
        change, so it is a biased random walk with linear growth.
        \item The state $q_0$ can always be reached with probability at least $1/4$, and 
        so it is reached infinitely many times.
        \item Going from $q_0$ to $q_2$ requires receiving two $b$s 
        without sending in-between.
        \item The probability to receive two $b$s is proportional to $1/n^2$, where 
        $n$ is the number of messages. Since the series $\sum_{i=1}^\infty 1/n^2$
        converges, so with probability $1$ the state $q_2$ 
        is only observed a finite number of times.
        \end{itemize} 

        Consider a step $C \rightarrow C'$.  If $C$ has no messages in transit, then the number or messages increases by $1$;
        otherwise there is probability $p>\frac{1}{2}$ to increase
        the number of messages by $1$ and probability $1-p$ 
        to decrease the number of messages.
        This is a biased random walk.
        Let $X_i$ be the random variable equal to the number of messages
        at the $i$-th step. The expected value of $X_i$ is $(2p-1)i$ and the standard
        deviation grows proportionally to $\sqrt{i}$, and so in particular
        $\lim_{c\to+\infty}\text{Pr}[\forall i>c: X_i > (p-\frac{1}{2})i]=1$.
        For any given configuration the probability of reaching 
        a configuration with state $q_0$ is $1$ (it is enough to 
        send a message two times in a row which has probability larger 
        than $\frac{1}{4}$ at each step), therefore
        state $q_0$ occurs infinitely often with probability $1$.
        To reach state $q_2$ from
        state $q_1$ before reaching $q_0$ the agent needs to
        receive messages without transmitting until it receives 
        the only message of type $b$.
        The probability of reaching $q_2$ from $q_1$
        before either returning to $q_0$ or getting below $k$ 
        messages is less than
        $\sum_{j=1}^{\infty}(1-p)^j\frac{1}{k}<
        \sum_{j=1}^{\infty}\left(\frac{1}{2}\right)^j\frac{1}{k}=\frac{1}{k}$.
        Note that probability of the transition from $q_0$ to 
        $q_1$ with at least $k$ messages in transit is at most 
        $(1-p)\frac{2}{k}<\frac{1}{k}$.
        Therefore the probability of the agent starting
        at $q_0$ and reaching $q_2$ before
        either returning to $q_2$ or having fewer than $k$ 
        messages is at most $\frac{1}{k^2}$.

        Let $N_{q_0\to{}q_2}^j$ be the random variable equal to $1$ if
        $q_2$ is visited after the $j$-th visit to $q_0$,
        before $q_0$  is visited again, and before the number of messages in transit
        goes below $(p-\frac{1}{2})j$ (and $0$ otherwise).
        Let $N_{q_0\to{}q_2}$ be the sum $\sum_{j=1}^{\infty}N_{q_0\to{}q_2}^j$.
        We have shown that $E\left(N_{q_0\to{}q_2}^j\right) \in O(\frac{1}{((p-\frac{1}{2})j)^2})$
        and therefore $E\left(N_{q_0\to{}q_2}\right) = c \cdot \sum_{j=0}^{\infty}\frac{1}{((p-\frac{1}{2})j)^2}$ for some constant $c$, which is finite.

        Consider executions that reach $q_2$ after at least $N$
        different returns to $q_0$.
        Such an execution must either have the number of messages go below 
        $(p-\frac{1}{2})i$ at the moment $i>\frac{N}{2}$,
        or have the value $N_{q_0\to{}q_2}$
        to be at least $\frac{N}{2}$.
        The probability of either option tends to zero when
        $N$ grows. Therefore $q_2$ occurs only a finite 
        number of times with probability $1$, and so the execution converges to $0$. So the
        protocol computes the value $0$ on input $\multiset{q_0}$.
 \end{proof}